\documentclass[pra,floatfix,twocolumn,showpacs,amsmath,amssymb]{revtex4}
\usepackage{graphicx}
\usepackage{dcolumn}
\usepackage{bm}
\usepackage{color}
\usepackage{ulem}

\begin{document}
\title{
Enhancements of the 3/2 and 5/2 frequencies of de Haas-van Alphen oscillations 
near the Lifshitz transition in the two-dimensional compensated metal with overtilted Dirac cones
}

\author
{Keita Kishigi}
\affiliation{Faculty of Education, Kumamoto University, Kurokami 2-40-1, 
Kumamoto, 860-8555, Japan}

\author{Yasumasa Hasegawa}
\affiliation{Department of Material Science, 
Graduate School of Material Science, 
University of Hyogo, Hyogo, 678-1297, Japan}

\date{\today}

\begin{abstract}
We study the de Haas-van Alphen (dHvA) oscillations in 
the two-dimensional compensated metal with overtilted Dirac cones near the Lifshitz transition. We employ the tight-binding model of $\alpha$-(BEDT-TTF)$_2$I$_3$, in which the massless Dirac fermions 
are realized. 
When a uniaxial pressure $(P)$ along 
the $y$-axis is applied above $P\simeq 0.2$ kbar in $\alpha$-(BEDT-TTF)$_2$I$_3$, one electron pocket, which encloses the overtilted Dirac points, is changed to two electron pockets, i.e., Lifshitz transition happens,  while a hole pocket does not change a topology. 
We show that the Fourier components corresponding to the $3/2$ and $5/2$ areas of the hole pocket are anomalously enhanced in the region of pressure where the Lifshitz transition occurs. 
This phenomenon will be observed in the two-dimensional overtilted Dirac fermions near the Lifshitz transition. 
\end{abstract}

\date{\today}

\maketitle

\section{Introduction}

The de Haas van Alphen (dHvA) oscillations are the
phenomena that the magnetization in metals oscillates
periodically as a function of the inverse of the magnetic
field ($1/H$) at low temperatures\cite{shoenberg}. 
Lifshitz and Kosevich\cite{LK,mineev} have derived semiclassically the standard formula called the LK formula, which is Eq. (\ref{LK_0}) in Appendix \ref{LKformula_dHvA}. 
The fundamental frequency of the dHvA oscillations corresponds to the extremal cross-sectional area of
the Fermi surface perpendicular to the applied field.


When the distance between two parts of a Fermi surface is narrow and the energy barrier between them is not high, 
electrons can tunnel from one part of the Fermi surface to another
in the magnetic field, which is called the magnetic breakdown. In that case, 
a new period of the dHvA oscillations appears, which corresponds to the area of the effective closed orbit of electrons. The dHvA oscillations in the system with 
magnetic breakdown have been studied in the semiclassical network model\cite{Pippard62,Falicov66}, in which the probability amplitudes of
the tunneling are introduced into the LK formula as parameters. 
The phenomenological probability amplitudes, however, are not necessary to
be introduced into the quantum-mechanical treatment in the tight-binding model, as we will show in this paper.

When the system has a three-dimensional Fermi surface (for example, a sphere),
the kinetic energy perpendicular to the magnetic field is quantized as Landau levels and
the kinetic energy parallel to the magnetic field is not affected. In that case the chemical potential
changes little as a function of the magnetic field. Then we can use the LK formula neglecting the
magnetic-field dependence of the chemical potential.
In two-dimensional systems or quasi-two-dimensional systems
with the interlayer coupling smaller than the spacings of the Landau levels,
however, the magnetic-field dependence of the chemical potential cannot be neglected
in general,
and the LK formula assuming the fixed chemical potential is not justified.
In the simple systems with two-dimensional free electrons,
the saw-tooth pattern of the dHvA oscillations is inverted depending on whether we fix the chemical potential
or electron number, although the frequency of the dHvA oscillations is the same\cite{shoenberg,champel,grigo}. 
When the quasi-two dimensional system has two or more Fermi surfaces
(electron pocket(s) and hole pocket as studied in this paper, or two-dimensional Fermi pocket and
quasi-one-dimensional (open) Fermi surface), the oscillation of the chemical potential
as a function of the inverse magnetic field causes the ``\textit{forbidden}'' frequencies
such as $f_{\beta} - f_{\alpha}$
\cite{Meyer1995,harrison,Uji1997,Steep1999,Honold,Audouard2005,Audouard2013}. 
The forbidden frequencies in the dHvA oscillations 
have been studied theoretically in the systems with isolated Fermi surfaces (without the magnetic breakdown) \cite{nakano,alex1996,alex2001,champel2002,KH,its2003,gvoz2003} and with Fermi surfaces connected by the magnetic breakdown
\cite{machida,kishigi_1995,harrison,kishigi1997,sandu,so,fortin1998,gvoz2004}.

We have studied the energy of the quasi-two-dimensional organic conductor, $\alpha$-(BEDT-TTF)$_2$I$_3$\cite{review,review2, Katayama2006}, in the external magnetic field at various uniaxial pressure in the recent paper\cite{KH2017}. 
We found that the magnetic-field dependence of the Landau levels
is proportional to $H^{4/5}$ at the pressure when the Dirac cone  is tilted critically (three-quarter Dirac points), i.e., 
the linear term disappears and the quadratic term becomes dominant in one direction, while the linear term is finite in other three directions\cite{KH2017,HK2019}. 
The Dirac cone is under-tilted when the pressure is higher than 
the critical pressure (2.3 kbar) and 
it is over-tilted when the pressure is lower.
Even at the critical pressure where the three-quarter Dirac 
points are realized, 
the Fermi energy is higher than the energy at the three-quarter 
Dirac point, since the three-quarter Dirac point is not the 
local maximum of the lower band due to the positive quadratic 
term. 
As a result, the system is a compensated metal with one hole pocket around the maximum of the lower band and two electron pockets surrounding two Dirac points below 3.0 kbar. 
It is known that the phase ($\gamma$)\cite{Igor2004PRL,Igor2011,Sharapov} in dHvA oscillations is $1/2$ in the topologically trivial parabolic band
(hole pocket in our case) but
$\gamma=0$ due to the Berry phase $\pi$ in the Dirac fermions (electron pockets in our case).

The dHvA oscillations in the two-dimensional
compensated metallic state have been studied semiclassically\cite{fortin2008,fortin2009} and 
quantum-mechanically\cite{KM1996,KH2016} in the systems such as (TMTSF)$_2$NO$_3$\cite{pouget,fisdw_no3,kang_2009}, where there are one hole pocket and one electron pocket with the same area. In these studies, both of 
the free hole pocket and the free electron pocket are topologically trivial. 
The compensated metallic state of $\alpha$-(BEDT-TTF)$_2$I$_3$ is a unique system, because there are coexisting trivial hole pocket and one or two electron pocket(s) with Dirac-fermion property near the Lifshitz transition\cite{Lifshitz,volv} [The similar situation will be realized in the systems with over-tilted Dirac points (type II Weyl semimetal)\cite{Solu2015,Yu2016}]. 
Therefore, it is interesting to study the dHvA oscillations near the Lifshitz transition in $\alpha$-(BEDT-TTF)$_2$I$_3$. 
The semiclassical picture will be difficult to be applied in that system, since the 
density of states diverges at the Fermi surface when the Lifshitz transition happens\cite{Its}. 

In this paper we study the dHvA oscillations at and near the Lifshitz transition
in the two-dimensional compensated metal with trivial hole pocket and
  two or one electron pocket(s) surrounding the over-tilted Dirac points.
We use the tight-binding model for $\alpha$-(BEDT-TTF)$_2$I$_3$ with pressure-dependent
transfer integrals.
The effect of the magnetic field is taken into account by the Peierls substitution and we do not
consider the semiclassical network model of the magnetic breakdown.

\begin{figure}[bt]
\begin{flushleft} \hspace{0.5cm}(a) \end{flushleft}\vspace{-0.0cm}
\includegraphics[width=0.52\textwidth]{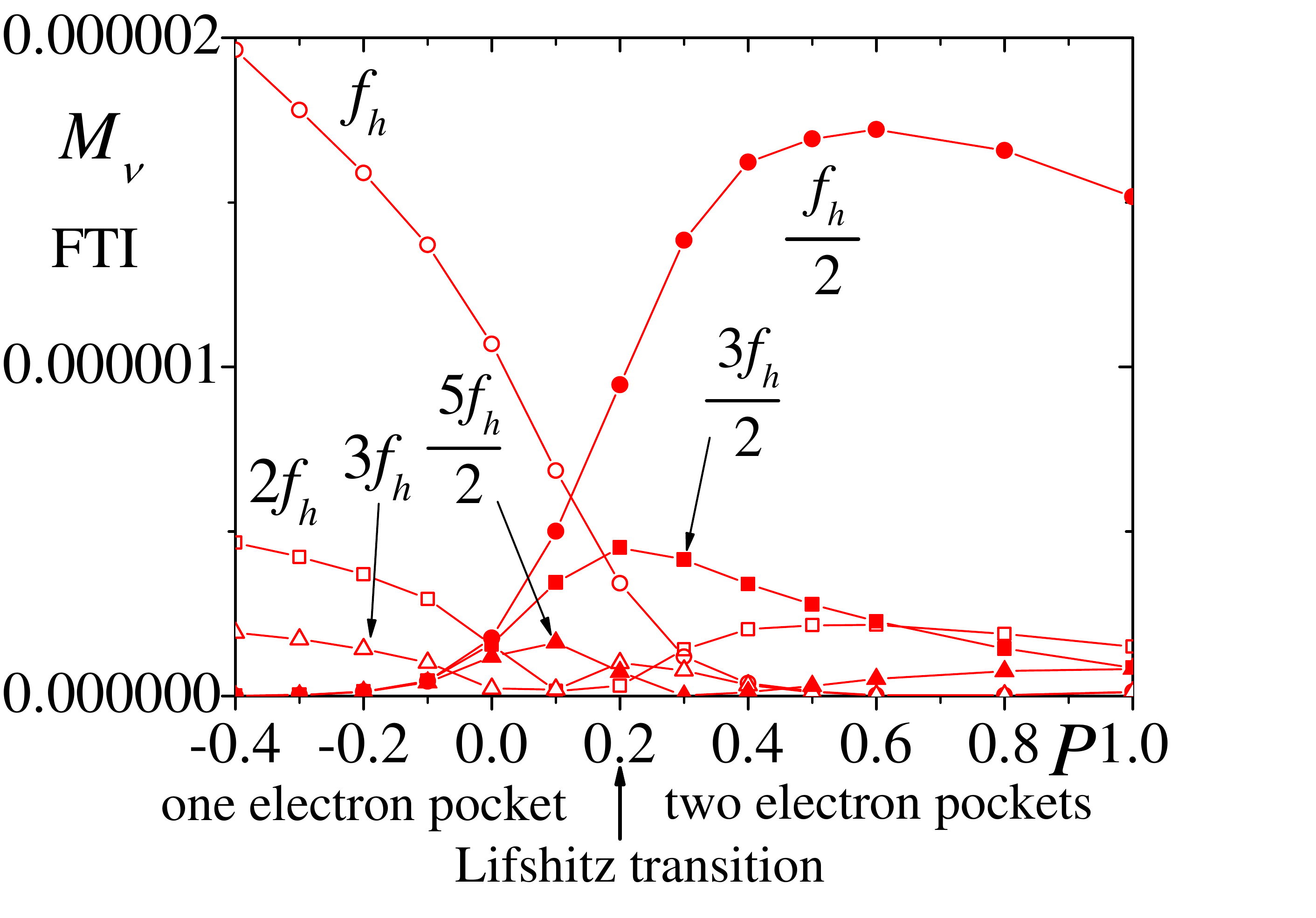}\vspace{-0.5cm}
\begin{flushleft} \hspace{0.5cm}(b) \end{flushleft}\vspace{-0.0cm}
\includegraphics[width=0.52\textwidth]{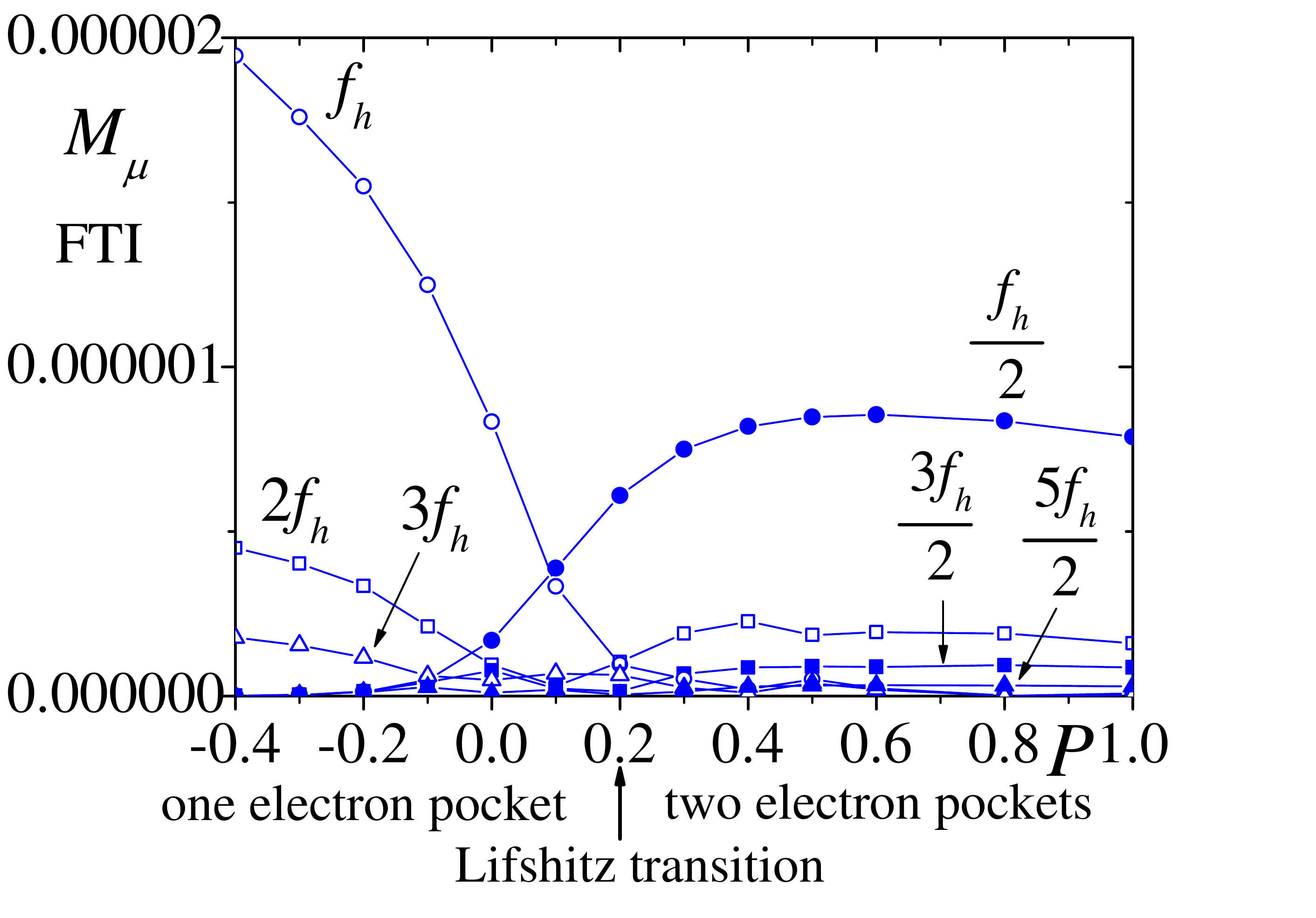}\vspace{-0.3cm}
\caption{
The $P$-dependences of the FTIs, where (a) and (b) are for $M_{\nu}$ in Fig. \ref{fig28_a} and $M_{\mu}$ in Fig. \ref{fig28_b}, respectively. 
The Lifshitz transition occurs at $P=0.2$. There exist one electron pocket and one hole pocket at $P<0.2$, while there exist two electron pockets and one hole pocket at $P>0.2$.
Dirac cones are overtilted at $P<2.3$.
}
\label{fig30}
\end{figure}

A main result in this study is shown in Fig. \ref{fig30}. 
The Fourier transform intensities (FTIs) for the 3/2 and 5/2 times of the 
area of a hole pocket are enhanced in the region of the pressure, where 
the Lifshitz transition occurs, when we take the condition that the electron number is fixed, as shown in Fig. \ref{fig30} (a). When the chemical potential is fixed, these enhancements do not appear, as shown in Fig. \ref{fig30} (b).

\section{Tight-binding model for $\alpha$-(BEDT-TTF)$_2$I$_3$; compensated metal and the Lifshitz transition}

\begin{figure}[bt]
\begin{flushleft} \hspace{-0.5cm}(a) \end{flushleft}\vspace{-0.0cm}
\includegraphics[width=0.33\textwidth]{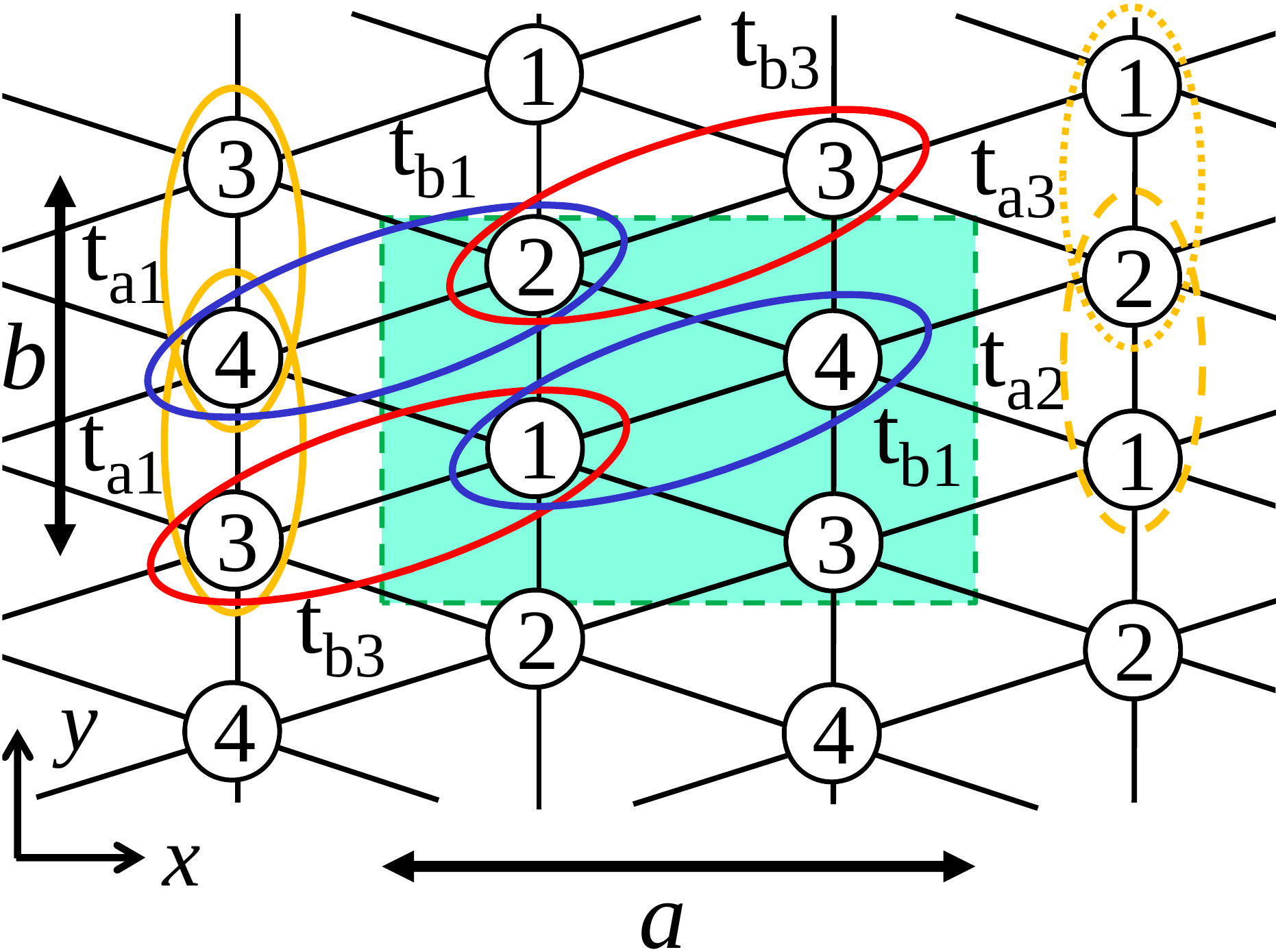}
\begin{flushleft} \hspace{0.0cm}(b) \end{flushleft}\vspace{0.1cm}
\includegraphics[width=0.31\textwidth]{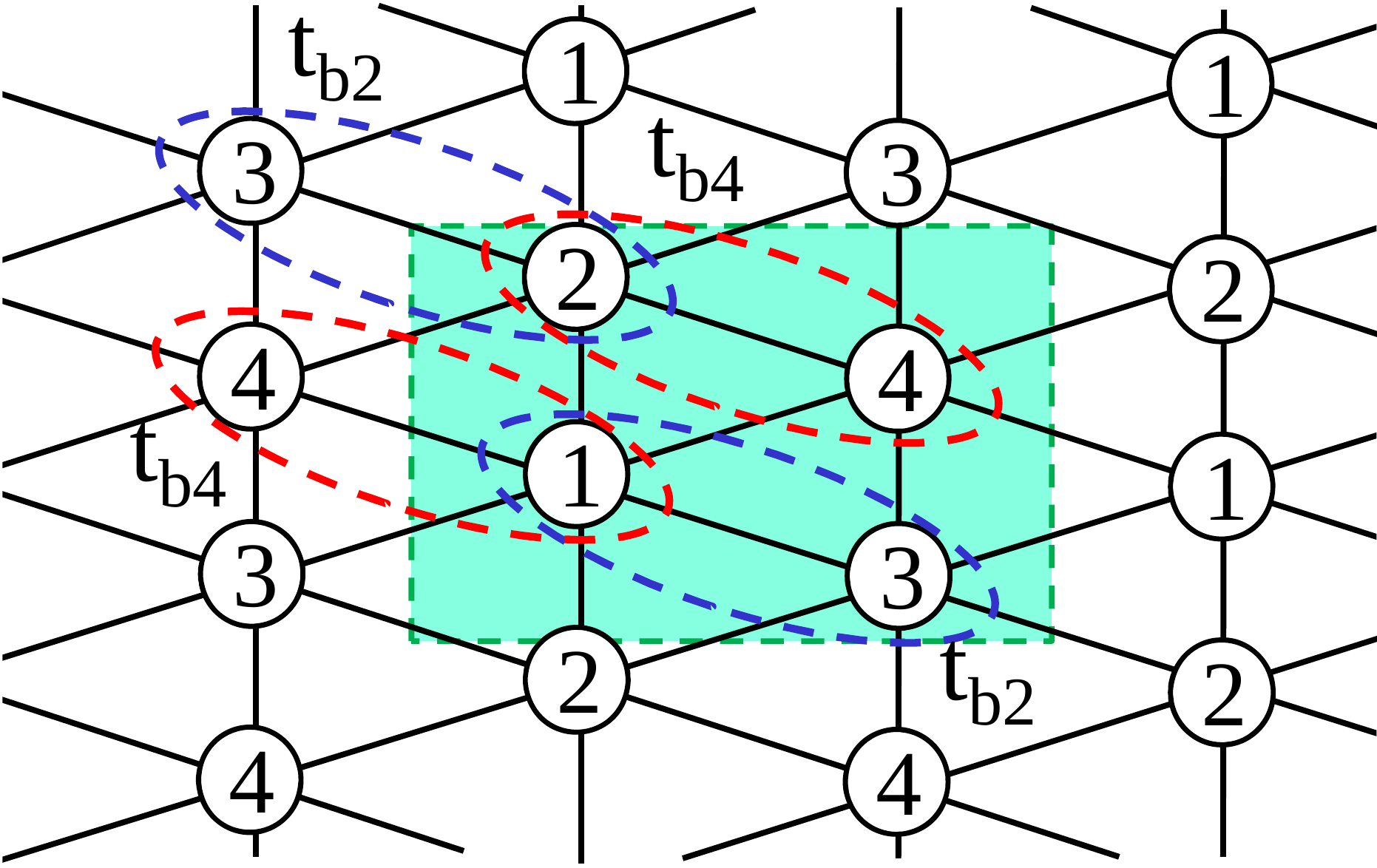}
\vspace{0.3cm}
\caption{
(Color online) The crystal lattice and the transfer integrals in the tight-binding model in $\alpha$-(BEDT-TTF)$_2$I$_3$.  
The transfer integrals
 ($t_{\mathrm{a1}}$,  $t_{\mathrm{a2}}$, $t_{\mathrm{a3}}$, $t_{\mathrm{b1}}$, 
$t_{\mathrm{b2}}$, $t_{\mathrm{b3}}$ and $t_{\mathrm{b4}}$) are shown as ovals.
The unit cell is the rectangle in (a) and (b), where 
$a$ and $b$ are the lattice constants along $x$-axis and $y$-axis, respectively. 
}
\label{fig1}
\end{figure}

$\alpha$-(BEDT-TTF)$_2$I$_3$\cite{review,review2,Kajita2014} is known as 
one of quasi-two-dimensional organic superconductors. 
Due to the smallness of the interlayer coupling, we ignore the three-dimensionality of $\alpha$-(BEDT-TTF)$_2$I$_3$. 
Four BEDT-TTF molecules exist in the unit cell, as shown in Fig. \ref{fig1}. 
The four energy bands are constructed by the highest occupied molecular orbits (HOMO) of BEDT-TTF molecules. 
Since one electron is removed from two BEDT-TTF molecules, the electron bands are 3/4-filled. 
The lower two bands are completely filled by electrons. 
In this paper, the tight-binding model for the HOMO having  
the transfer integrals between neighboring sites is used, which are shown in Fig.~\ref{fig1}. 
The tight-binding model has been explained in the previous study\cite{KH2017}. 



Although $\alpha$-(BEDT-TTF)$_2$I$_3$ exhibits a metal-insulator transition due to the charge ordering at low temperatures and low pressures\cite{KF1995,Seo2000,takano2001,Woj2003}, we employ the model without interactions in order to study the effects of the Lifshitz transition. 
The charge ordering phase in $\alpha$-(BEDT-TTF)$_2$I$_3$ has been observed at $P\lesssim 3.0$ kbar under the uniaxial pressure along $x$-axis and $P\lesssim 5.0$ kbar along $y$-axis in the conductivity\cite{Tajima2002} and under the hydrostatic pressure at $P\lesssim 17$ kbar from the magneto conductivity\cite{Tajima2013} and at $P\lesssim 11-12$ kbar from the optical investigation\cite{Beyer} and conductivity\cite{Dong}. 
Even in the field induced charge-density-wave state, magnetic quantum oscillations have been observed in the similar system of $\alpha$-(BEDT-TTF)$_2$MHg(SCN)$_4$ with 
$M$=K and Tl\cite{Kart2011}. 

Recently, massless Dirac fermions have 
been observed\cite{kajita1992,tajima2000,Hirata2011,Konoike2012,Osada2008} in $\alpha$-(BEDT-TTF)$_2$I$_3$ under the pressure. The appearance of the massless Dirac fermions has been  theoretically shown by Katayama, Kobayashi and Suzumura\cite{Katayama2006}. They have used the interpolation formula\cite{Kobayashi2004}
for the transfer integrals based on the extended H\"uckel method\cite{Mori1984,Kondo2005}. These are given by 
\begin{equation}
\begin{split}
t_{a1} &= -0.028  (1.0 + 0.089P), \\
t_{a2} &= -0.048  (1.0 + 0.167P), \\
t_{a3} &=   0.020  (1.0-0.025P), \\ 
t_{b1} &=   0.123,               \\
t_{b2} &=   0.140  (1.0 + 0.011P),  \\
t_{b3} &=   0.062  (1.0 + 0.032P),  \\
t_{b4} &=   0.025, 
\end{split} 
\label{eqhoppings}
\end{equation}
where $P$ is the uniaxial strain along the $y$-axis. 
Hereafter, we employ eV and kbar as the units of transfer integrals and the pressure, respectively. In this study, we use Eq. (\ref{eqhoppings}). 


When $P\geq 3.0$, the Fermi energy is located at the Dirac points. 
The similar band structure has been shown in the first-principle band calculations by Kino and Miyazaki\cite{Kino,Kino2009} and Alemany, Jean-Pouget and Canadell\cite{Alemany2012}. 


\begin{figure}[bt]
\begin{center}
\begin{flushleft} \hspace{0.5cm}(a) \end{flushleft}\vspace{-0.8cm} 
\includegraphics[width=0.40\textwidth]{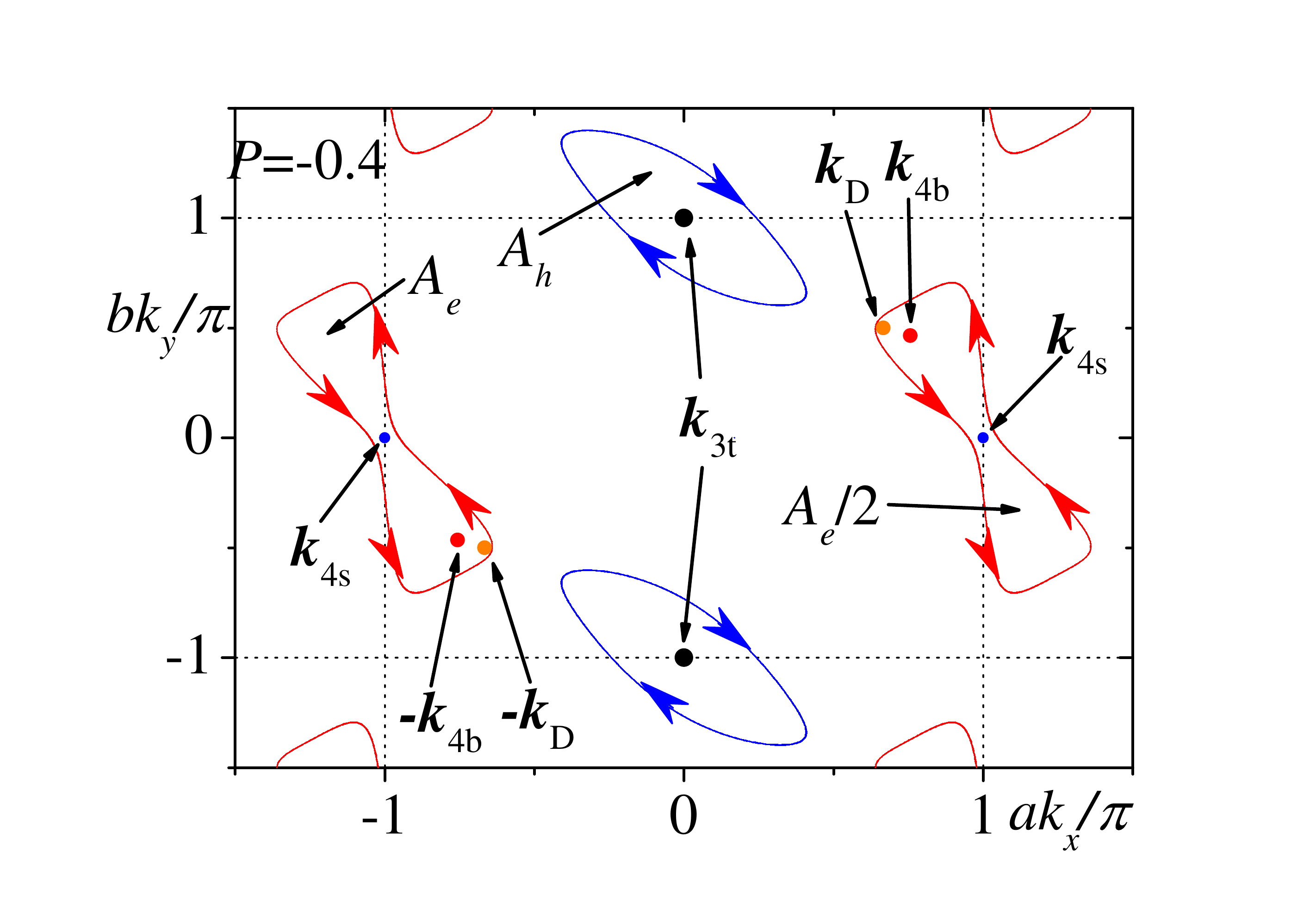}\vspace{-1.0cm}
\begin{flushleft} \hspace{0.5cm}(b) \end{flushleft}\vspace{-0.8cm} 
\includegraphics[width=0.40\textwidth]{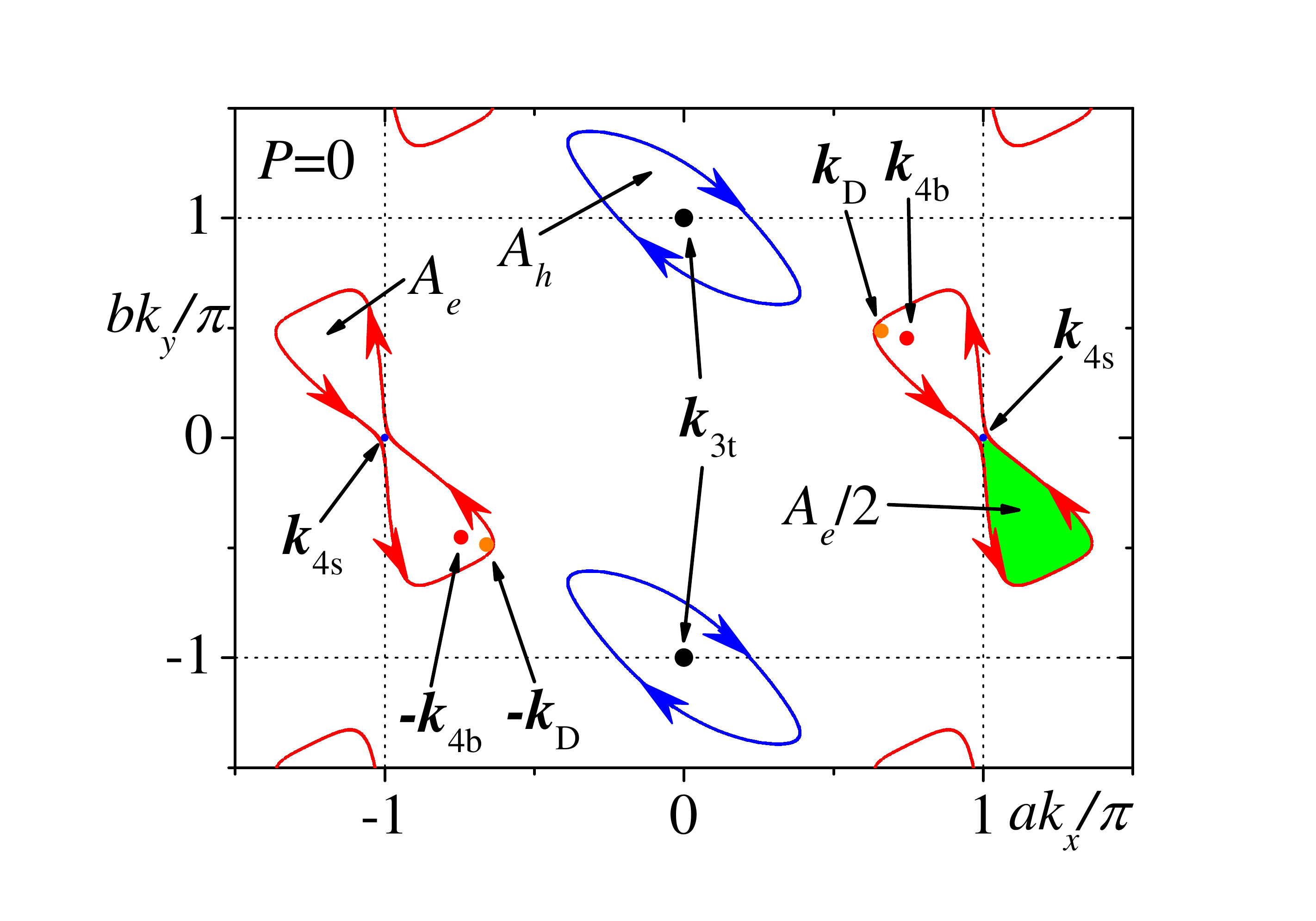}\vspace{-1.0cm}
\begin{flushleft} \hspace{0.5cm}(c) \end{flushleft}\vspace{-0.8cm}
\includegraphics[width=0.40\textwidth]{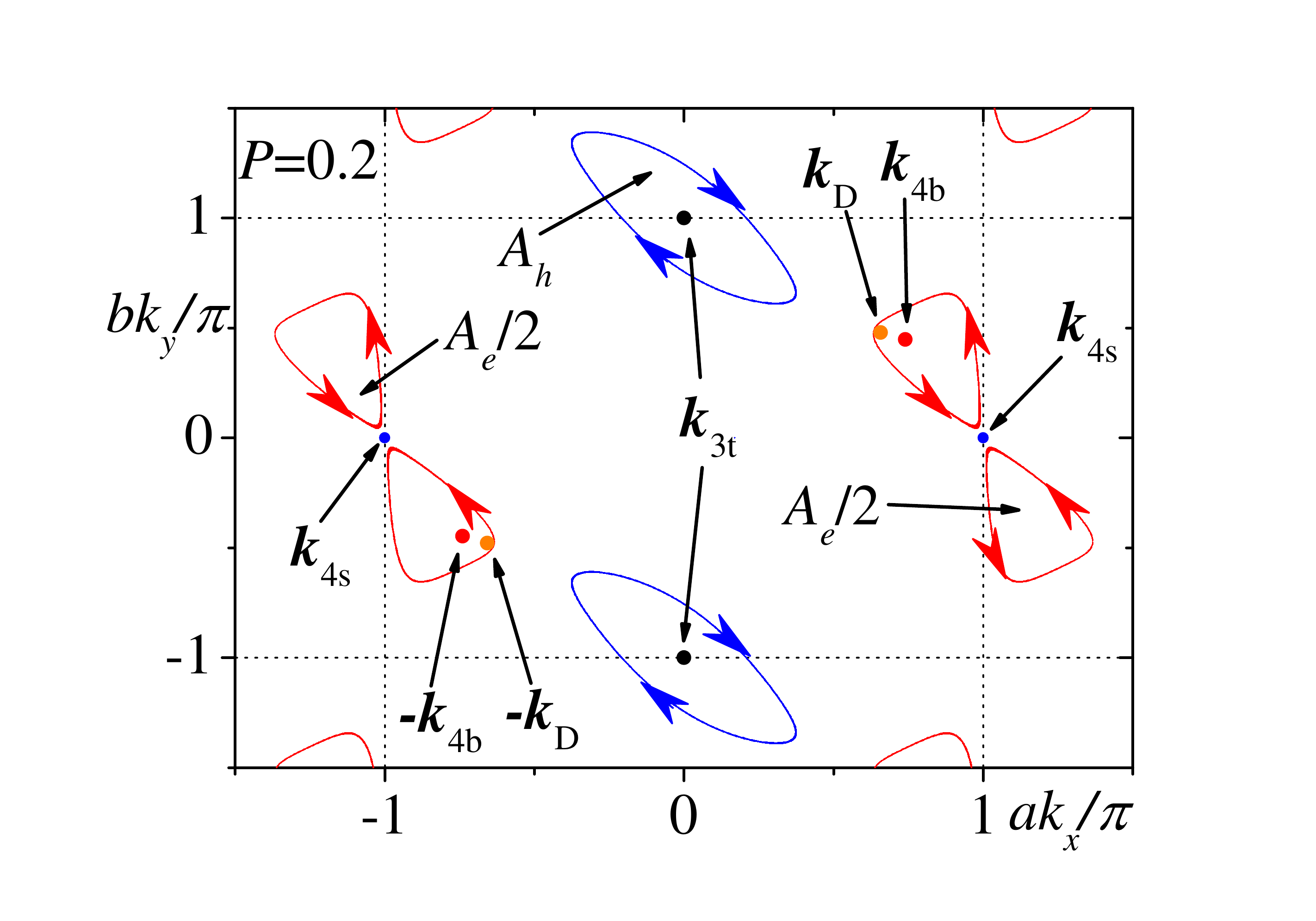}\vspace{-1.0cm}
\begin{flushleft} \hspace{0.5cm}(d) \end{flushleft}\vspace{-0.8cm}
\includegraphics[width=0.40\textwidth]{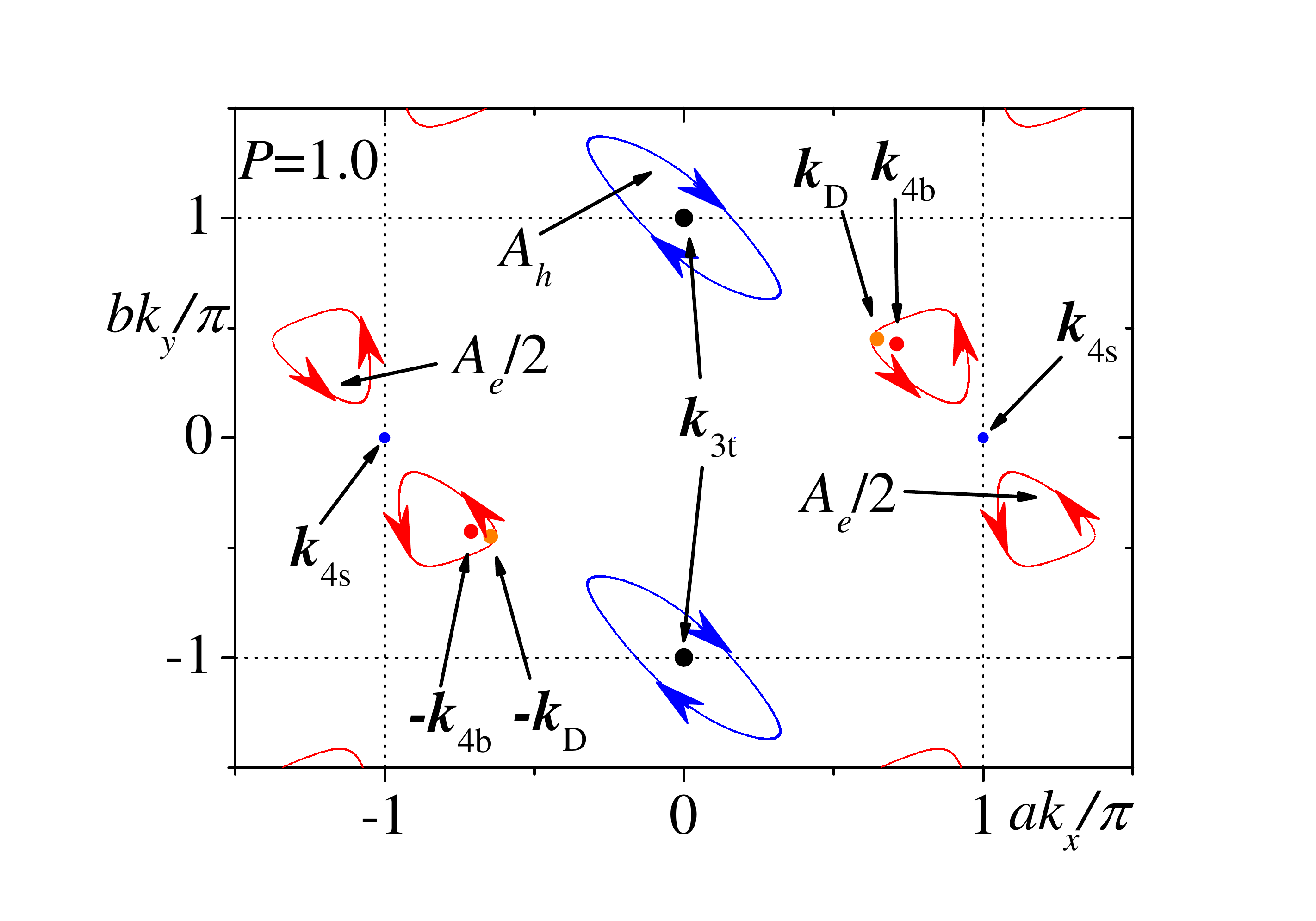}\vspace{-0.9cm}
\end{center}
\caption{
 (Color online) 
Fermi surfaces at $P=-0.4$ (a), 
$P=0$ (b), $P=0.2$ (c) and $P=1.0$ (d) in the extended zone, 
where ${\bf k}_{\mathrm{3t}}$ is the wave vector for the top of the third band, $\pm{\bf k}_{\rm D}$ are Dirac points, $\pm{\bf k}_{\rm 4b}$ are the points with the bottom energy in the fourth band, 
${\bf k}_{\rm 4s}$ is the saddle point of the fourth band. Arrows indicate the direction of the orbital motion for electrons in the magnetic field in the semiclassical picture. In (b), a green area is a half of $A_e$. We obtain $|A_h|/A_{\rm BZ} \simeq 0.0797$ at $P=-0.4$ (a), 
$A_e/A_{\rm BZ}=|A_h|/A_{\rm BZ}\simeq 0.0714$ at $P=0$ (b), $|A_h|/A_{\rm BZ}\simeq 0.0666$ at $P=0.2$ (c) and 
$|A_h|/A_{\rm BZ}\simeq 0.0480$ at $P=1.0$ (d), where $A_{\mathrm{BZ}}$ is the area of the Brillouin zone, $|A_{{h}}|$ is the area of a hole pocket and $A_{{e}}$ is the area of an electron pocket [(a) and (b)] or the sum of the area of two electron pockets [(c) and (d)]. Note that the sign of the area of the hole pocket is opposite of that of the electron pocket\cite{Falicov66,fortin2008}. 
}
\label{fig8_N}
\end{figure}

At $P<3.0$, the top of the third band from the bottom band becomes a higher energy than that at the Dirac points. 
In this case there are one hole pocket and one or two electron pockets. 
For example, the Fermi surfaces at $P=-0.4, 0, 0.2$ and 1.0 are shown in Fig. \ref{fig8_N}, where the use of the negative $P$ is allowed in Eq. (\ref{eqhoppings}). We extrapolate the pressure to a negative value. The area of the hole pocket is the same as the sum of the areas of each electron pocket, i.e., the system is a compensated metal. The hole pocket has a parabolic dispersion around $\mathbf{k}_{\mathrm{3t}}$. The saddle point exists near the Dirac points in the fourth band at $\mathbf{k}_{\mathrm{4s}}=(\pi/a,0)$.
This point is the time reversal invariant momentum (TRIM) which may become one of a local maximum point, a local minimum point, an inflection point, or a saddle point. This is explained in Appendix \ref{TRIM}.




\begin{figure}[bt]
\begin{center}
\begin{flushleft} \hspace{0.5cm}
(a) 
\end{flushleft}\vspace{-0.3cm} 
\includegraphics[width=0.42\textwidth]{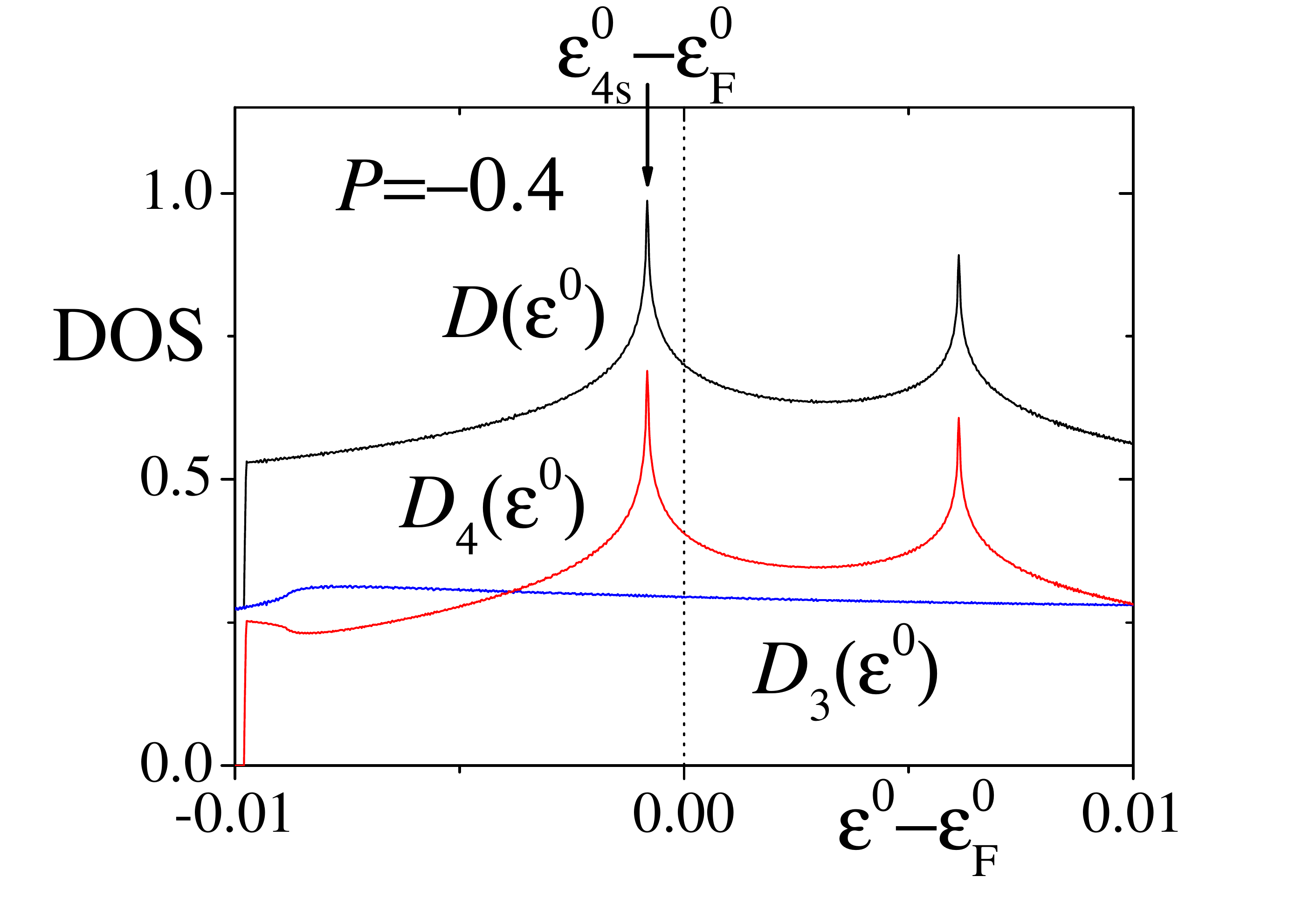}\vspace{-0.8cm}
\begin{flushleft} \hspace{0.5cm}(b) \end{flushleft}\vspace{-0.5cm} 
\includegraphics[width=0.42\textwidth]{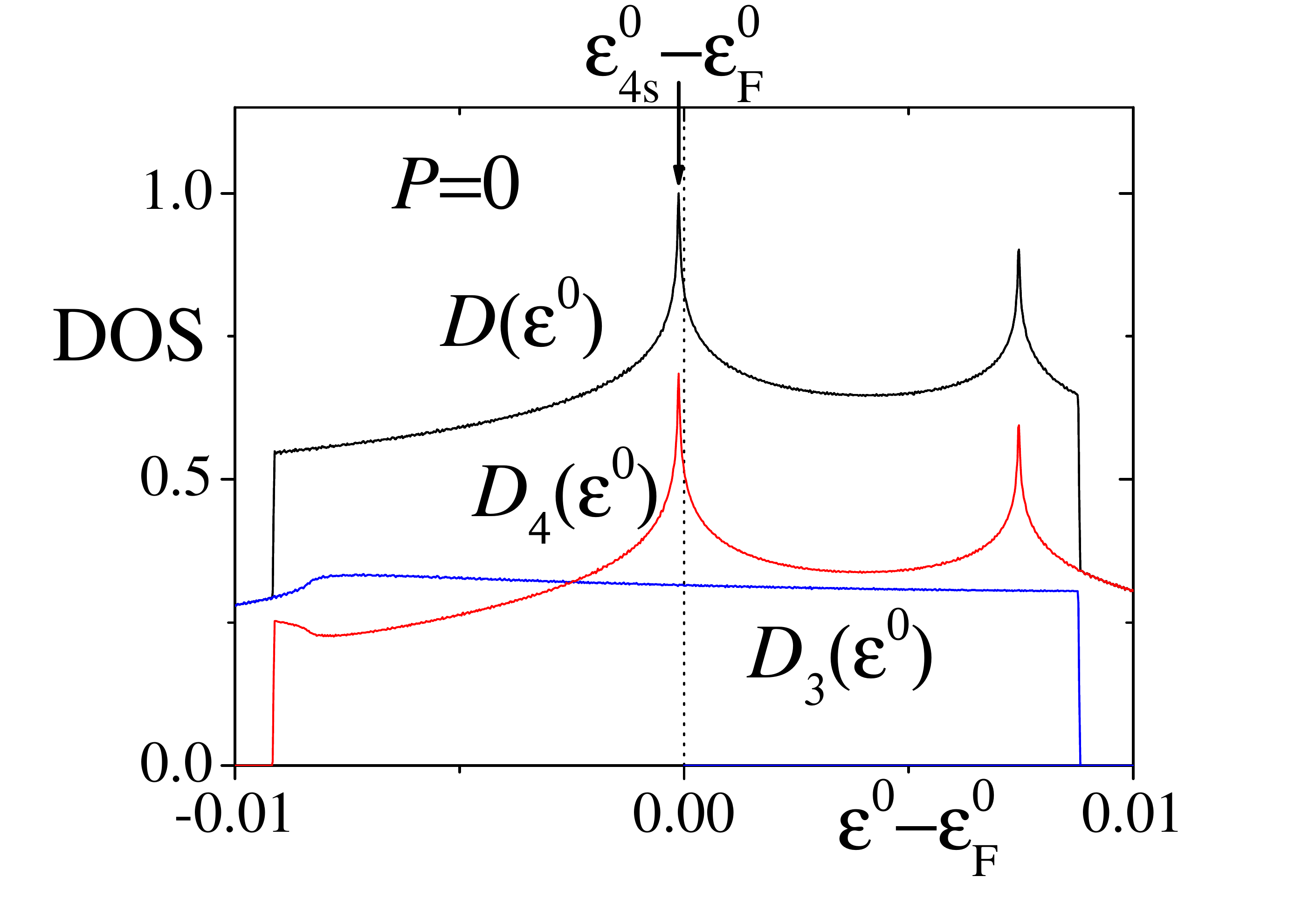}\vspace{-0.8cm}
\begin{flushleft} \hspace{0.5cm}(c) \end{flushleft}\vspace{-0.5cm}
\includegraphics[width=0.42\textwidth]{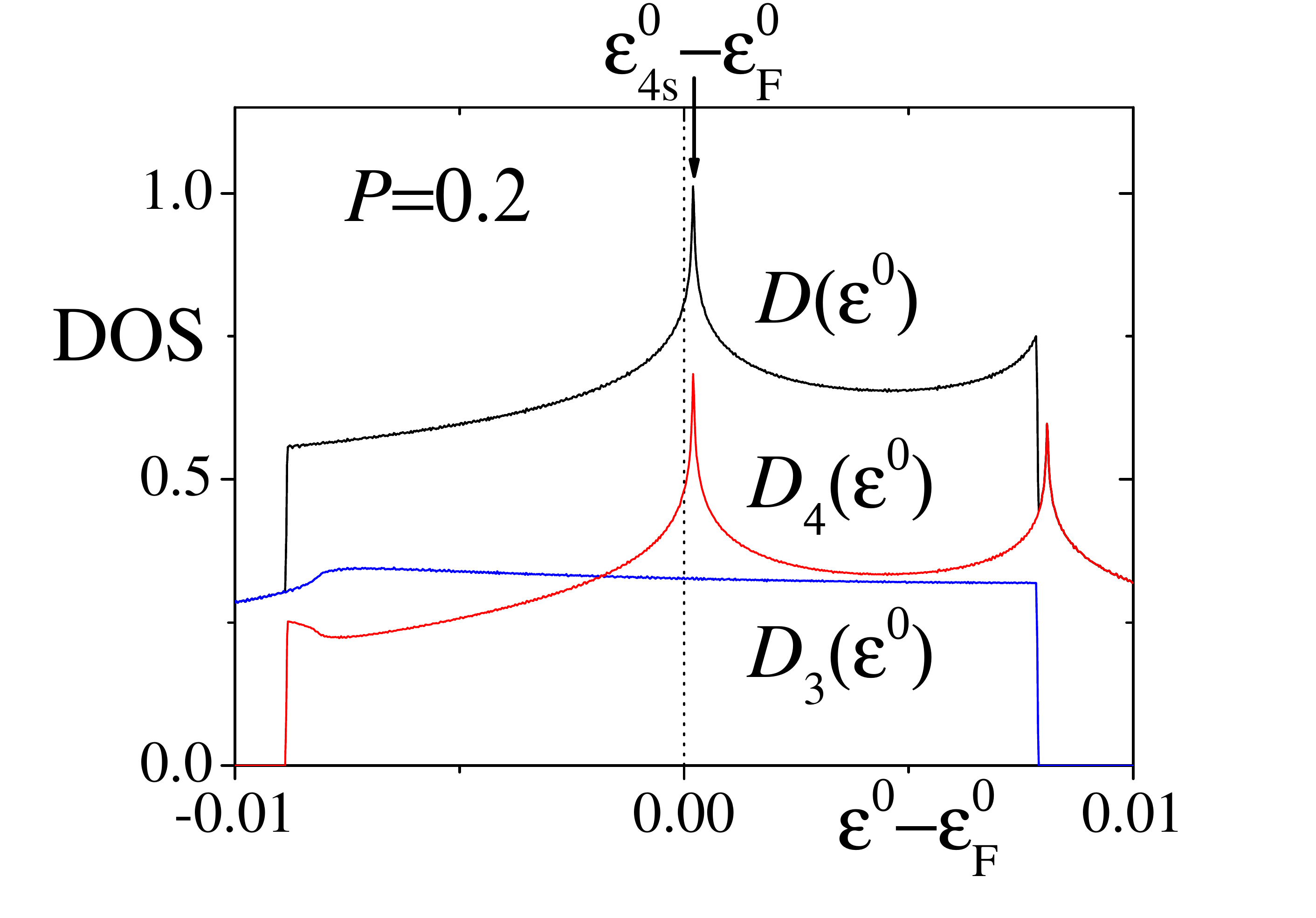}\vspace{-0.8cm}
\begin{flushleft} \hspace{0.45cm}(d) \end{flushleft}\vspace{-0.5cm}
\includegraphics[width=0.42\textwidth]{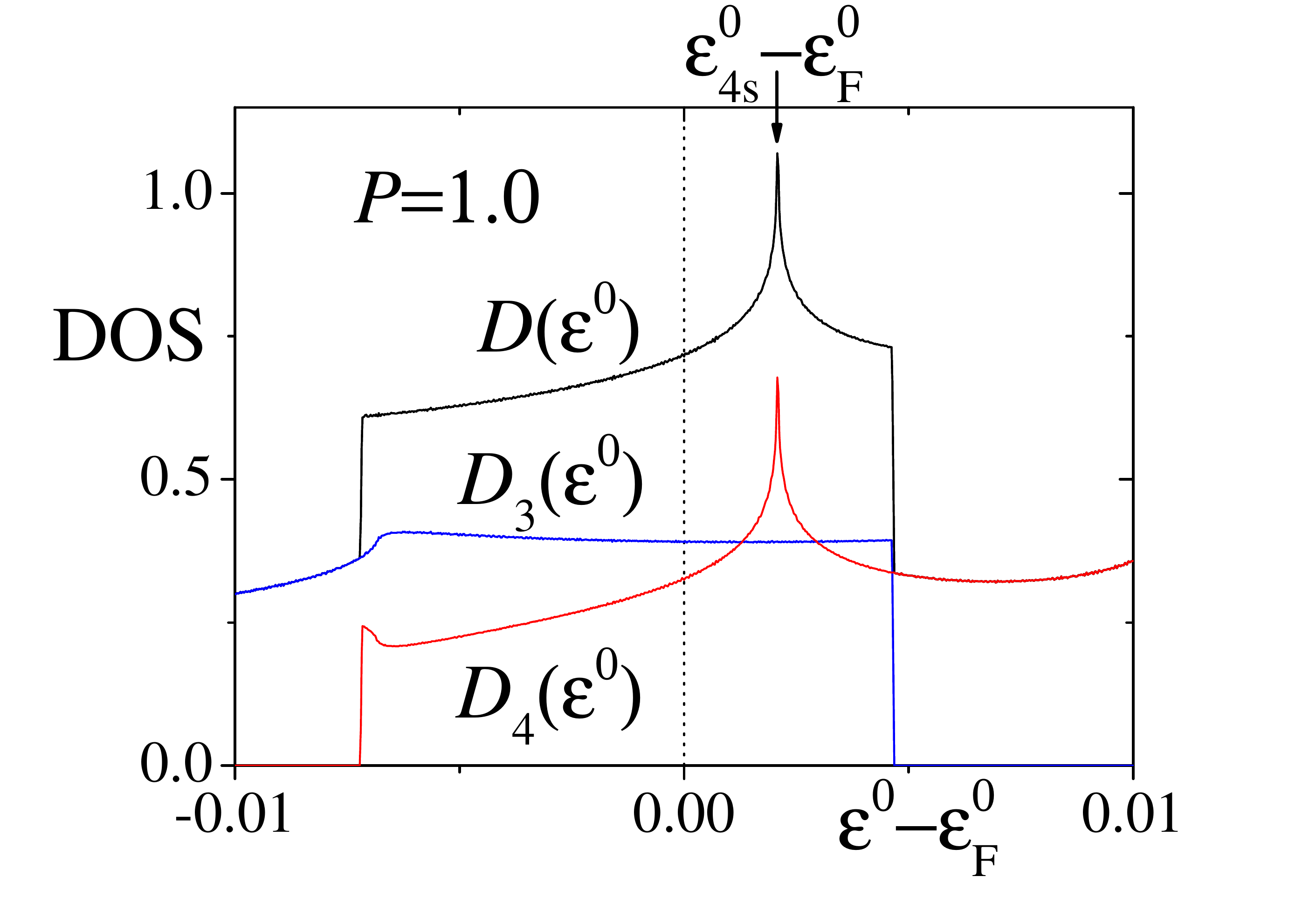}\vspace{-0.7cm} 
\end{center}
\caption{
 (Color online) 
Density of states [$D(\varepsilon ^0)$] at $H=0$ as a function of the energy $(\varepsilon^0)$ measured from $\varepsilon^0_{\rm F}$ (the Fermi energy at $H=0$) at (a) $P=-0.4$, (b) $P=0$,
(c) $P=0.2$, and (d) $P=1.0$.
The density of states originated from the fourth band (red lines) and third band (blue lines) are also plotted. 
}
\label{fig9_N}
\end{figure}

The densities of states at $P=-0.4, 0, 0.2$ and $1.0$ are shown in Fig.~\ref{fig9_N}. We can see the logarithmic divergence caused by the fourth band. The divergence near the Fermi energy at $P=0$ and 0.2 are due to the saddle point at $\mathbf{k}_{\mathrm{4s}}$. The peak of the density of states crosses the Fermi energy at the pressure of the Lifshitz transition. 



\begin{figure}[bt]
\begin{flushleft} \hspace{0.5cm}
\end{flushleft}\vspace{-0.8cm}
\includegraphics[width=0.45\textwidth]{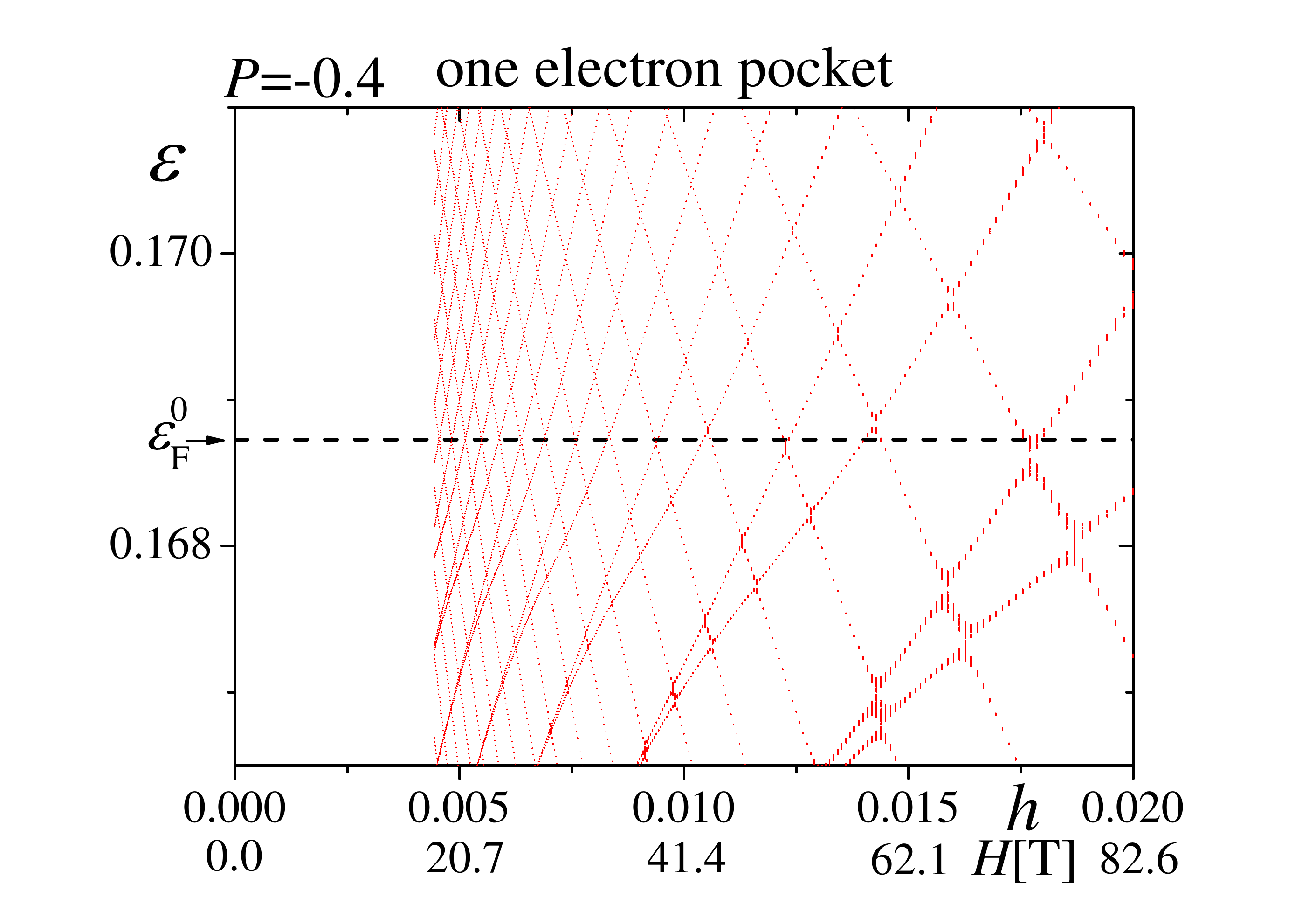}\vspace{-0.5cm}
\caption{
Energies near the Fermi energy as a function of $h$ 
at $P=-0.4$  (one electron pocket). 
We choose $p=2$ and $100\leq q 
\leq 450$ ($q=100, 101, \cdots, 449, 450$). The magnetic Brillouin zone are $-\frac{\pi}{a}\leq k_x<\frac{\pi}{a}$ and $-\frac{\pi}{qb}\leq k_y<\frac{\pi}{qb}$. Since the energies are independent of the gauge of the vector potential, we set the increments in the magnetic Brillouin zone along $k_x$ and $k_y$ are taken as $2 \pi/(aq) \times 1/7$ and $2 \pi/(bq) \times 1/5$, respectively. A black dotted line is $\varepsilon_{\rm F}^0$. 
}
\label{fig18_0}
\end{figure}

\begin{figure}[bt]
\begin{flushleft} \hspace{0.5cm}(a) \end{flushleft}\vspace{-0.7cm}
\includegraphics[width=0.40\textwidth]{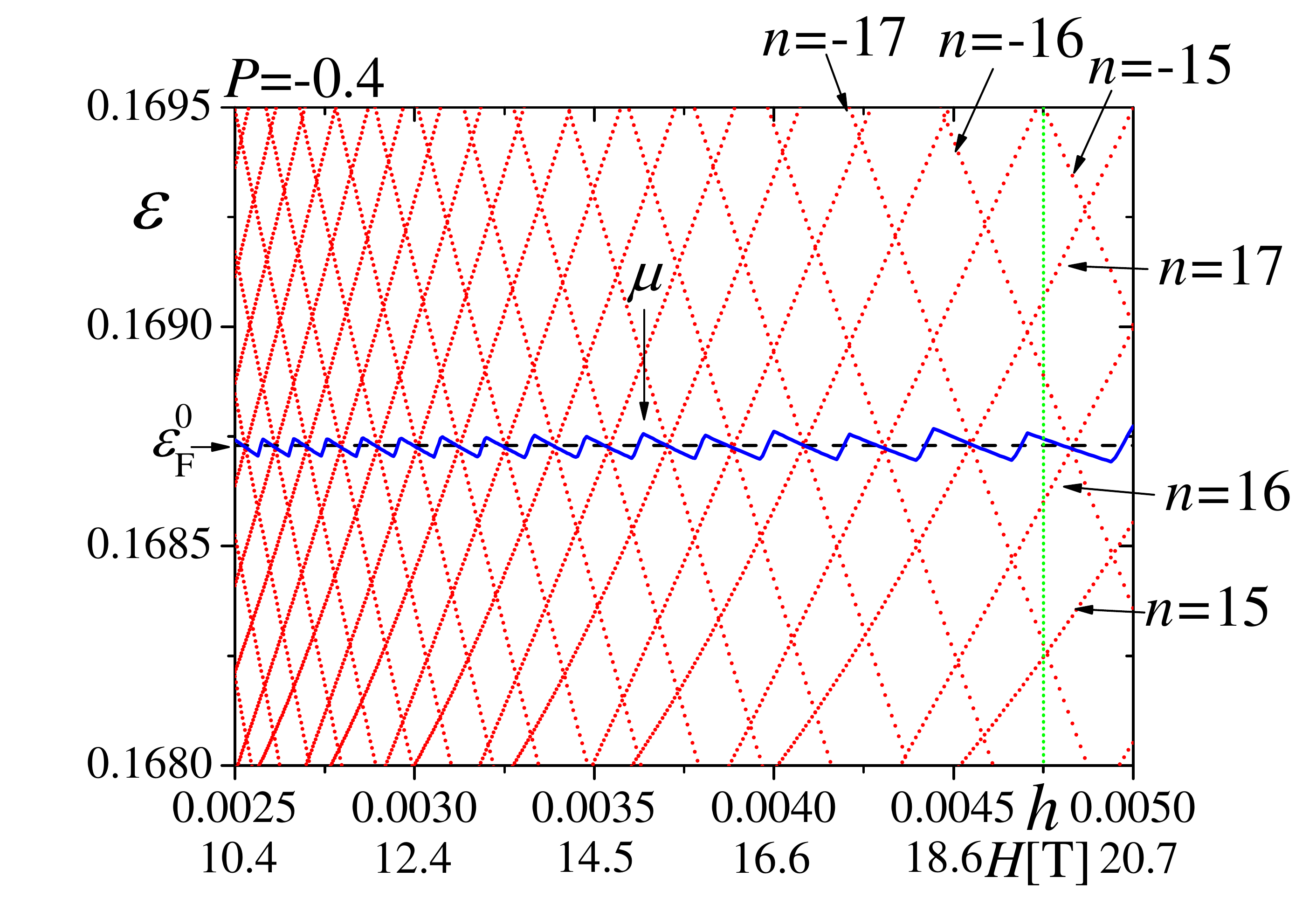}\vspace{-0.4cm}
\begin{flushleft} \hspace{0.5cm}(b) \end{flushleft}\vspace{-0.8cm}
\includegraphics[width=0.40\textwidth]{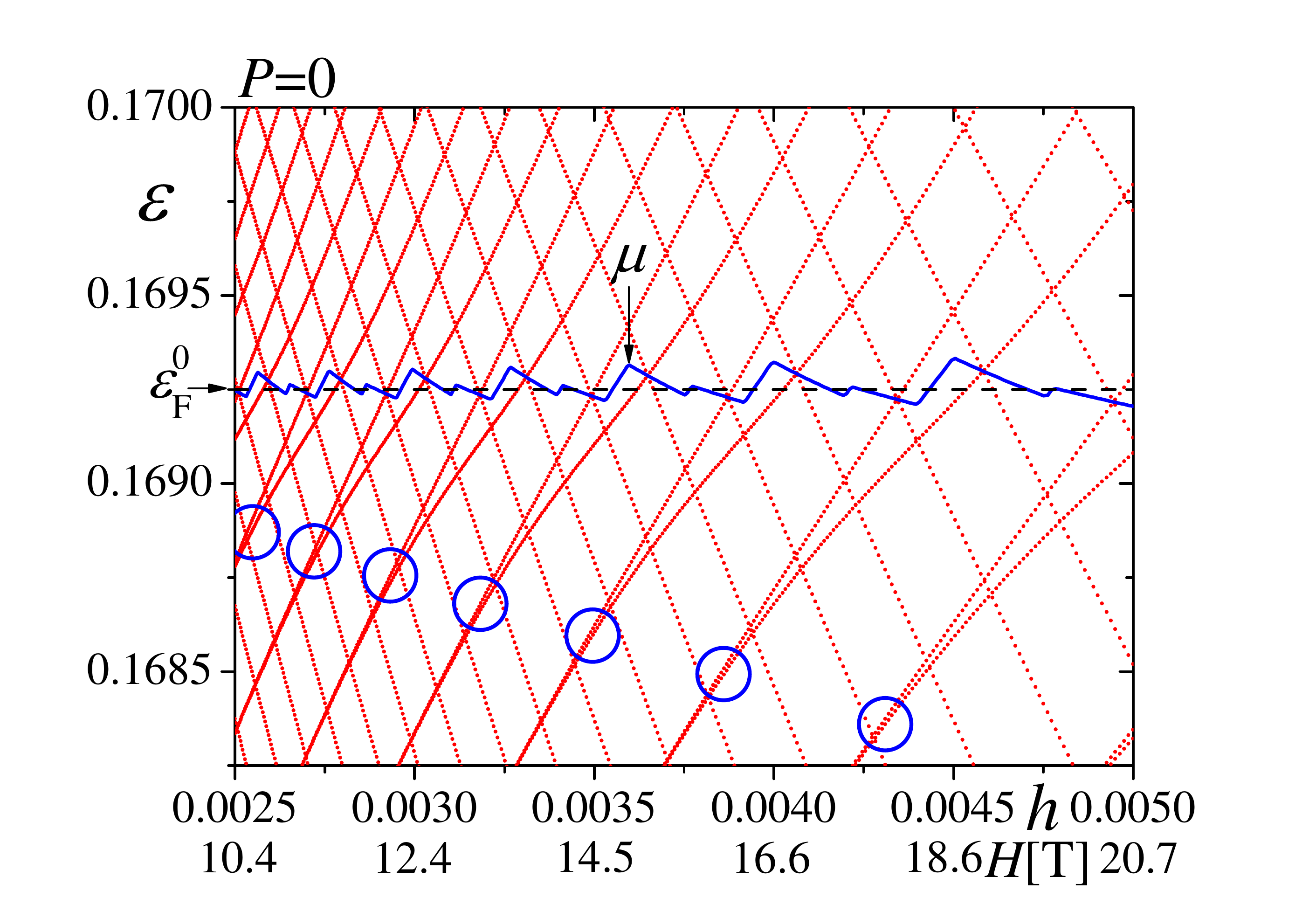}\vspace{-0.4cm}
\begin{flushleft} \hspace{0.5cm}(c) \end{flushleft}\vspace{-0.8cm}
\includegraphics[width=0.40\textwidth]{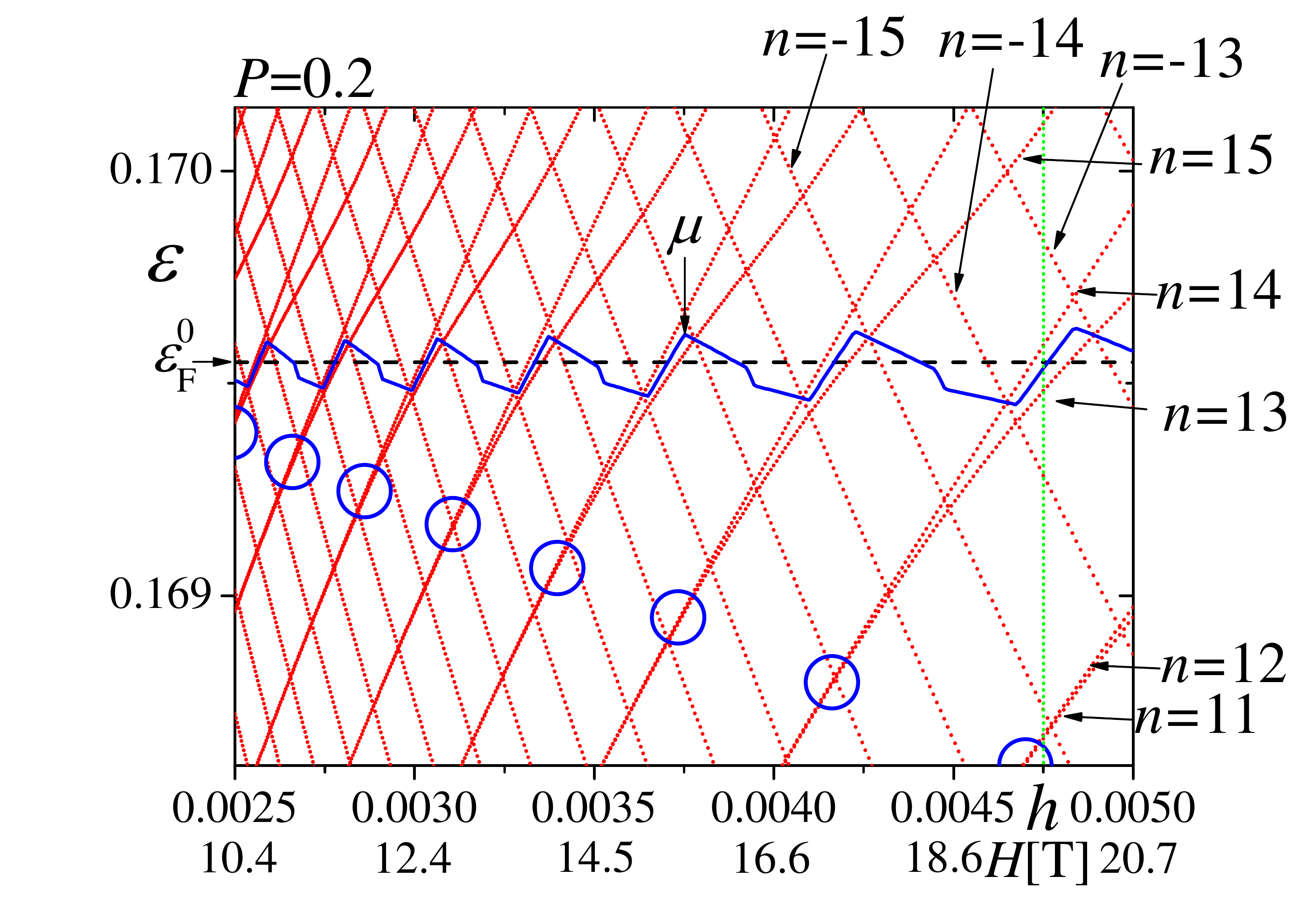}\vspace{-0.4cm}
\begin{flushleft} \hspace{0.5cm}(d) \end{flushleft}\vspace{-0.8cm}
\includegraphics[width=0.40\textwidth]{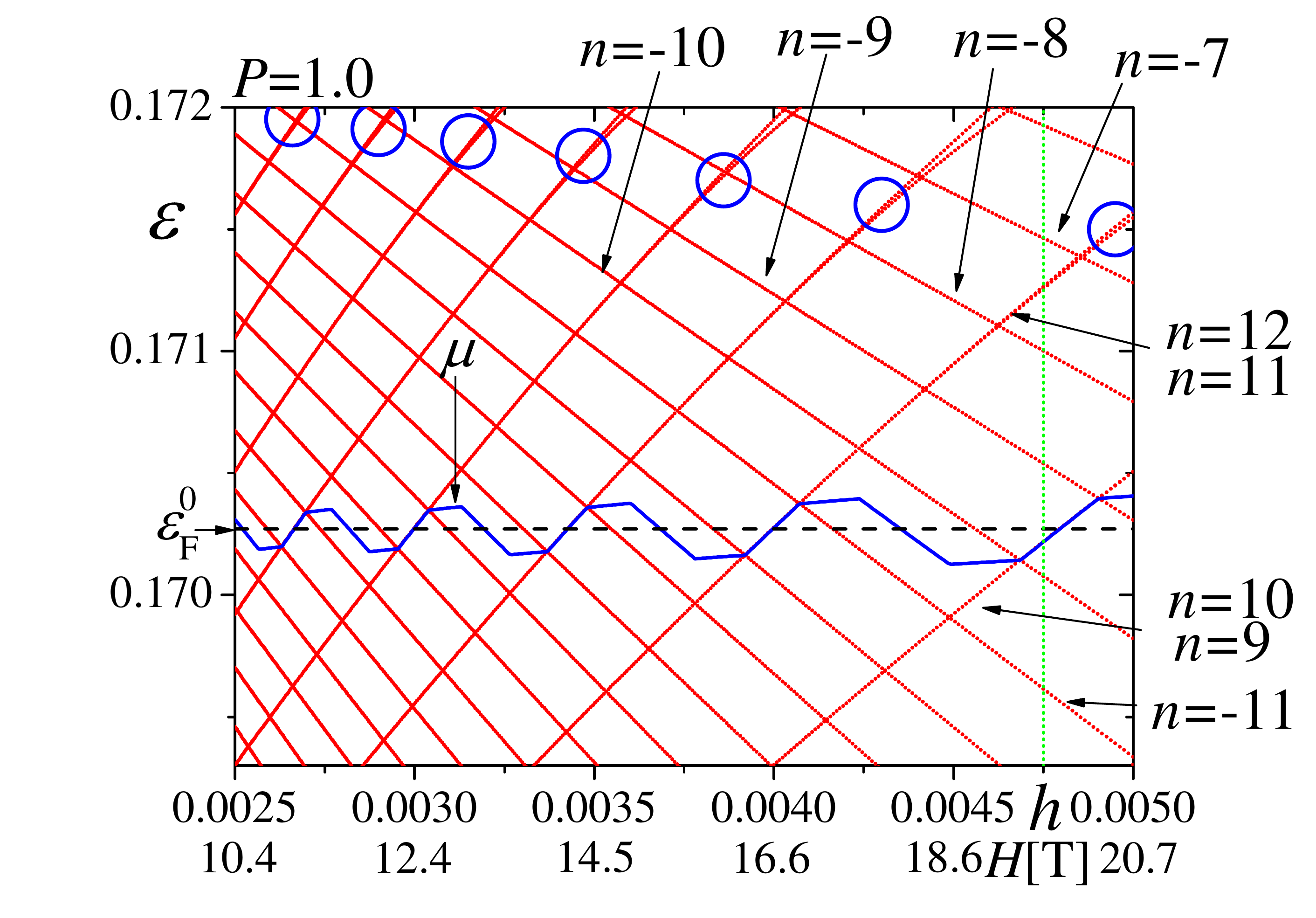}\vspace{-0.2cm}
\caption{
Energies near the Fermi energy as a function of $h$ 
at $P=-0.4$ (a), $P=0$ (b), $P=0.2$ (c) and $P=1.0$ (d). 
The region of the magnetic field is smaller than that in Fig. \ref{fig18_0}, where we take $p=2$ and $400\leq q \leq 800$ ($q=400, 401, \cdots, 799, 800$). We indicate $\varepsilon_{\rm F}^0$ at $h=0$ by the black broken lines and $\mu$ at $h\neq 0$ in fixed $\nu$ by the blue lines, respectively. Blue circles indicate the effective merging points of two Landau levels for electron pockets.
}
\label{fig18}
\end{figure}

\begin{figure}[bt]
\begin{flushleft} \hspace{0.5cm}(a) \end{flushleft}\vspace{-0.0 cm}
\includegraphics[width=0.26\textwidth]{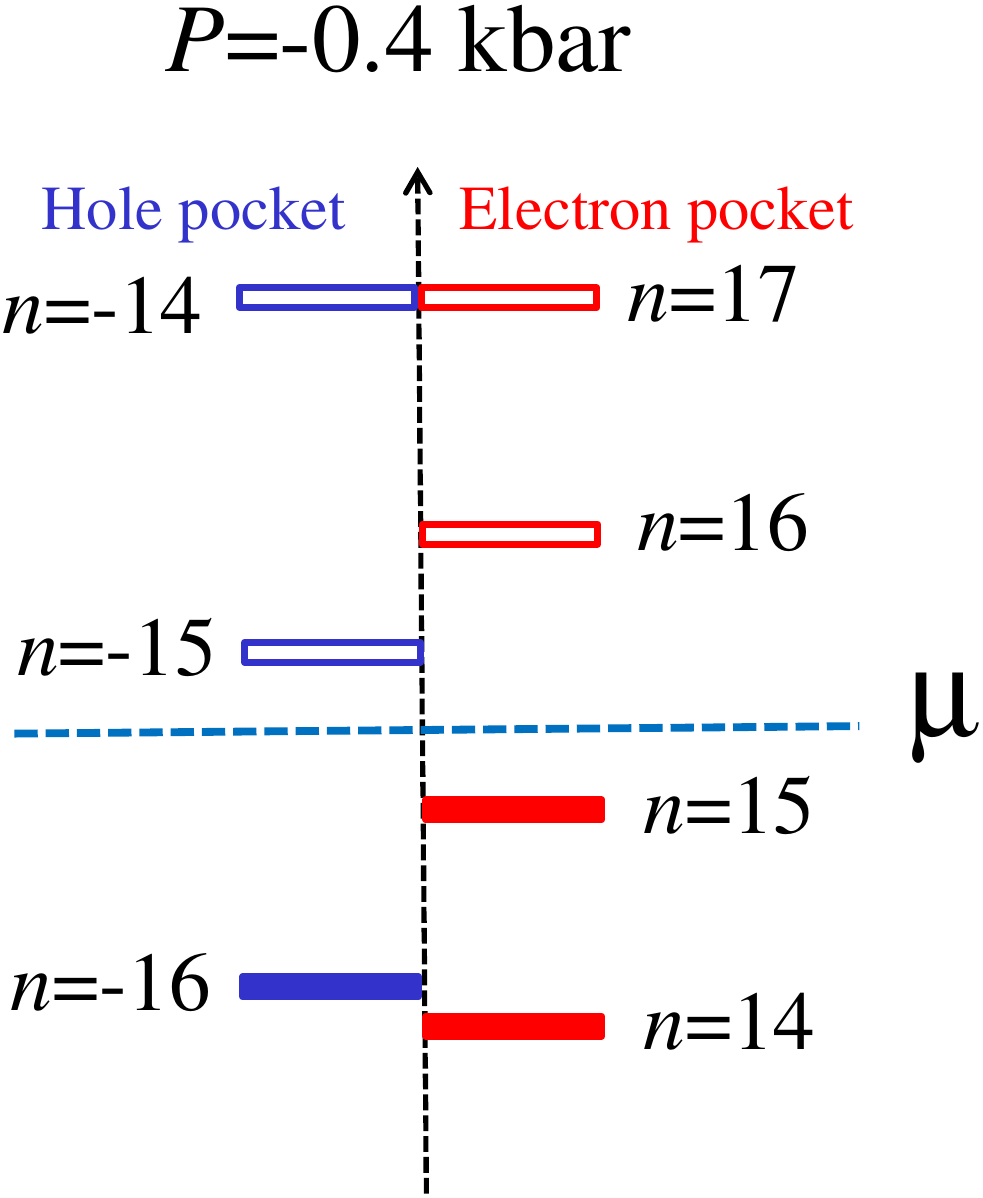}\vspace{-0.0cm}
\begin{flushleft} \hspace{0.5cm}(b) \end{flushleft}\vspace{-0.0cm}
\includegraphics[width=0.26\textwidth]{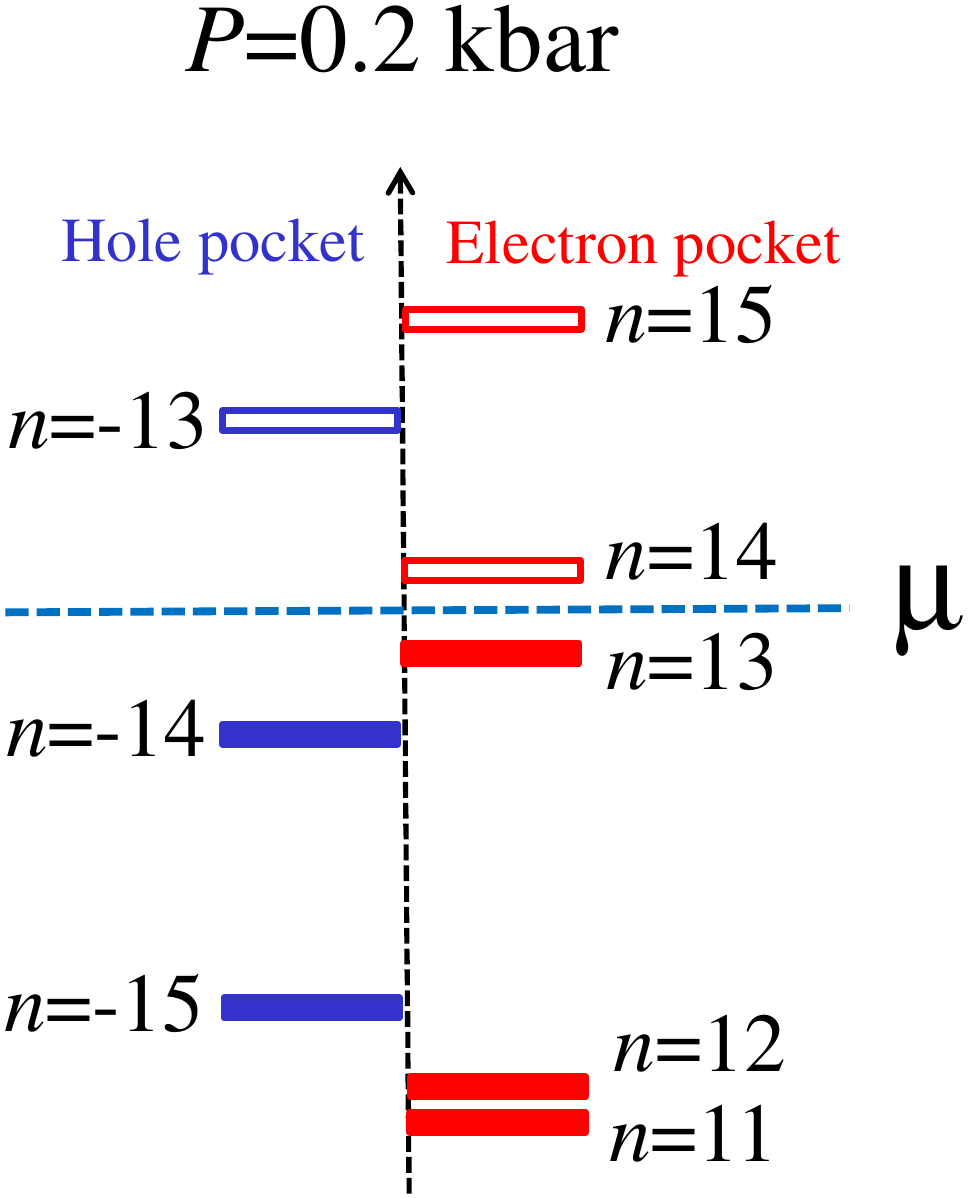}\vspace{-0.3cm}
\begin{flushleft} \hspace{0.5cm}(c) \end{flushleft}\vspace{-0.0cm}
\includegraphics[width=0.26\textwidth]{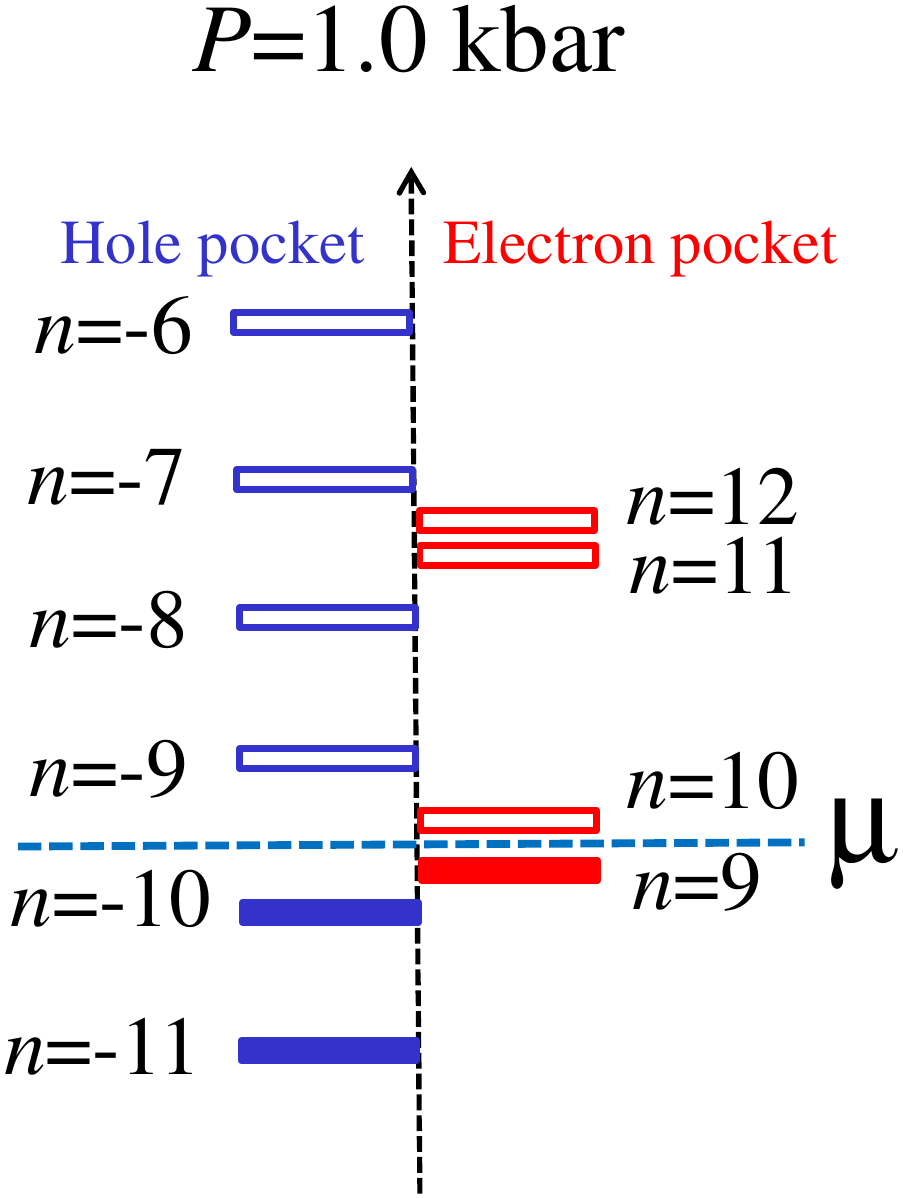}\vspace{-0.0cm}
\caption{(a) , (b) and (c) are the Landau levels at $P=-0.4$, 0.2 and 1.0 at $h$=0.00475 [dotted green lines in Figs. \ref{fig18} (a), (c) and (d)], respectively. Filled red (blue) boxes are for the completely occupied electron's (hole's) Landau levels, and blank red (blue) boxes are for the unoccupied electron's (hole's) Landau levels, respectively. 
}
\label{fig19}
\end{figure}

\begin{figure}[bt]
\begin{flushleft} \hspace{0.5cm}(a)
\end{flushleft}\vspace{-0.8cm}
\includegraphics[width=0.51\textwidth]{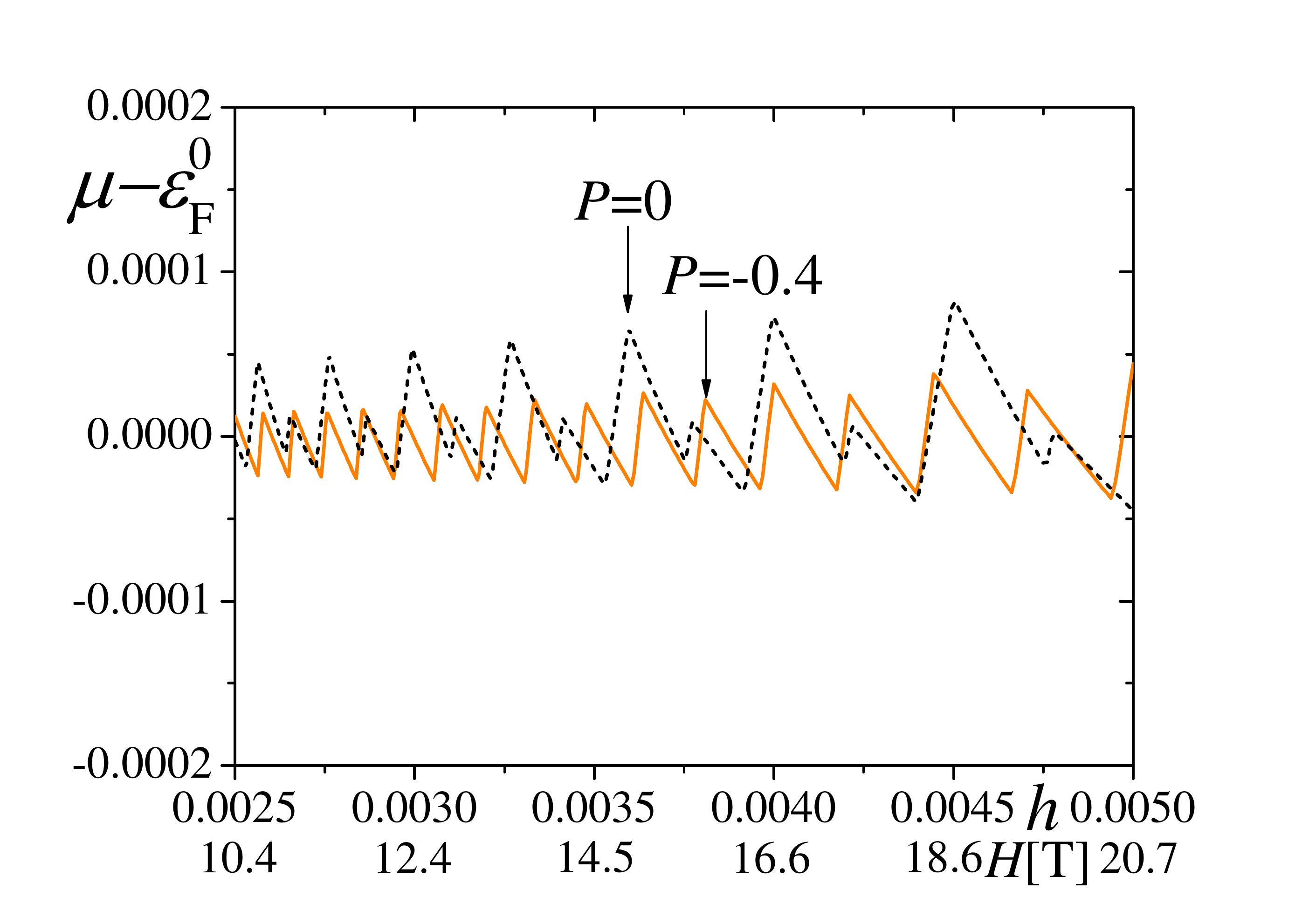}\vspace{-0.5cm}
\begin{flushleft} \hspace{0.5cm}(b) \end{flushleft}\vspace{-0.8cm}
\includegraphics[width=0.51\textwidth]{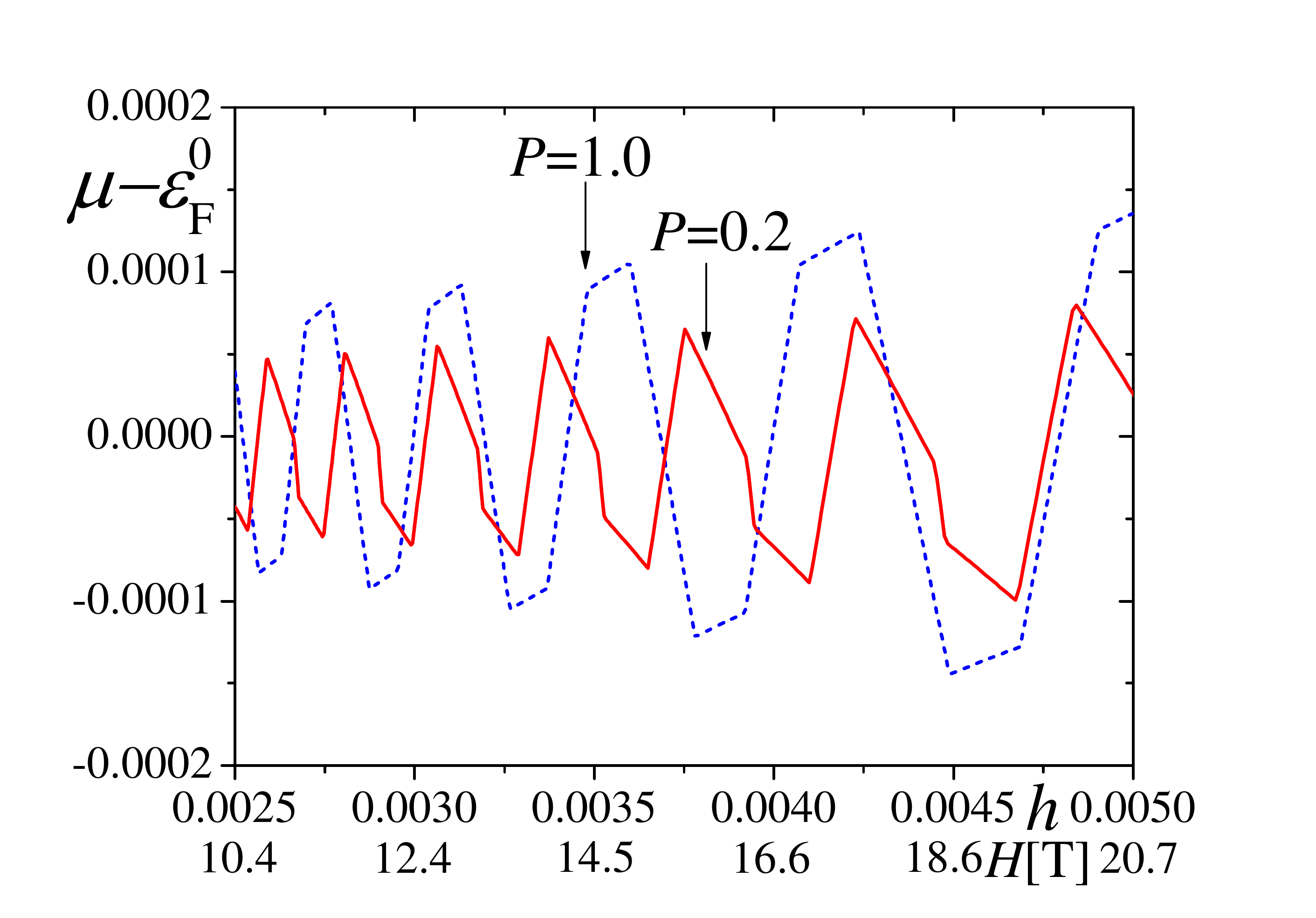}\vspace{-0.0cm}
\caption{
The chemical potential $\mu$ for the 3/4-filling as a function of $h$ at $P=-0.4$ and 0 (a) and at $P=0.2$ and 1.0 (b). 
}
\label{fig18_2}
\end{figure}

\section{energy in the magnetic field}

We study the case that the uniform magnetic field is applied perpendicular to 
the $x-y$ plane. We neglect spins for simplicity. If we consider effect of spin, it has been known the amplitudes for some frequencies in the fixed electron number are different from those of LK formula\cite{nakano2000}. 

We take the ordinary Landau gauge 
\begin{equation}
{\bf A}=(Hy,0,0).
\label{ordinaryLandau}
\end{equation} 
The flux through the unit cell is given by 
\begin{eqnarray}
\Phi=abH, 
\end{eqnarray}
where $a$ and $b$ are the lattice constants. 
We use the Peierls substitution as done before\cite{KH2017}. We obtain numerical solutions when the magnetic field is commensurate with the lattice period, i.e.,
\begin{eqnarray}
\frac{\Phi}{\phi_0}=\frac{p}{q}\equiv h, 
\label{Phi}
\end{eqnarray}
where $\phi_0=2\pi\hbar c/e\simeq 4.14\times 10^{-15}$ Tm$^2$ is a flux quantum, $e$ is the absolute value of the electron charge ($e > 0$), 
$c$ is the speed of light, $\hbar$ is the Planck constant divided by $2\pi$, 
$p$ and $q$ are integers. 
Hereafter, we represent the strength of the magnetic field 
by $h$ of Eq. (\ref{Phi}). Since $a\simeq 9.211$~\AA \ and 
$b\simeq 10.85$\AA \ in $\alpha$-(BEDT-TTF)$_2$I$_3$\cite{review}, 
$h=1$ corresponds to $H\simeq 4.14\times 10^{3}$~T.

%

At $P=-0.4$, 
the energies near the Fermi energy are shown in Fig. \ref{fig18_0}. 
At the higher field ($h \gtrsim 0.010$), the broadening of the Landau levels (the Harper broadening\cite{harper}) are seen. 
These broadenings are very small at the lower field ($h\lesssim 0.010$). Thus, the energies at the low field regions can be understood in terms of the Landau quantization with no broadenings. The similar results are obtained other pressure values such as $P=0, 0.2$ and $1.0$. 
Hereafter, we ignore the wave number dependences in the magnetic Brillouin zone along $k_x$ and $k_y$, because the Harper broadening is negligibly small. 

At the lower field ($h\leq 0.005$), the energies as a function of $h$ at $P=-0.4, 0, 0.2$ and 1.0 are shown in Figs. \ref{fig18} (a), (b), (c) and (d). 
We consider that the system is at the low temperature ($T$) much smaller
than the spacing of the Landau levels. Then we take $T=0$.
The chemical potential ($\mu$, i.e. the Fermi energy) changes as a function of $h$, when
the electron number is fixed in the compensated metal without the electron-hole symmetry. 
Since all Landau levels for electron pocket(s) and hole pocket have the same degeneracy
at the fixed magnetic field, the numbers of occupied Landau levels for electrons and holes
are the same in the compensated metal.  The chemical potential is
taken as the middle value between the energy of the highest occupied states and
that of the lowest unoccupied state, as shown in blue lines in
Fig. \ref{fig18} as a function of $h$. Two Landau levels are effectively
degenerated at the blue circles in Figs. \ref{fig18} (b), (c) and (d), although there are negligible energy gaps between them. 
This separation of the Landau levels is a quantum mechanical picture of the semiclassical magnetic breakdown. Montambaux, Piechon, Fuchs and Goerbig\cite{Montambaux2009} have also studied 
the degeneracy and the separation of Landau levels  
 in the model with two electron pockets. 


The Landau levels and $\mu$ at $h=0.00475$ [dotted green lines in Figs. \ref{fig18} (a), (c) and (d)] are shown in Figs. \ref{fig19} (a), (b) and (c), respectively, where $n$ is the index of Landau levels. The Landau levels with $n$=11 and 12 at $P$=0.2 [Fig. \ref{fig19} (b)] and those with $n$=(7,8), (9,10), and (11,12) 
at $P$=1.0 [Fig. \ref{fig19} (c)] are degenerated or effectively degenerated, which can be interpreted as the Landau quantization of almost-independent two electron pockets in a semiclassical picture. 
The degeneracy of the Landau levels with $n$=13 and 14 at $P$=0.2 is lifted [Fig. \ref{fig18} (c) and Fig. \ref{fig19} (b)], which can be 
understood as a result of weakly coupled two electron pockets near the Fermi energy as shown in Fig. \ref{fig8_N} (c). 

%


We show $\mu-\varepsilon^0_{\rm F}$ at $P=-0.4$, 0, $0.2$ and 1.0 in Fig. \ref{fig18_2}. 
Since the symmetry of the electron band and the hole band near the Fermi energy is not bad at $P=-0.4$ [see Fig. \ref{fig18} (a)], the $h$-dependence of $\mu$ is small, as shown in Fig. \ref{fig18_2} (a). 
However, the electron-hole symmetry becomes worse as $P$ increases, as shown in Figs. \ref{fig18} (a), (b), (c) and (d). Therefore, as $P$ increase, the amplitude of the oscillation of $\mu$ increases, as shown in Fig. \ref{fig18_2}.



\begin{figure}[bt]
\begin{flushleft} \hspace{0.5cm}(a) \end{flushleft}\vspace{-0.8cm}
\includegraphics[width=0.41\textwidth]{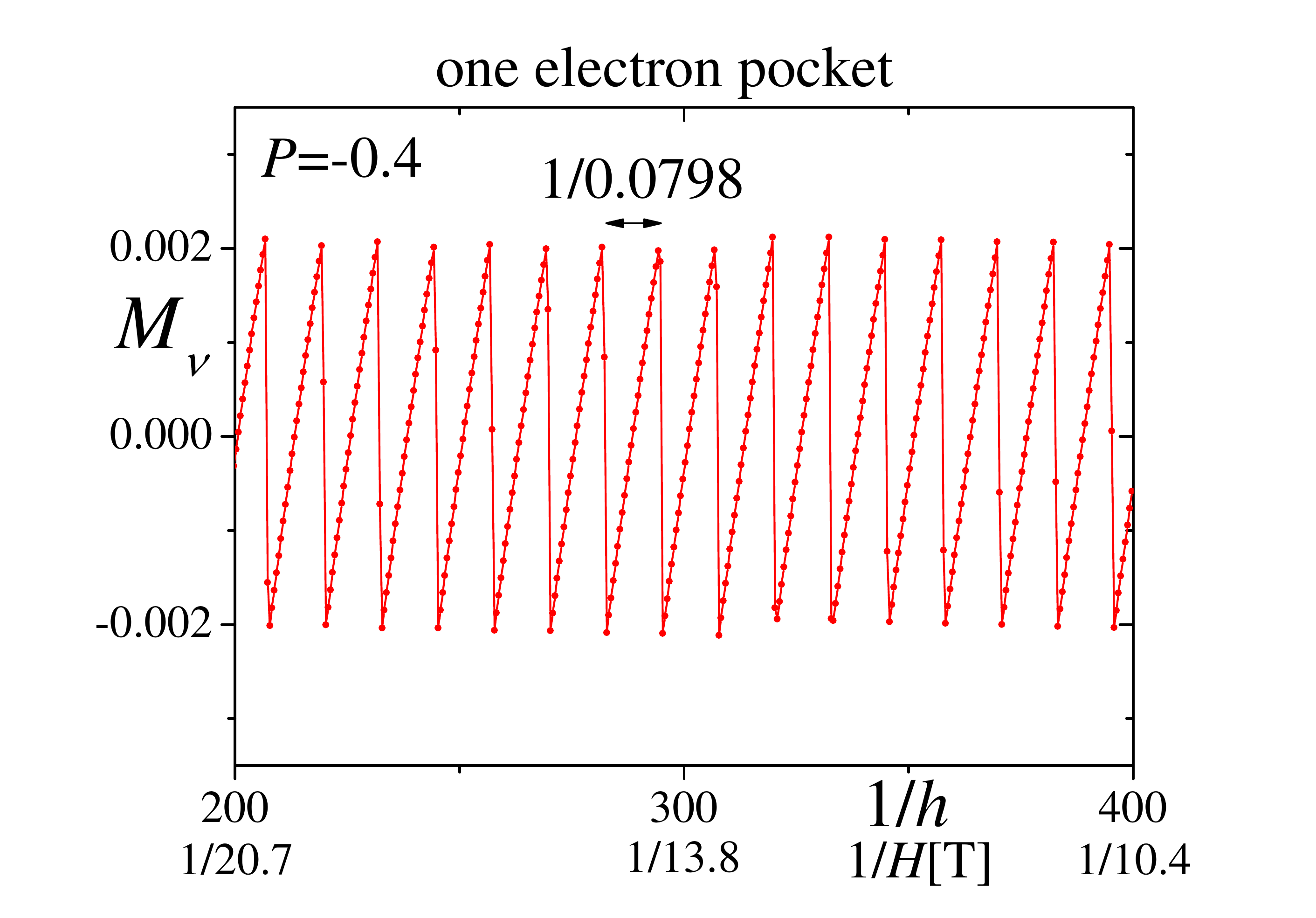}\vspace{-0.5cm}
\begin{flushleft} \hspace{0.5cm}(b) \end{flushleft}\vspace{-0.6cm}
\includegraphics[width=0.41\textwidth]{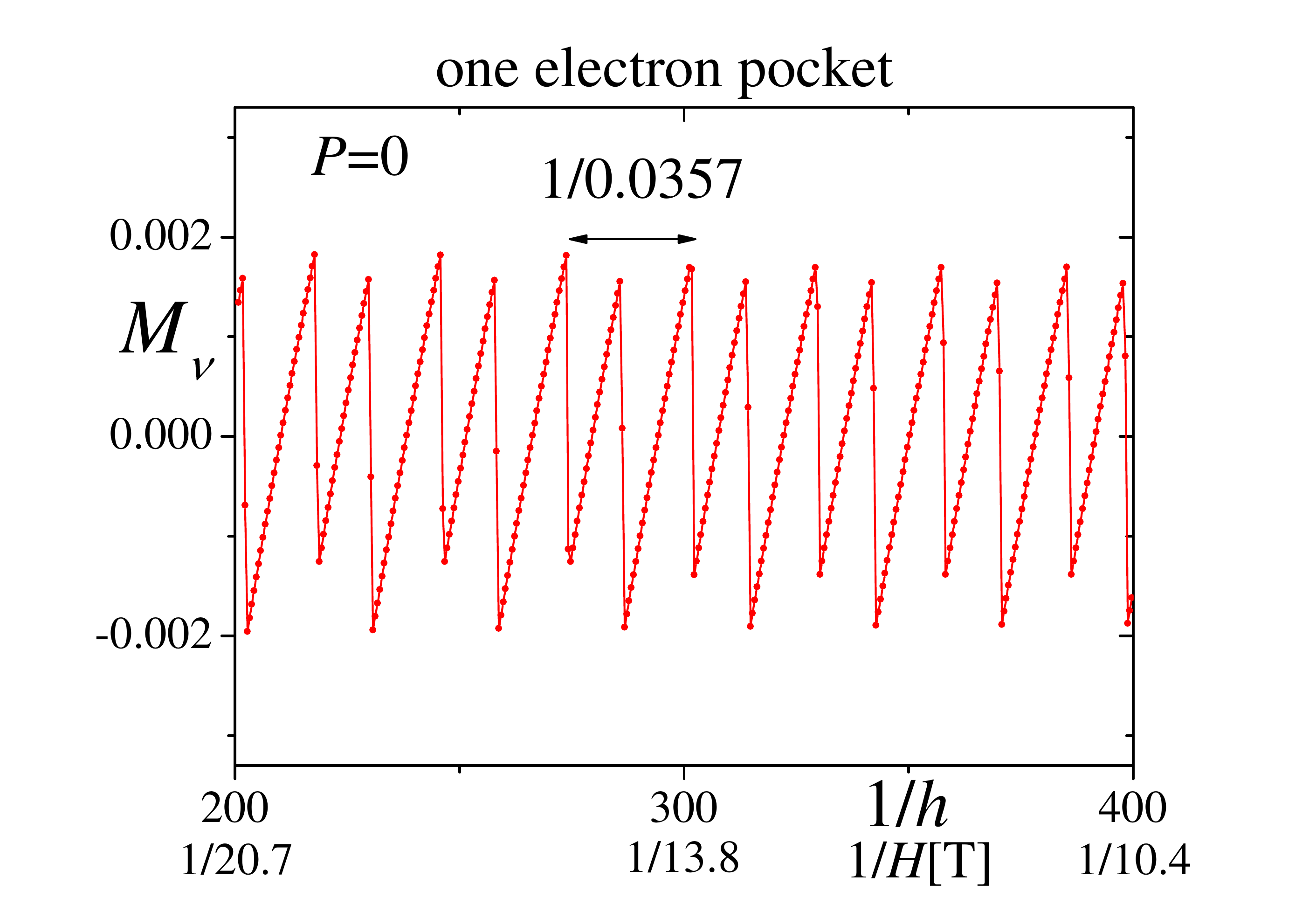}\vspace{-0.5cm}
\begin{flushleft} \hspace{0.5cm}(c) \end{flushleft}\vspace{-0.6cm}
\includegraphics[width=0.41\textwidth]{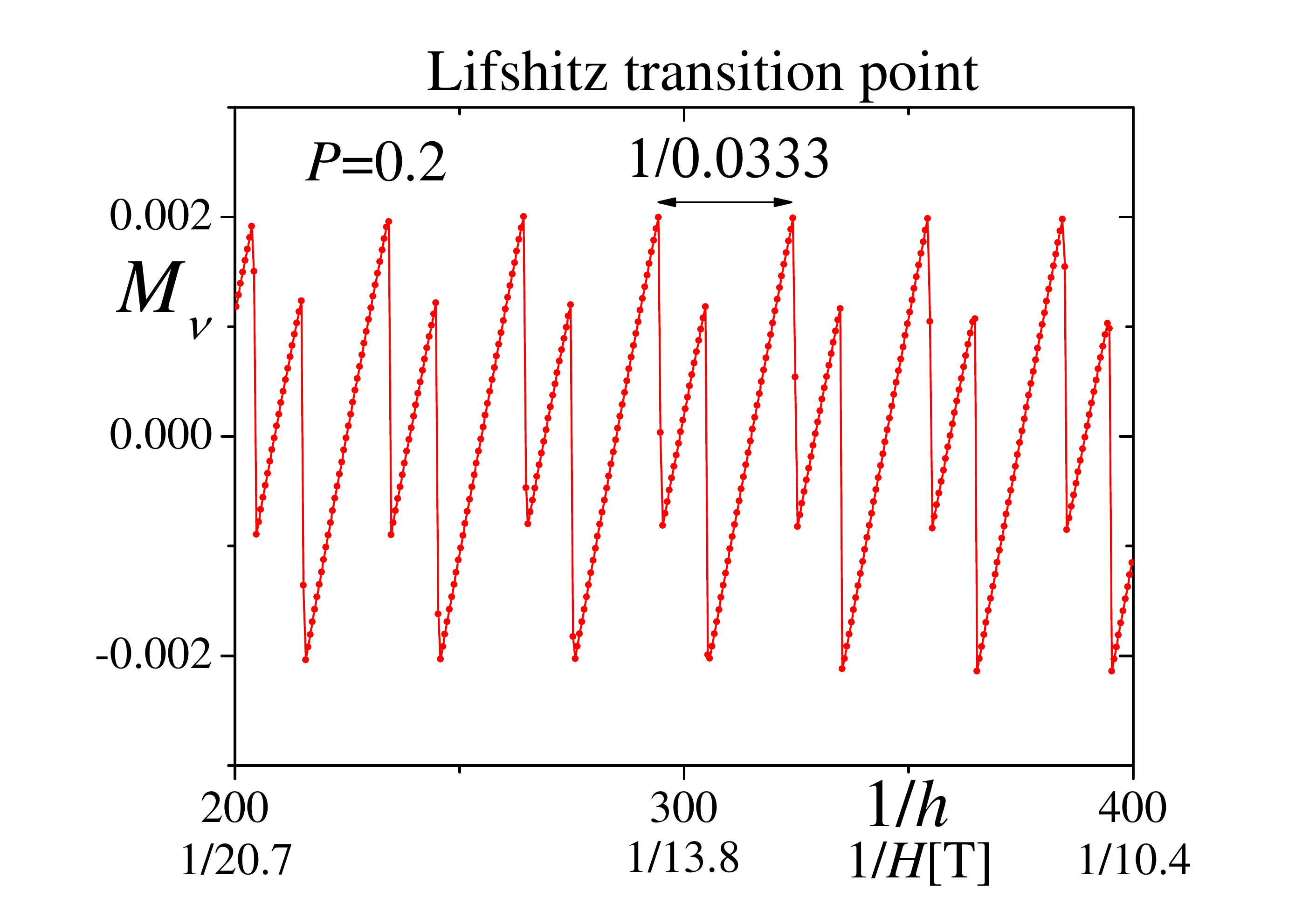}\vspace{-0.5cm}
\begin{flushleft} \hspace{0.5cm}(d) \end{flushleft}\vspace{-0.6cm}
\includegraphics[width=0.41\textwidth]{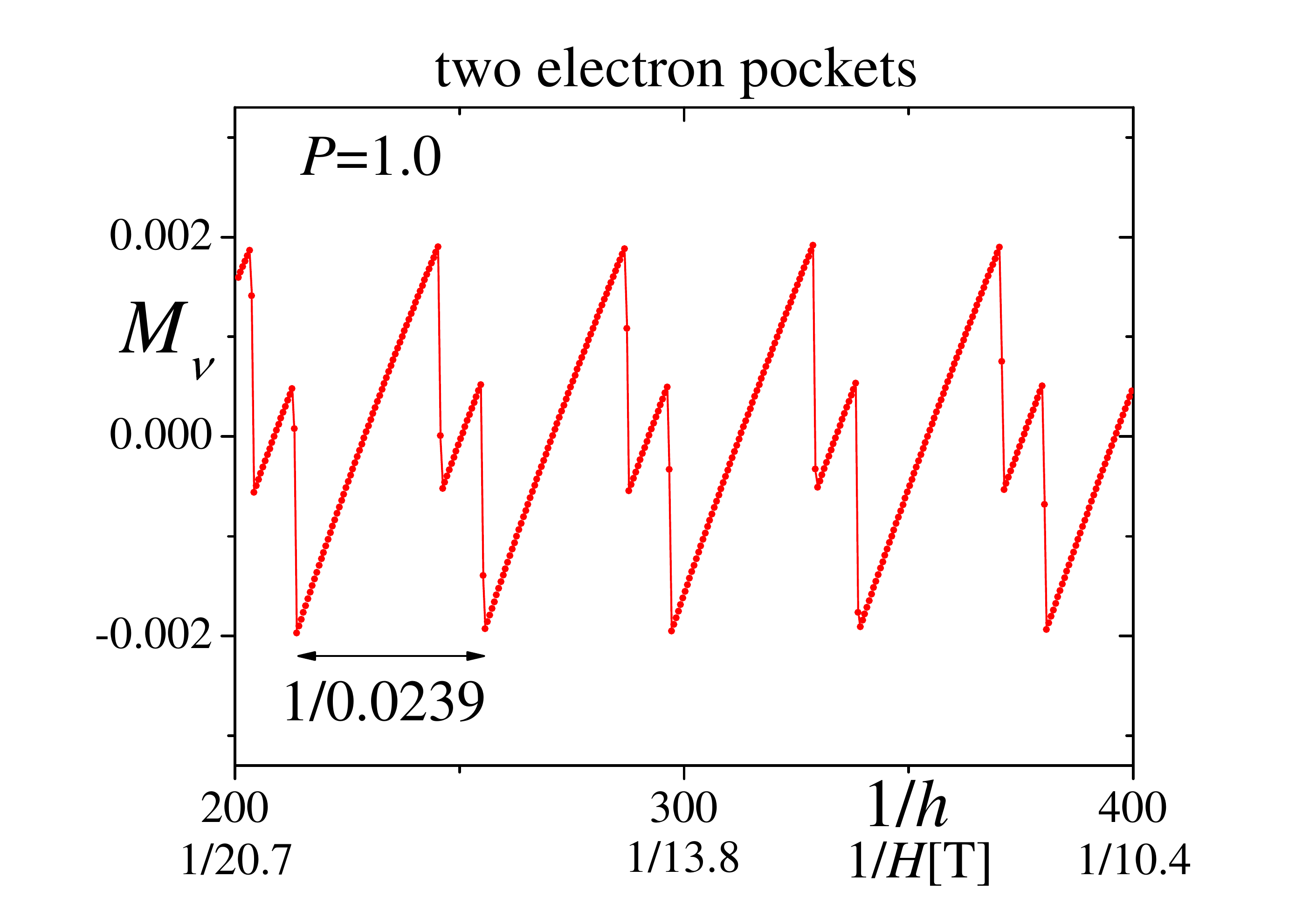}\vspace{0.4cm}
\caption{
Magnetization as a function of $1/h$ at $P=-0.4$  (one electron pocket) (a), $P=0$  (one electron pocket) (b), $P=0.2$ (Lifshitz transition point) (c) and $P=1.0$ (two electron pockets) (d) obtained with fixed electron filling $\nu$. 
}
\label{fig28_a}
\end{figure}

\begin{figure}[bt]
\begin{flushleft} \hspace{0.5cm}(a) \end{flushleft}\vspace{-0.8cm}
\includegraphics[width=0.41\textwidth]{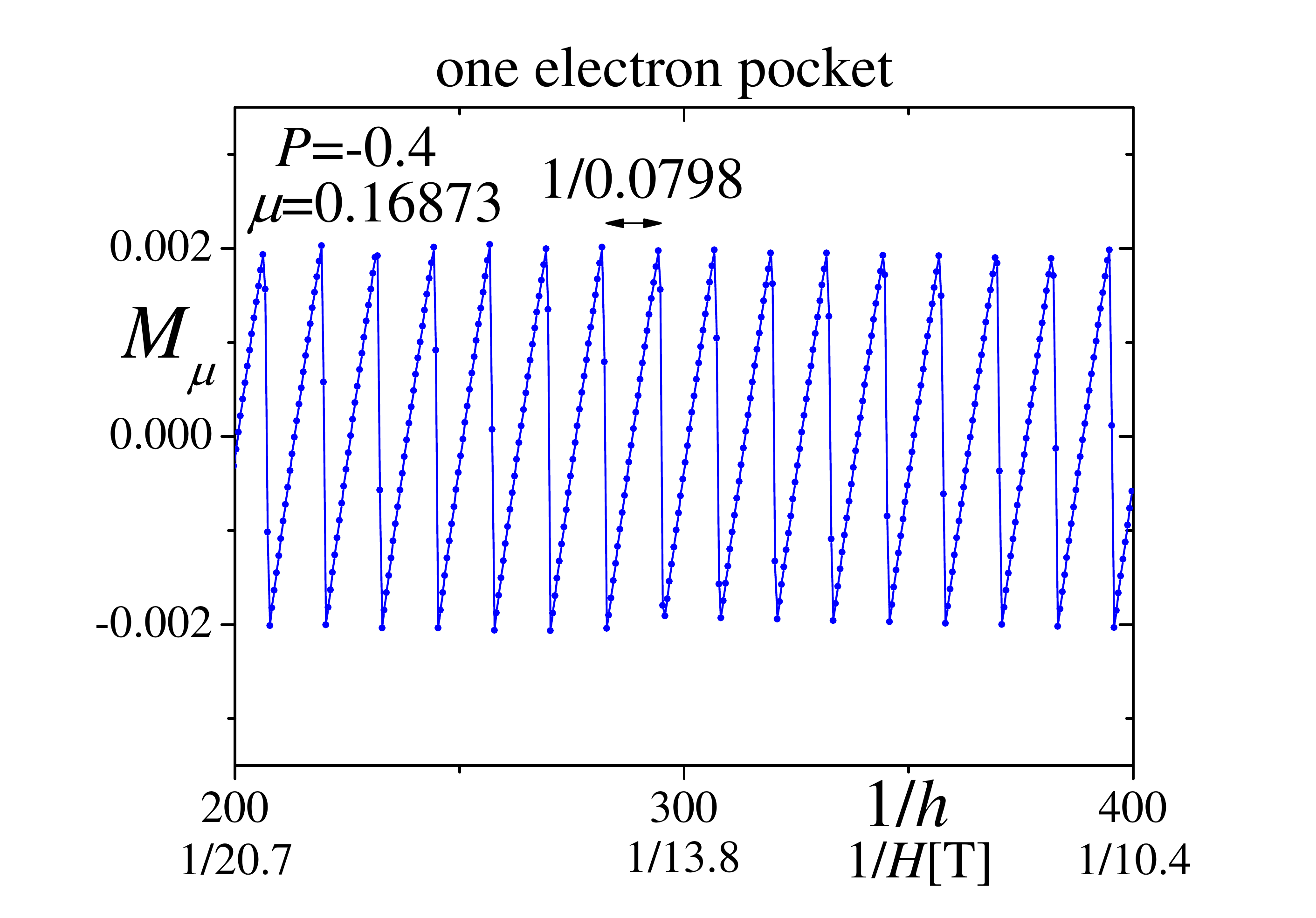}\vspace{-0.5cm}
\begin{flushleft} \hspace{0.5cm}(b) \end{flushleft}\vspace{-0.6cm}
\includegraphics[width=0.41\textwidth]{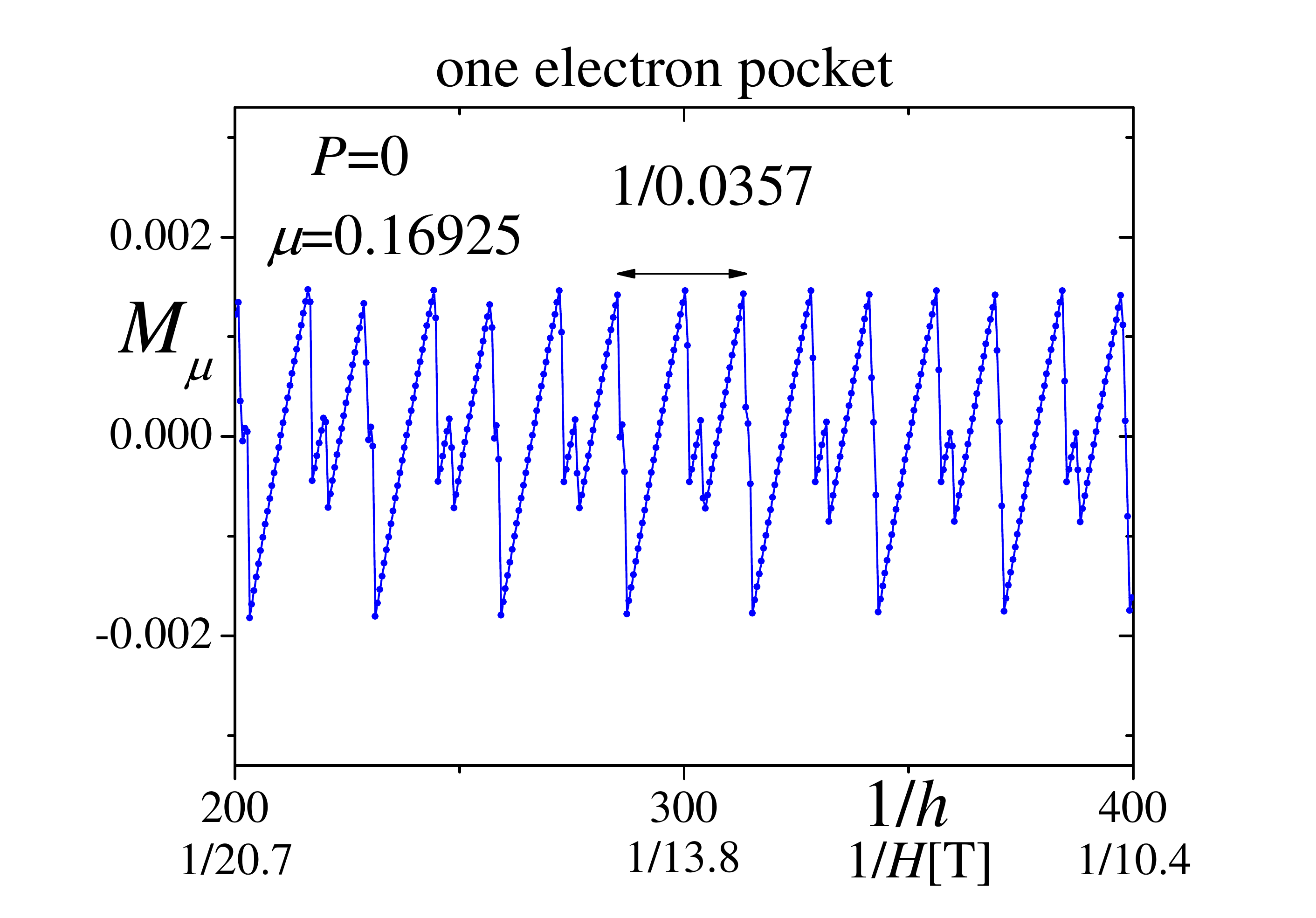}\vspace{-0.5cm}
\begin{flushleft} \hspace{0.5cm}(c) \end{flushleft}\vspace{-0.6cm}
\includegraphics[width=0.41\textwidth]{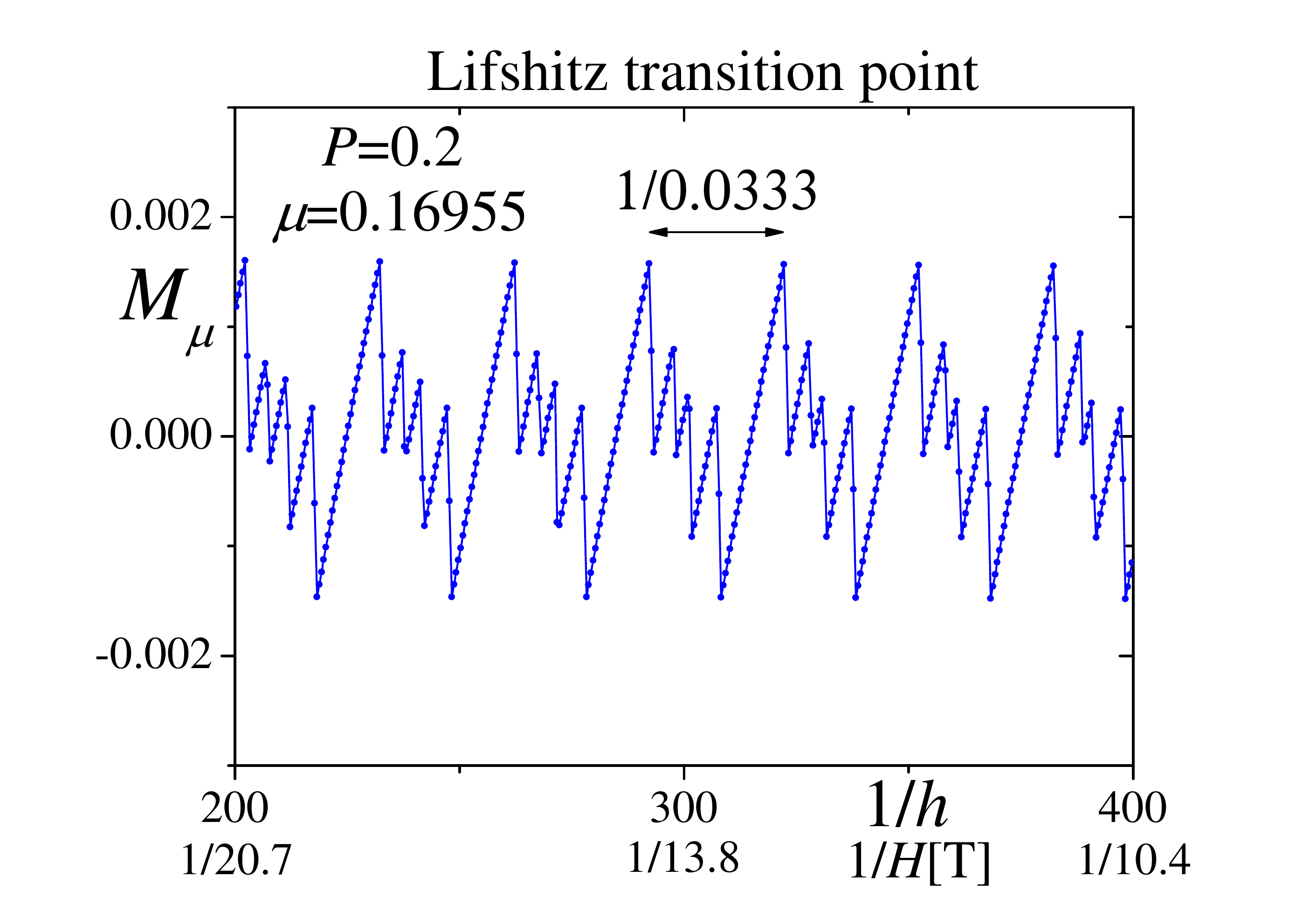}\vspace{-0.5cm}
\begin{flushleft} \hspace{0.5cm}(d) \end{flushleft}\vspace{-0.6cm}
\includegraphics[width=0.41\textwidth]{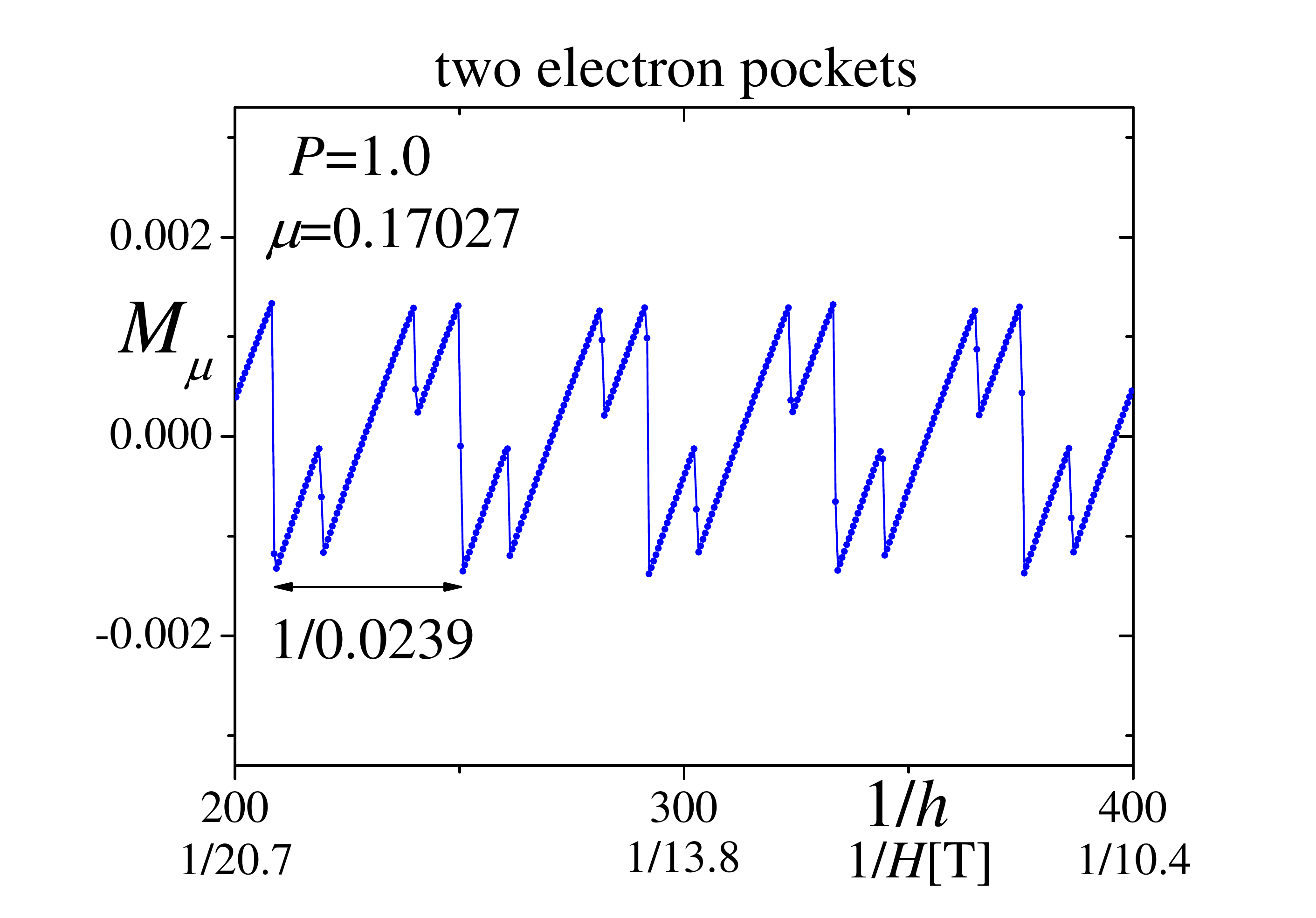}\vspace{0.2cm}
\caption{
Magnetization as a function of $1/h$ at $P=-0.4$  (one electron pocket) (a), $P=0$  (one electron pocket) (b), $P=0.2$ (Lifshitz transition point) (c) and $P=1.0$ (two electron pockets) (d) obtained with fixed chemical potential $\mu$.}
\label{fig28_b}
\end{figure}
\section{Total energy and magnetization in the magnetic field}
%

%

In this section we explain the calculations of the total energy and magnetization as a function of 
external magnetic field in two situations; one is the case of the fixed electron number and the other is the case of the fixed chemical potential. The first case, the fixed electron number or fixed electron filling to be 3/4, is plausible in the isolated two-dimensional systems\cite{shoenberg,nakano,kishigi1997,grigo,harrison,alex2001,champel}. The fixed chemical potential is realized if there exist the electron reservoirs, the three-dimensionality or the thermal broadening.

At $T=0$, the total energy ($E_{\nu}$) under the condition of the fixed electron number [i.e., the fixed electron filling ($\nu$)] is calculated by 
\begin{equation}
E_{\nu}=\frac{1}{4qN_k}\sum_{i=1}^{4qN_k\nu}
\varepsilon(i,{\bf k}), \label{E_nu}
\end{equation}
where $N_k$ is the number of $\mathbf{k}$ points taken in the magnetic Brillouin zone. 
In this system, $\nu=3/4$. 

The total energy ($E_{\mu}$) under the condition of fixed $\mu$ is calculated by 
\begin{equation}
E_{\mu}=\frac{1}{4qN_k}\sum_{\varepsilon(i,{\bf k})\leq\mu}  (\varepsilon(i,{\bf k})-\mu), \label{E_mu}
\end{equation} 
where $4q$ is the number of bands in the presence of magnetic field, and $\varepsilon(i,{\bf k})$ 
is the eigenvalues of $4q \times 4q$ matrix. 
The fixed chemical potential in Eq. (\ref{E_mu}) is given by 
\begin{equation}
\mu=\varepsilon_{\rm F}^0, 
\end{equation}
where $\varepsilon_{\rm F}^0$ is the Fermi energy at $h=0$.

We have checked 
that if $q$ is large enough as taken in the present study, 
 the wave-number dependence of the eigenvalues $\varepsilon({i,\mathbf{k}})$ 
is very small. Therefore, we can take $N_k=1$. 

The magnetizations for fixed $\nu$ and fixed $\mu$ are numerically obtained by 
\begin{eqnarray}
M_{\nu}&=& -\frac{E_{\nu}(h+\Delta h)- E_{\nu}(h)}{\Delta h}, \\
M_{\mu}&=& -\frac{E_{\mu}(h+\Delta h)- E_{\mu}(h)}{\Delta h}, 
\end{eqnarray}
respectively. 
If the $h$-dependence of $\mu$ is negligibly small, 
we obtain 
\begin{eqnarray}
M_{\nu}=M_{\mu}.
\end{eqnarray}

\section{de Haas van Alphen oscillations and its pressure dependence}
\label{results}
%


At $P=-0.4$, the wave form of $M_\nu$ of Fig.~\ref{fig28_a} (a) is almost the same as that of $M_\mu$ of Fig.~\ref{fig28_b} (a), because the $h$-dependence of $\mu$ is very small [see Fig. \ref{fig18_2} (a)]. 
The saw-tooth shapes of $M_\nu$ and $M_\mu$ are the same as 
that of the LK formula. 
As $P$ increases, the amplitude of the oscillation of $\mu$ increases [see Figs. \ref{fig18_2} (a) and (b)], and 
the wave forms of $M_\nu$ at $P=0, 0.2$ and $1.0$ deviate from those of $M_\mu$, as shown in Figs.~\ref{fig28_a} (b), (c) and (d) and Figs.~\ref{fig28_b} (b), (c) and (d).

\begin{figure}[bt]
\begin{flushleft} \hspace{0.5cm}(a) \end{flushleft}\vspace{-0.5cm}
\includegraphics[width=0.48\textwidth]{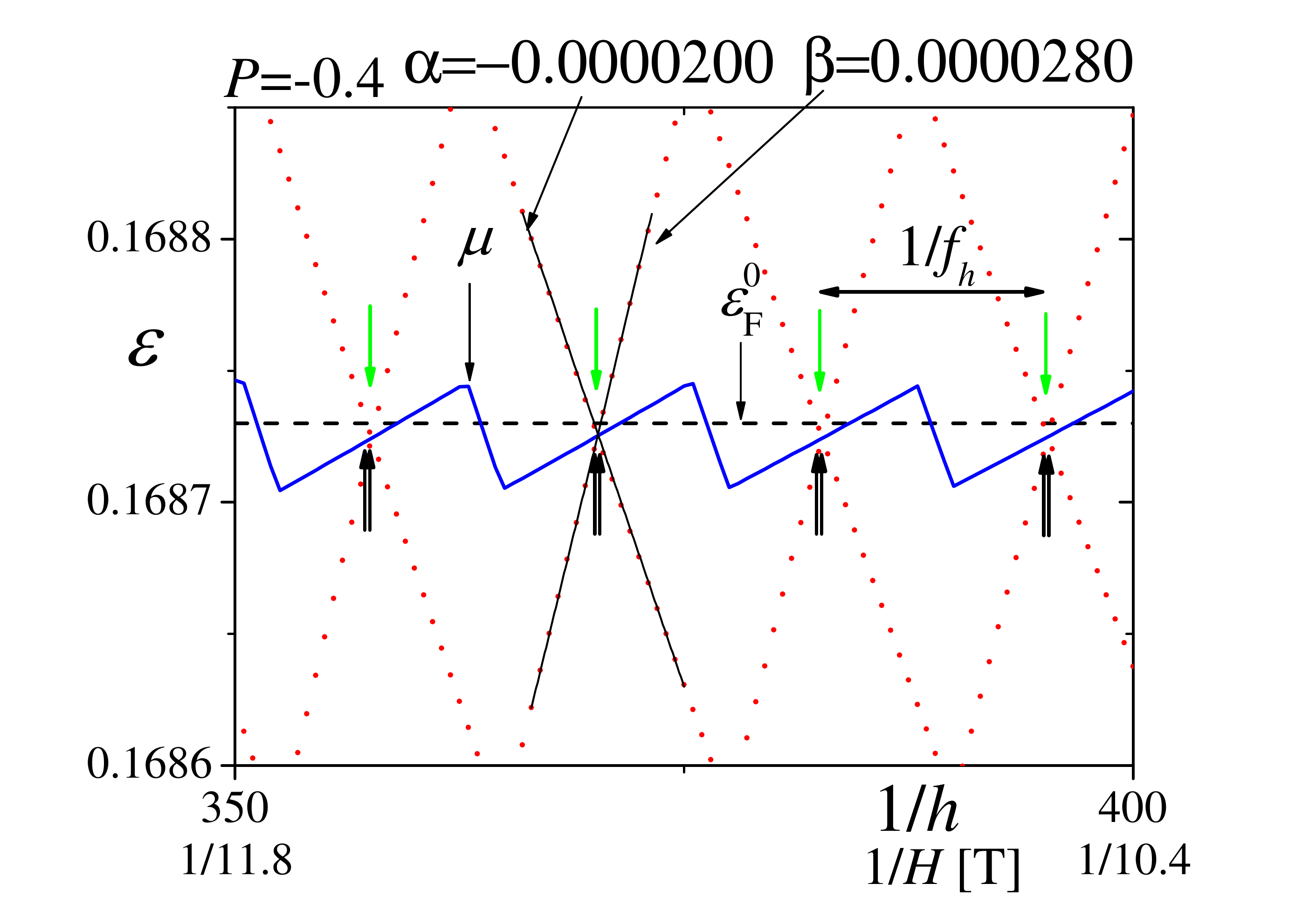}\vspace{-0.1cm}
\begin{flushleft} \hspace{0.5cm}(b) \end{flushleft}\vspace{-0.4cm}
\includegraphics[width=0.48\textwidth]{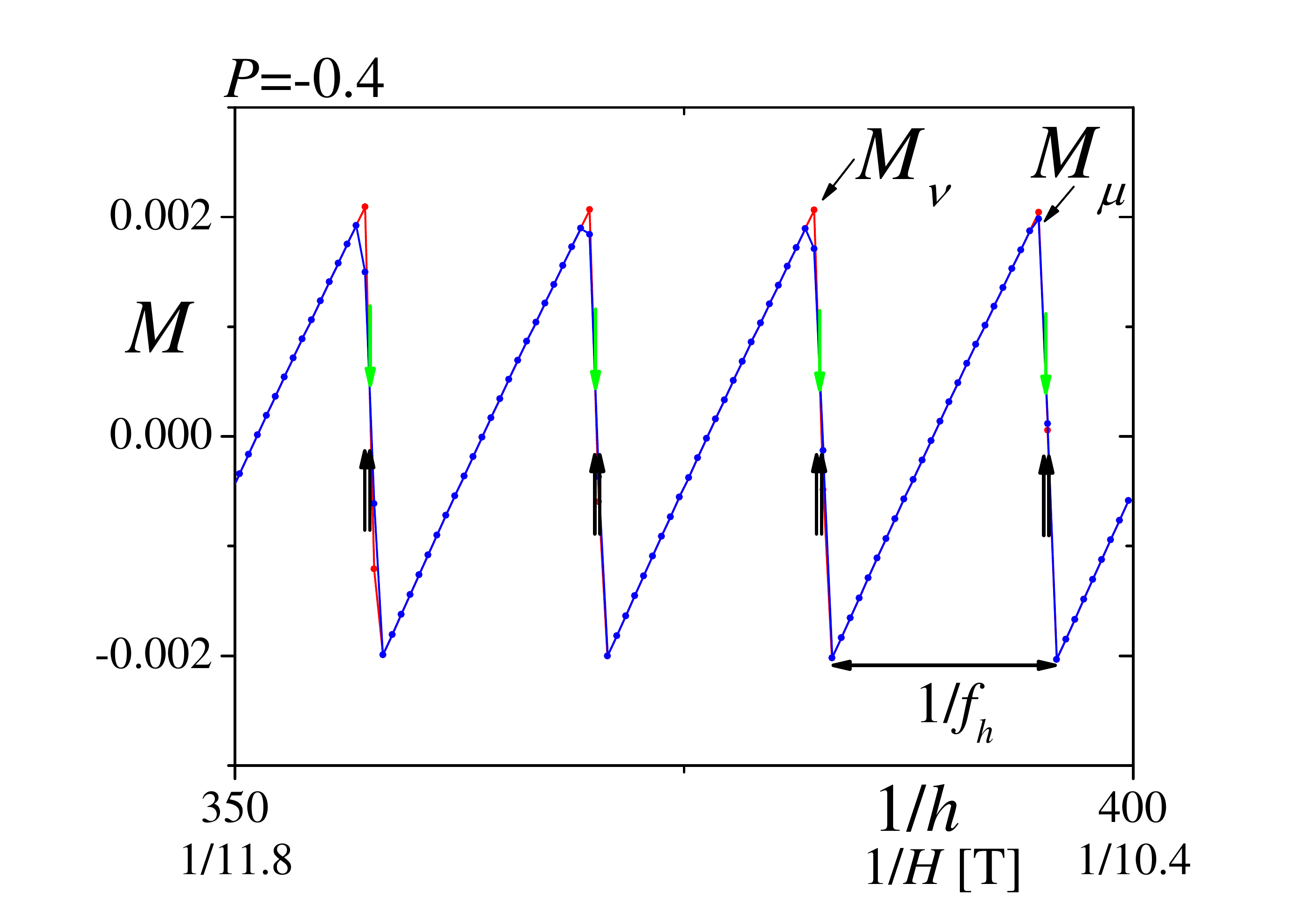}\vspace{-0.0cm}
\caption{
(a) Energy as a function
of $1/h$ [an enlarged view in the region near $\varepsilon^0_{\mathrm{F}}$ and small $h$ in Fig. \ref{fig18} (a) plotted $\varepsilon$ as a function of $1/h$].
(b) The enlarged figures of $M_{\nu}$ (a red line) 
in Fig. \ref{fig28_a} (a) and $M_{\mu}$ (a blue line) in Fig. \ref{fig28_b} (a)
in the same region of $1/h$ as (a). 
The green arrows in (a) are the crossing points of the chemical potential (for fixed $\nu$, a blue line) and the red dotted lines [Landau levels of the electron pockets (negatively-sloped lines) and those of the hole pockets (positively-sloped lines)]. 
The red dotted lines for the hole pockets or the electron pockets are almost parallel in this enlarged figure, respectively. 
The Landau levels can be approximated as straight lines 
[a negatively-sloped black line and a positively-sloped black line in (a)]. 
The slopes of Landau levels of electron pocket are 
$\alpha=-0.0000200$ and those of hole pocket are $\beta=0.0000280$. The black arrows in (a) are the crossing points of the chemical potential (for fixed $\mu$, a black dotted line) and the red dotted lines. 
Note that four lines cross almost simultaneously. 
The spacings of the green arrows are the same and $1/f_h$. Therefore, there appear in $M_{\nu}$ and $M_{\mu}$ the dHvA oscillations with the fundamental frequency of $f_h$, which corresponds to the area of the hole pocket (equal to that of the electron pocket).
Due to the difference of the density of states of electrons (fourth band) and holes (third band) [see Fig.\ref{fig9_N} (a)],
the absolute values of the slopes ($|\alpha|$ and $|\beta|$) in the electron pocket's Landau levels and the hole pocket's Landau levels are different. 
}
\label{fig34}
\end{figure}

\begin{figure}[bt]
\begin{flushleft} \hspace{0.5cm}(a) \end{flushleft}\vspace{-0.5cm}
\includegraphics[width=0.48\textwidth]{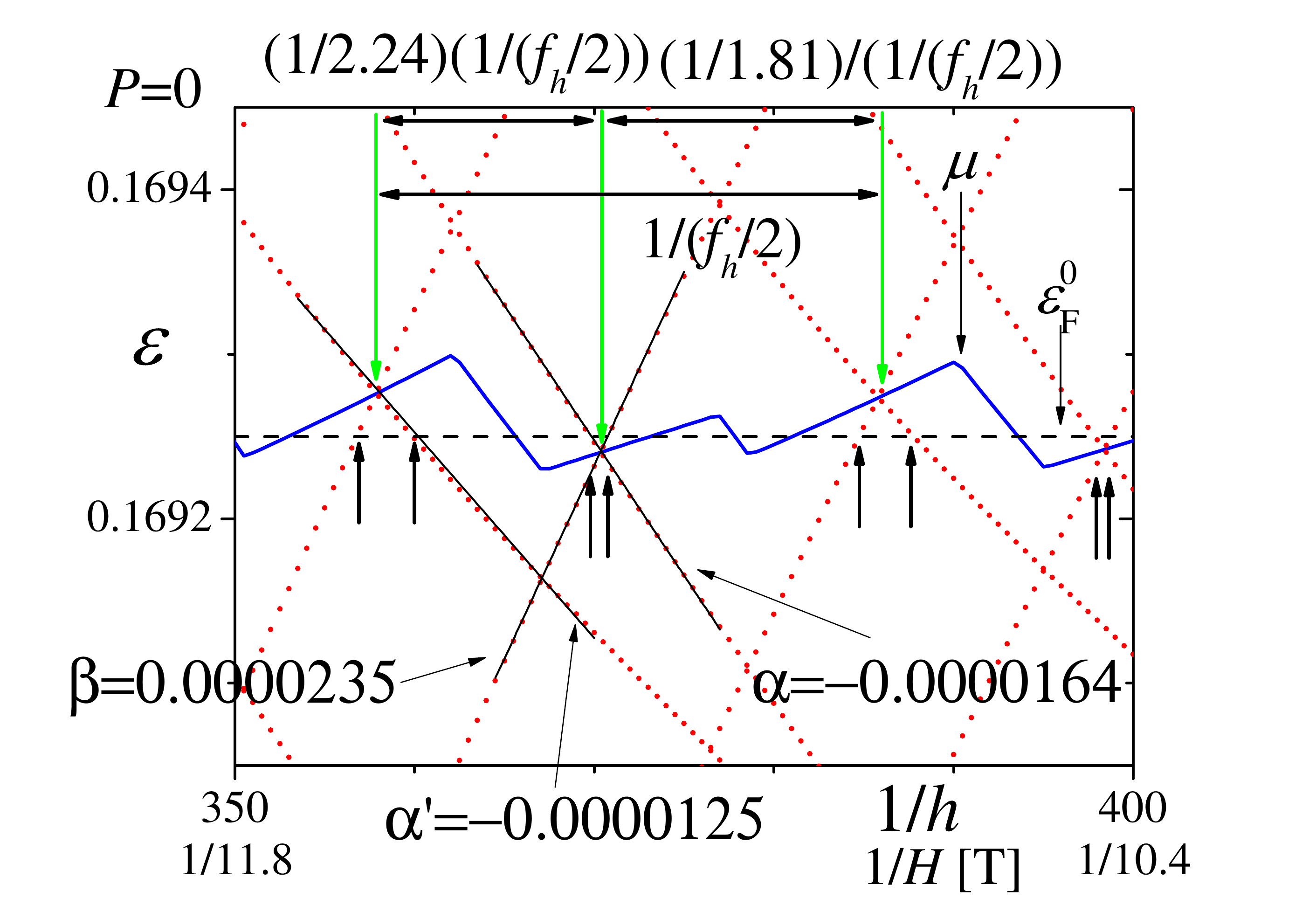}\vspace{-0.1cm}
\begin{flushleft} \hspace{0.5cm}(b) \end{flushleft}\vspace{-0.4cm}
\includegraphics[width=0.48\textwidth]{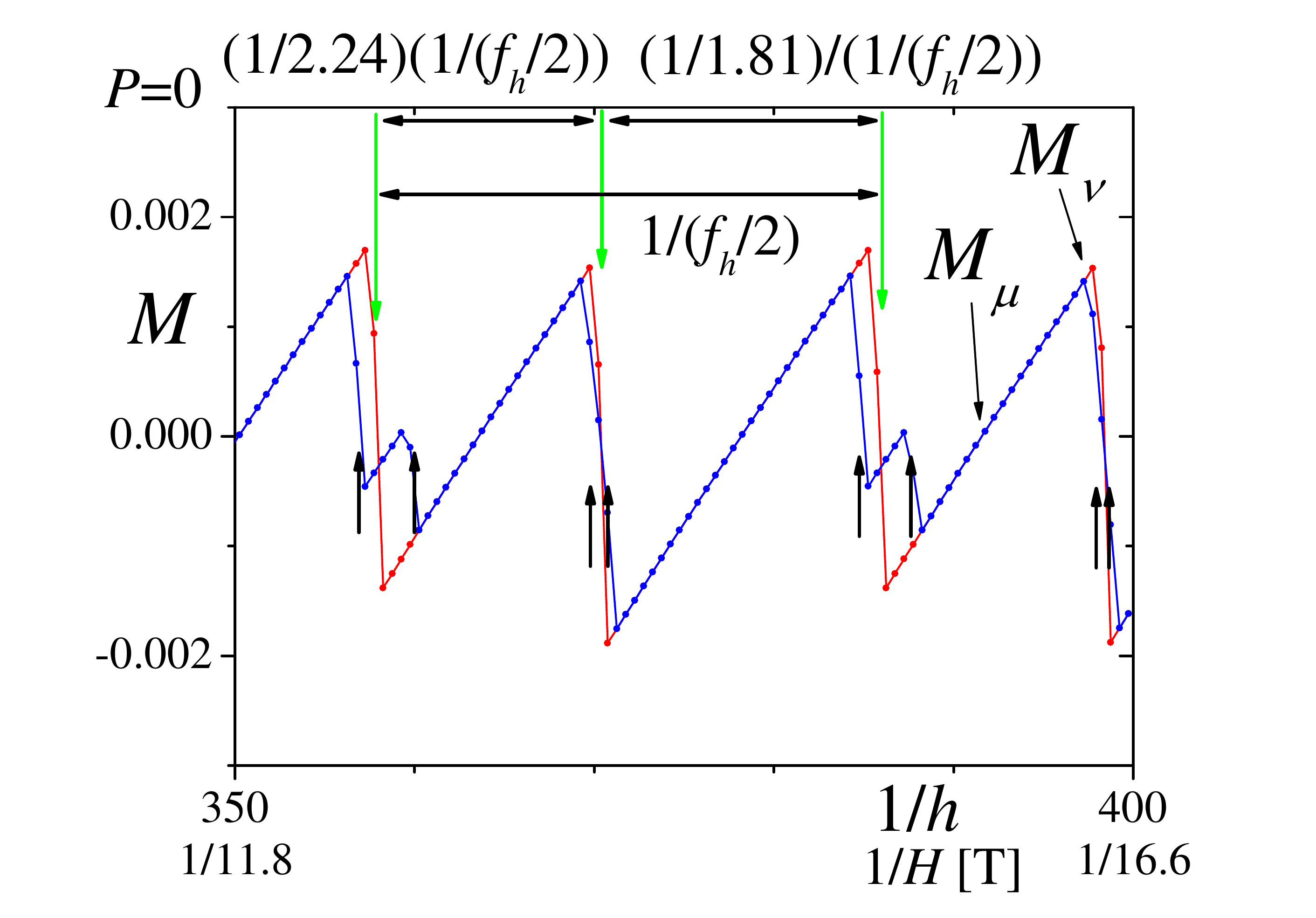}\vspace{-0.0cm}
\caption{
The same figures as Fig. \ref{fig34} at $P=0$, where there is a narrow neck in the electron pocket.
The Landau levels of the electron pocket with the slope $\alpha$ and with $\alpha^{\prime}$ (negatively-sloped lines) are not parallel each other, but the Landau levels of the electron pocket with the same slope value are parallel. 
The Landau levels of the hole pocket (positively-sloped lines) are also parallel each other. The values of slopes are $\alpha=-0.0000164$, $\alpha^{\prime}=-0.0000125$ and $\beta=0.0000235$, respectively [see (a)]. 
The spacing of $1/(f_h/2)$ of the green arrows is periodical. 
Similarly, one pattern of four crossings of $\varepsilon_{\rm F}^0$ and the Landau levels (four black arrows) is repeated periodically with $1/(f_h/2)$. As a result, there appear in $M_{\nu}$ and $M_{\mu}$ the dHvA oscillations with the fundamental frequency of $f_h/2$, which corresponds to the half area of the hole pocket
[equal to the area of green regions of electron pocket in Fig. \ref{fig8_N} (b)].
Since the Landau levels due to electron pockets are not parallel, 
jumps in $M_{\nu}$ (green arrows) and $M_{\mu}$ (black arrows)
occur at incommensurate points in the period [$1/(f_h/2)$] [see (b)]. These incommensurate jumps in $M_{\nu}$ and
$M_{\mu}$ mean that the FTIs are not given by a simple function of $f$.
}
\label{fig35}
\end{figure}

\begin{figure}[bt]
\begin{flushleft} \hspace{0.5cm}(a) \end{flushleft}\vspace{-0.5cm}
\includegraphics[width=0.48\textwidth]{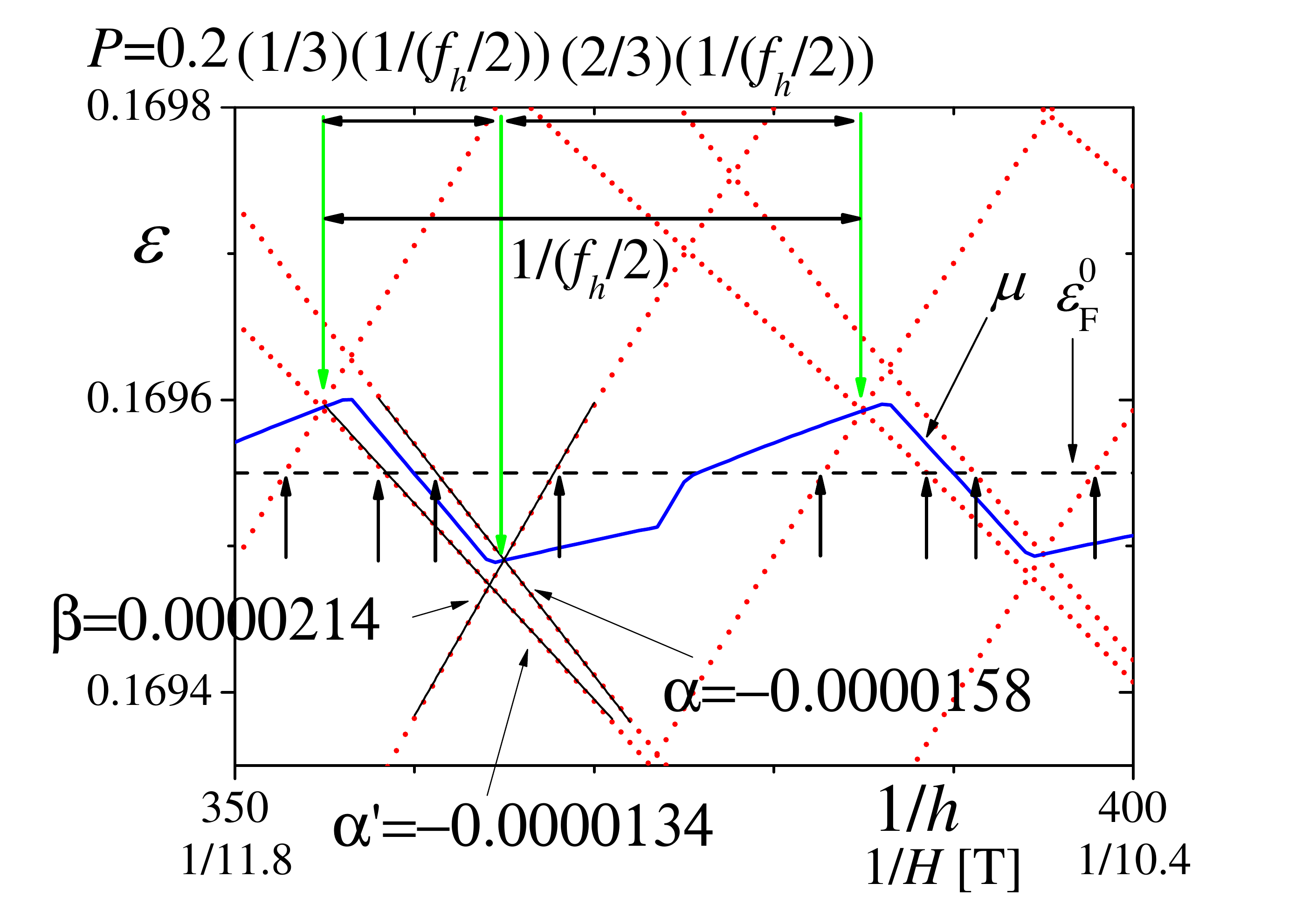}\vspace{-0.1cm}
\begin{flushleft} \hspace{0.5cm}(b) \end{flushleft}\vspace{-0.4cm}
\includegraphics[width=0.48\textwidth]{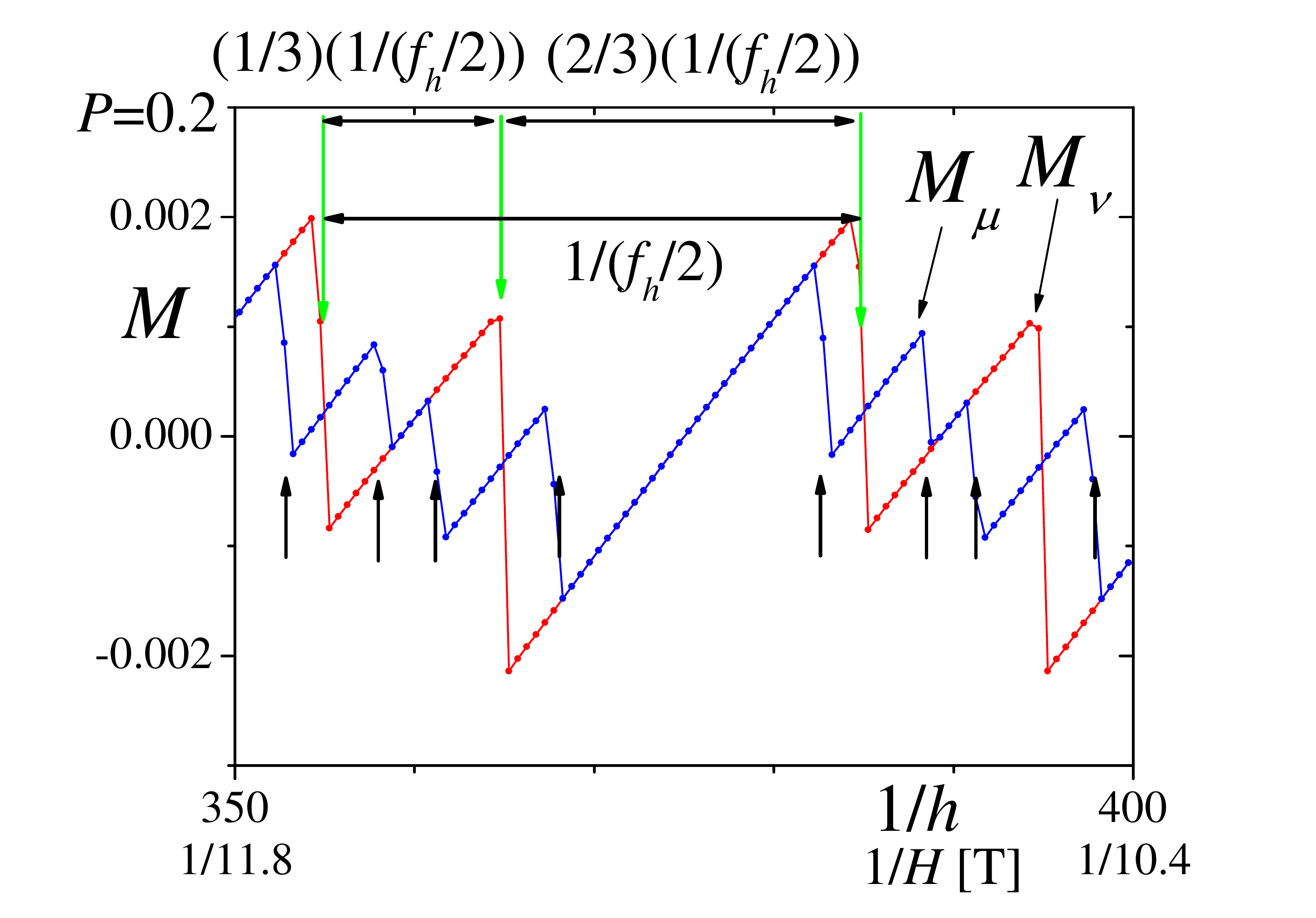}\vspace{-0.0cm}
\caption{
The same figures as Fig. \ref{fig34} at $P=0.2$, where two electron pockets are separated by narrow region in
the momentum space (close to the Lifshitz transition, 
see Fig. \ref{fig8_N} (c)). 
The slope of the Landau levels are $\alpha=-0.0000158$, $\alpha^{\prime}=-0.0000134$ and $\beta=0.0000214$, respectively [see (a)]. The Landau levels of the electron pockets come near as a pair but there is the small difference within the pair. This situation can be interpreted as that almost degenerate Landau levels in two electron pockets couple via tunneling (magnetic breakdown in other word). Although the electron pockets' Landau levels with $\alpha$ and $\alpha^{\prime}$ are not parallel each other, small lifting of the degeneracy of the electron pockets' Landau levels makes the almost-commensurate separation of the green arrows at $1/3:2/3$. As a result, the jumps in $M_{\nu}$ occur at commensurate point $[(1/3)(1/(f_h/2))]$ in the period [$1/(f_h/2)$], whereas jumps in $M_{\mu}$ occur at $(1/2)(1/(f_h/2))$ and incommensurate points [see (b)]. This fact explains why $3f_h/2$ frequency of the dHvA oscillations is enhanced when $\nu$ is fixed, but not when $\mu$ is fixed.
}
\label{fig36}
\end{figure}

\begin{figure}[bt]
\begin{flushleft} \hspace{0.5cm}(a) \end{flushleft}\vspace{-0.5cm}
\includegraphics[width=0.48\textwidth]{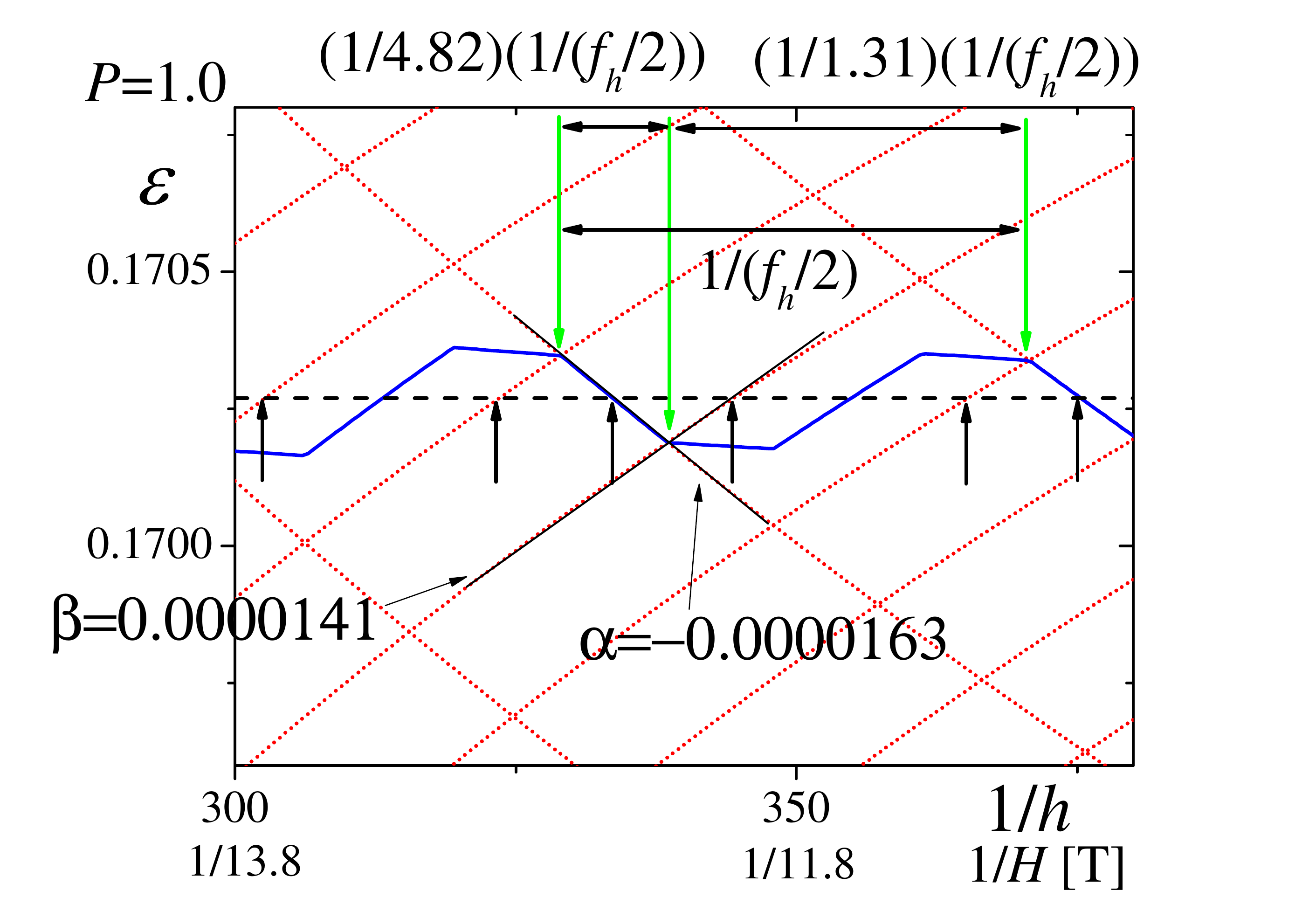}\vspace{-0.1cm}
\begin{flushleft} \hspace{0.5cm}(b) \end{flushleft}\vspace{-0.4cm}
\includegraphics[width=0.48\textwidth]{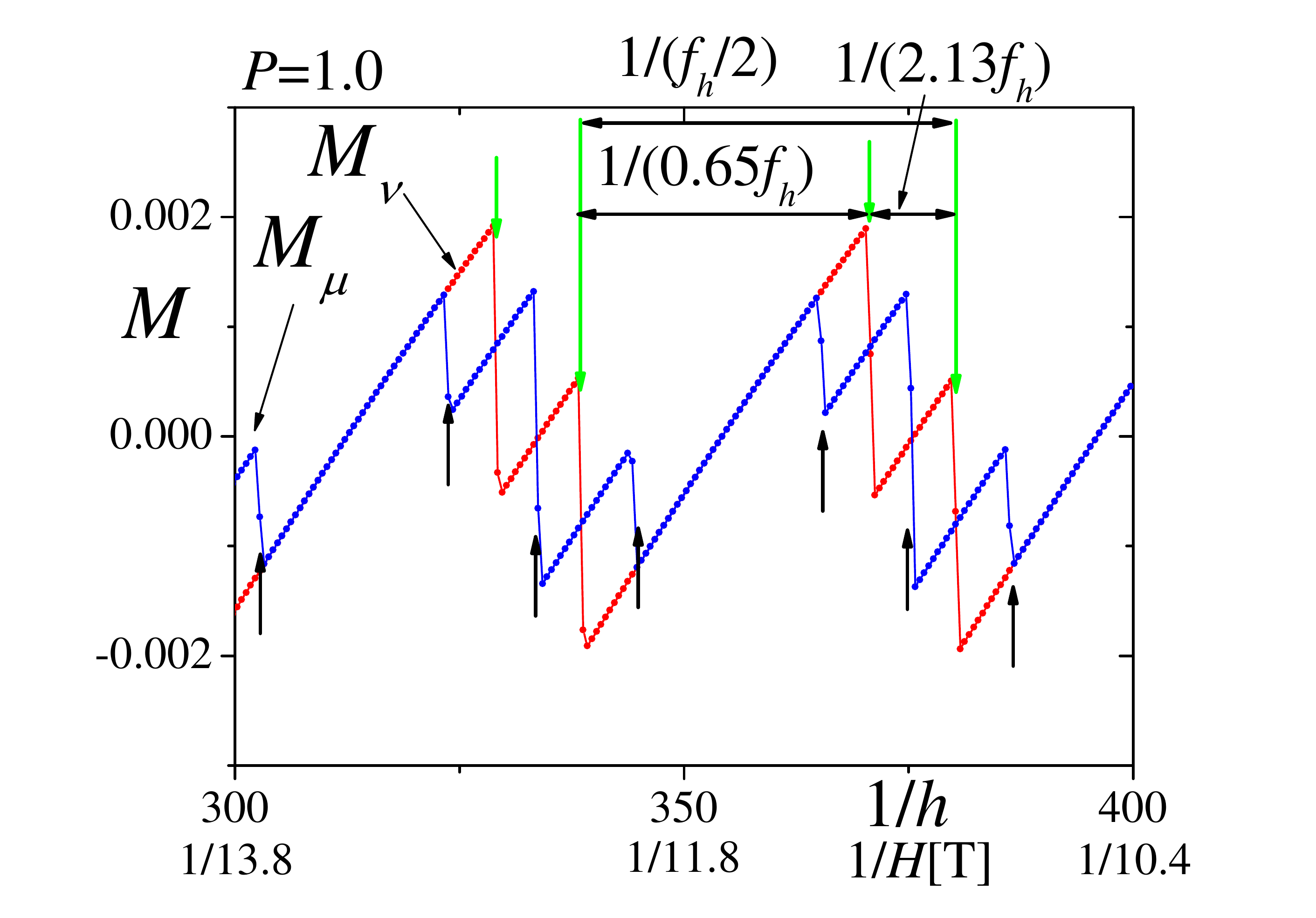}\vspace{-0.0cm}
\caption{
The same figures as Fig. \ref{fig34} at $P=1.0$, where two electron pockets are separated enough in the momentum space [see Fig. \ref{fig8_N} (d)]. The Landau levels of the electron pockets are degenerated as a pair in this scale of figures. The absolute value of the slope ($|\alpha|=0.0000141$) of the electron pockets' Landau levels is smaller than that ($|\beta|=0.0000163$) of the hole pocket's Landau levels. Jumps in $M_{\nu}$ and $M_{\mu}$ occur at incommensurate points in the period [$1/(f_h/2)$] [see (b)]. 
}
\label{fig36_b}
\end{figure}

We show an enlarged view of Fig. \ref{fig18} (a) at $P=-0.4$ in Fig. \ref{fig34} (a), where the crossing points (green arrows) of the Landau levels (red dotted lines) and $\mu$ (a blue line) are almost the same as those (black arrows) of the Landau levels and $\varepsilon_{\rm F}^0$ (a dotted black line). The jumps of $M_{\nu}$ and $M_{\mu}$ also occur at these crossing points, as shown in Fig. \ref{fig34} (b). 
We show the enlarged views of Figs. \ref{fig18} (b), (c) and (d) at $P=0$, 0.2 and 1.0 in Figs. \ref{fig35} (a), \ref{fig36} (a) and \ref{fig36_b} (a). In Figs. \ref{fig34} (a), \ref{fig35} (a), \ref{fig36} (a) 
and  \ref{fig36_b} (a), $\mu$ always intersects the points where the occupied electron's Landau levels and the unoccupied hole's Landau levels cross. 
The crossing points (green arrows) of the Landau levels and $\mu$ are different from those (black arrows) of the Landau 
levels and $\varepsilon_{\rm F}^0$. The jumps of $M_{\nu}$ and $M_{\mu}$ happen at green arrows and black arrows, respectively, as shown in Figs. \ref{fig35} (b), \ref{fig36} (b) and \ref{fig36_b} (b). 


\begin{figure}[bt]
\begin{flushleft} \hspace{0.5cm}(a) \end{flushleft}\vspace{-0.8cm}
\includegraphics[width=0.39\textwidth]{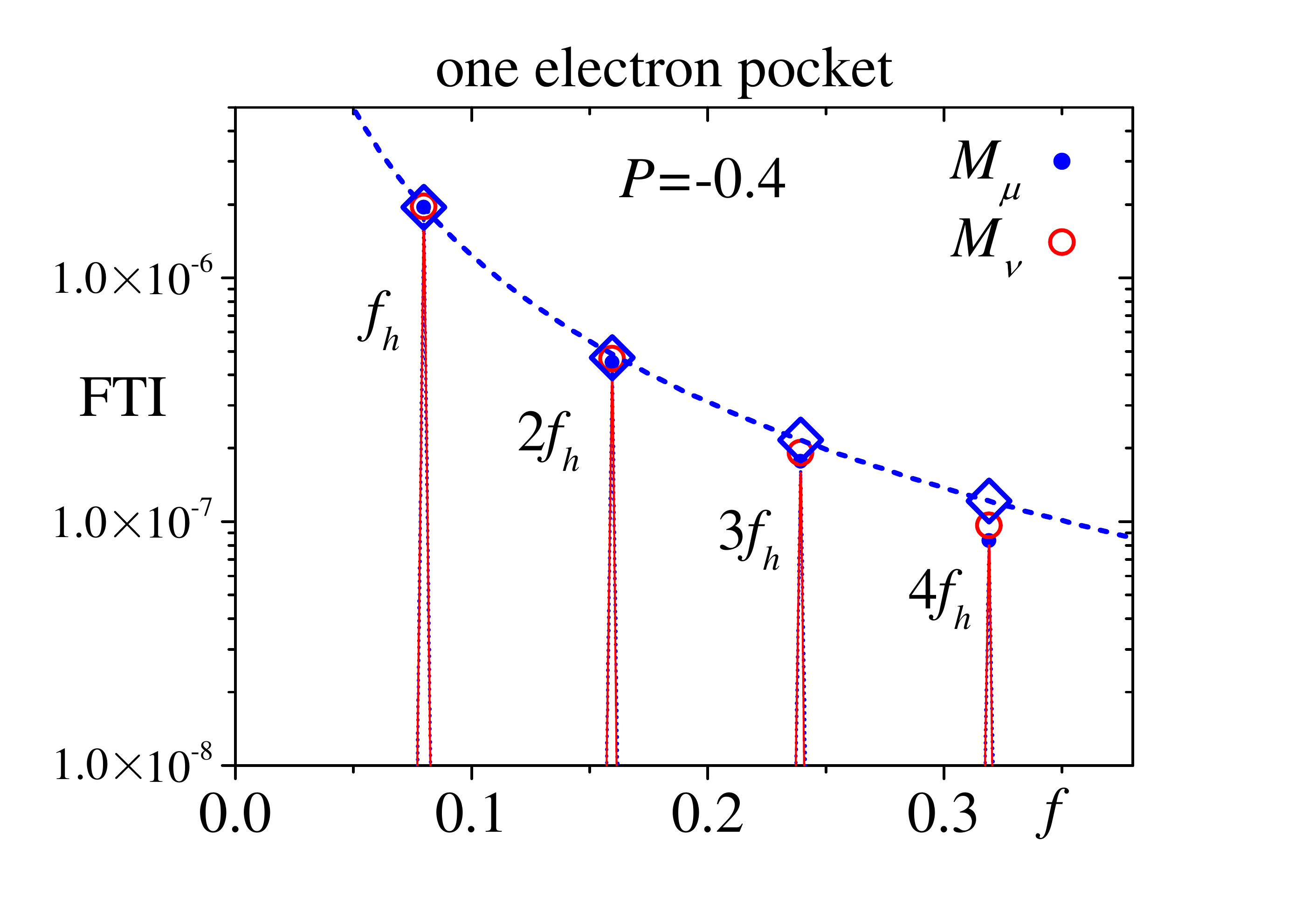}\vspace{-0.7cm}
\begin{flushleft} \hspace{0.5cm}(b) \end{flushleft}\vspace{-0.7cm}
\includegraphics[width=0.39\textwidth]{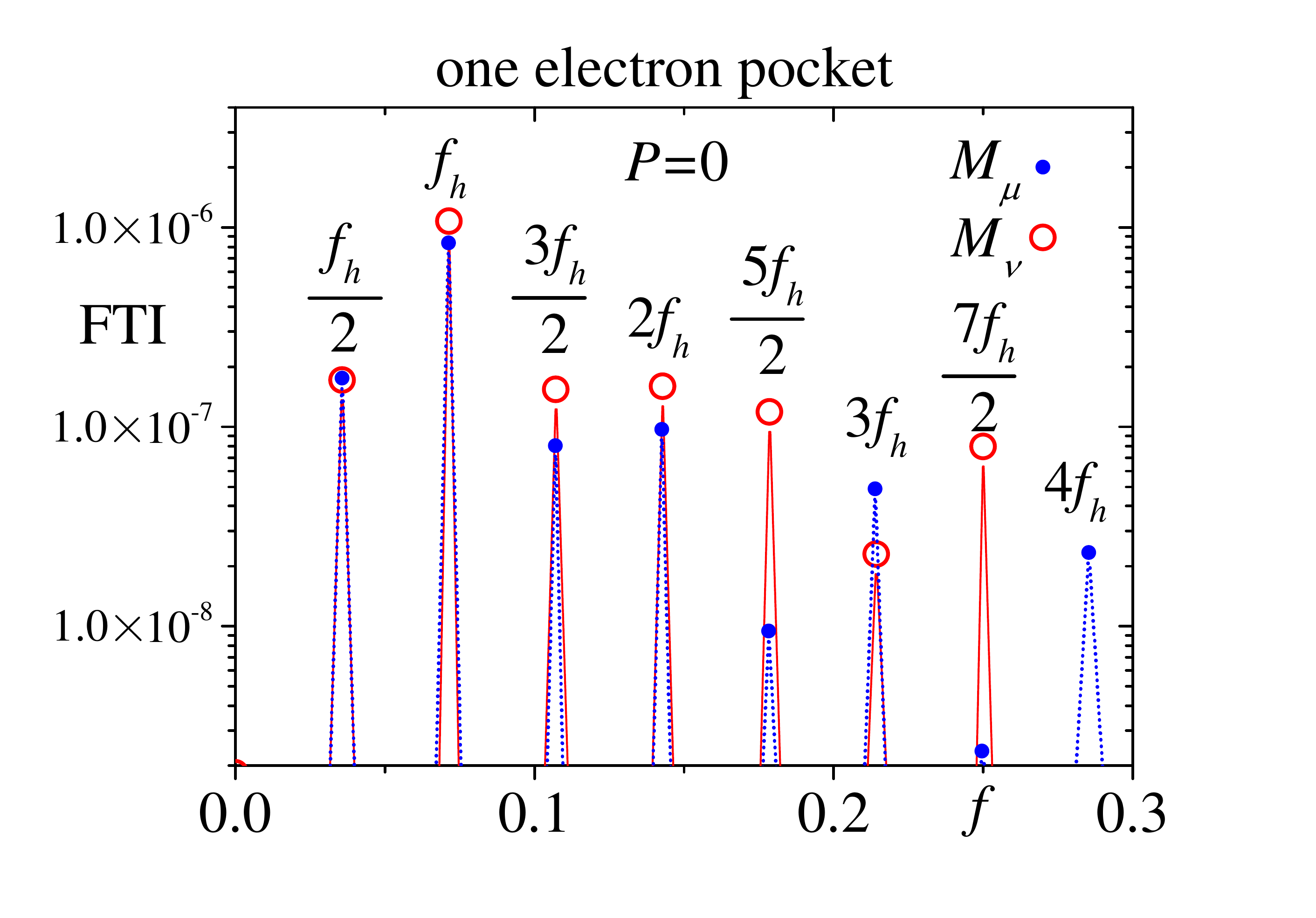}\vspace{-0.7cm}
\begin{flushleft} \hspace{0.5cm}(c) \end{flushleft}\vspace{-0.7cm}
\includegraphics[width=0.39\textwidth]{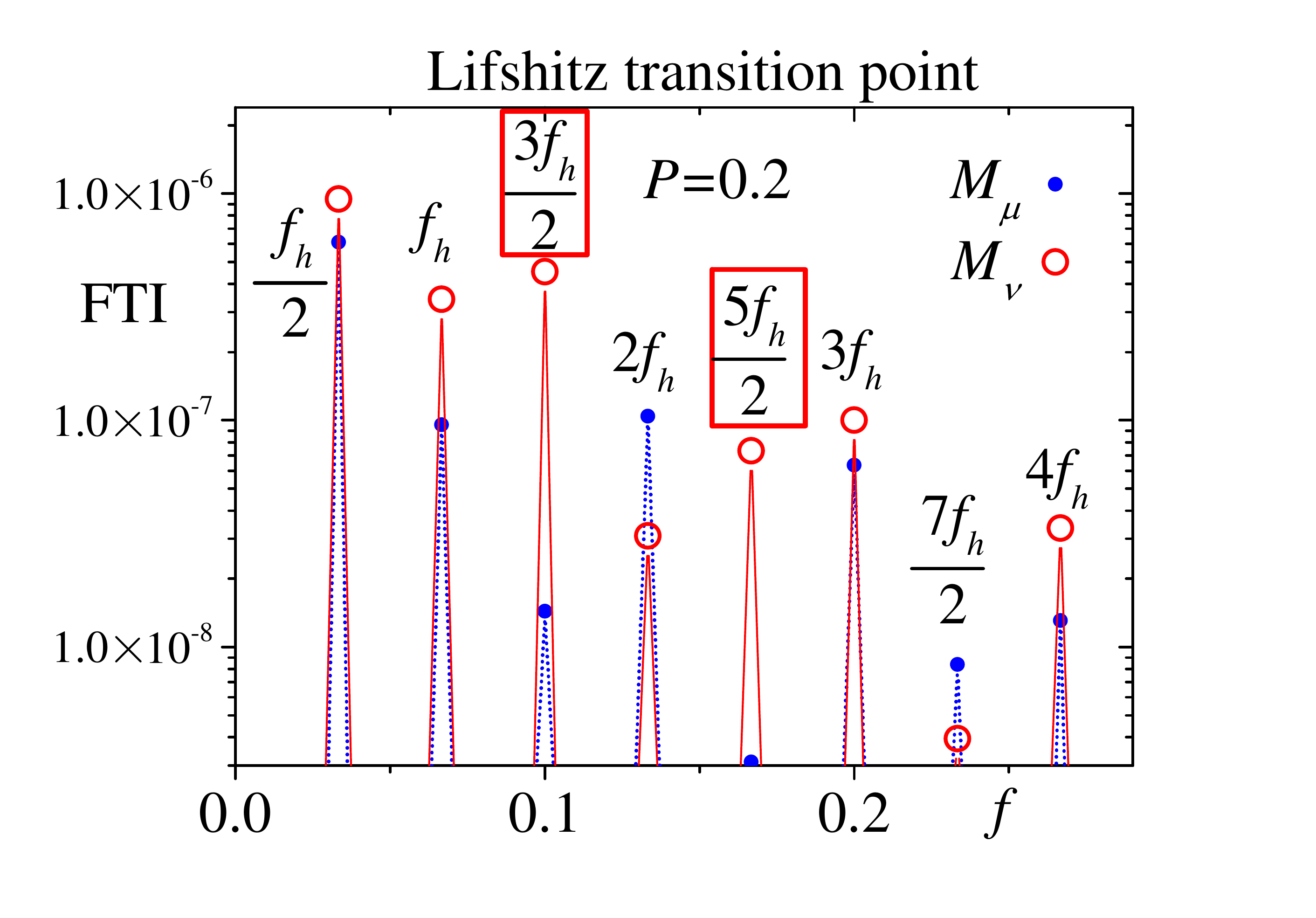}\vspace{-0.7cm}
\begin{flushleft} \hspace{0.5cm}(d) \end{flushleft}\vspace{-0.7cm}
\includegraphics[width=0.39\textwidth]{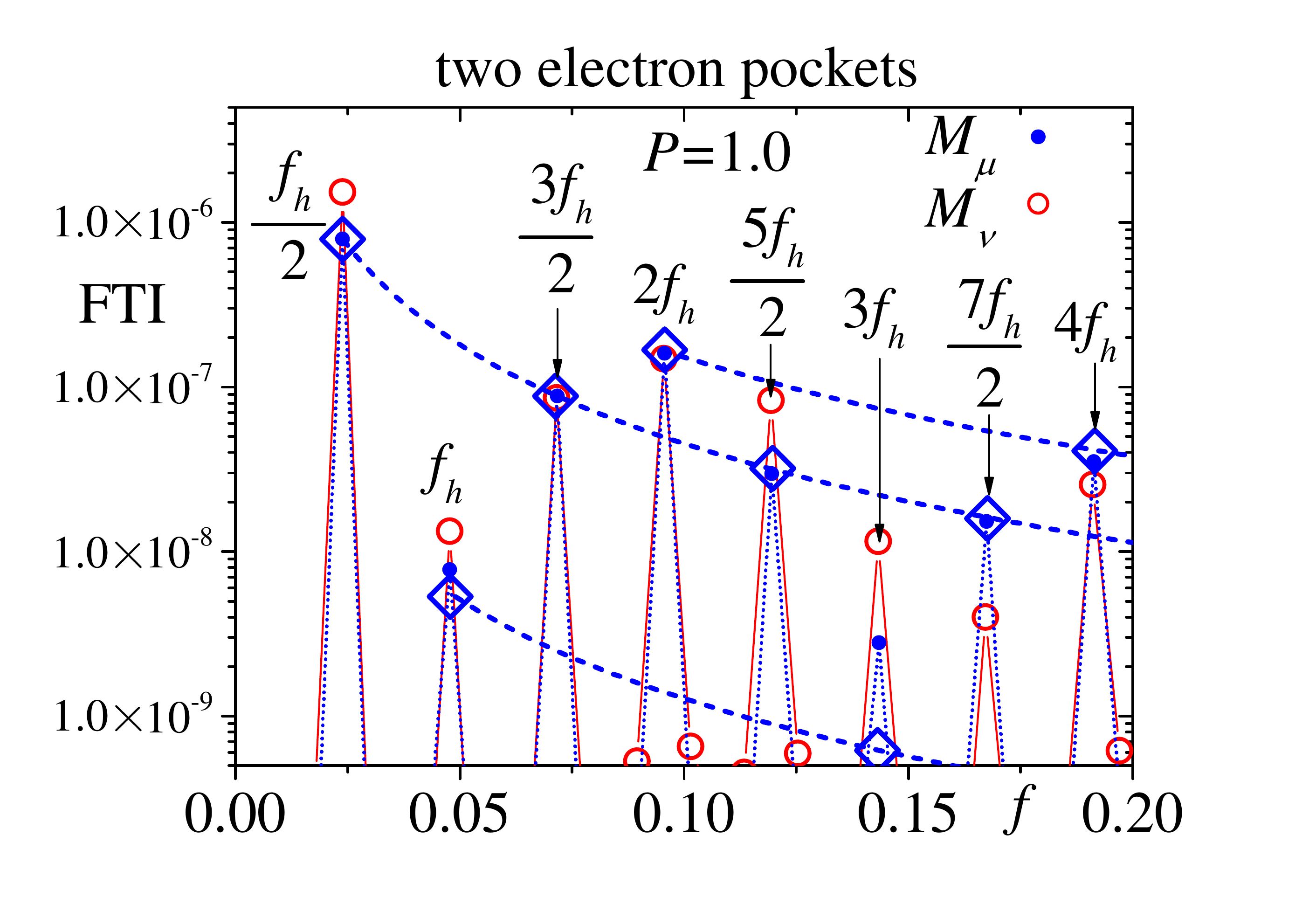}\vspace{-0.1cm}
\caption{
The FTIs of $M_{\nu}$ and $M_{\mu}$ at $P=-0.4$  (one electron pocket) (a), $P=0$  (one electron pocket) (b), $P=0.2$ (Lifshitz transition point) (c) and $P=1.0$ (two electron pockets) (d).
The regions of the Fourier transform are $200\leq 1/h\leq 388$ in (a), 
$202.75\leq 1/h\leq 398.75$ for $M_{\nu}$ and $203.25\leq 1/h\leq 399.25$ for $M_{\mu}$ in (b), 
$215.75\leq 1/h\leq 395.25$ for $M_{\nu}$ and $218.25\leq 1/h\leq 398.25$ for $M_{\mu}$ in (c) and 
$213.5\leq 1/h\leq 381$ for $M_{\nu}$ and $209.25\leq 1/h\leq 376.5$ for $M_{\mu}$ in (d). 
In (a) and (d), blue diamonds are the FTIs of the LK formula. 
The dotted blue lines are the guides to the eye. In (c), the FTIs at $3f_h/2$ and $5f_h/2$ are enhanced. 
}
\label{fig29}
\end{figure}

The FTIs of $M_\nu$ and $M_\mu$ are shown in Fig.~\ref{fig29} at $P=-0.4, 0, 0.2$ and 1.0. 
When $P=-0.4$, there are peaks at $f_h \simeq 0.0798$ and the higher harmonics $\ell f_h$, and they are proportional to $1/\ell$ [blue dotted line in Fig. \ref{fig29} (a)]. This is due to the saw-tooth waveform of $M_{\nu}$ and $M_{\mu}$ [see Fig. \ref{fig34} (b)]. The value of $f_h$ corresponds to 
the area of an electron pocket and a hole pocket ($A_e/A_{\textrm{BZ}}=|A_h|/A_{\textrm{BZ}}\simeq 0.0797$), as shown in Fig.~\ref{fig8_N} (a). This coincidence is expected in the LK formula. 
Furthermore, the FTIs at $f_h, 2f_h, 3f_h, \cdots, lf_h, \cdots$ are well fitted by the LK formula [Eq. (\ref{LK_03})] [Blue diamonds in Fig.~\ref{fig29} (a)], where the value of the FTI at $f_h$ in $M_{\mu}$ is used as $M_0\widetilde{A}_e$. In that fitting, the amplitude modified by the network model is not used. Thus, at $P=-0.4$ and 10.4 T$\leq H \leq 20.7$ T, we can neglect the inter-band coupling by the magnetic breakdown.

At $P=0$, there are the peaks at $f_h\simeq 0.0714$, $2f_h$ and $3f_h$ in the FTIs of Fig.~\ref{fig29} (b), where $f_h$ corresponds to 
the area of an electron pocket and a hole pocket 
($A_e/A_{\textrm{BZ}}=|A_h|/A_{\textrm{BZ}}\simeq 0.0714$), as shown in Fig.~\ref{fig8_N} (b). 
These peaks are expected in the LK formula. 
In addition, we can see the peaks at $f_h/2$, 
$3f_h/2$, $5f_h/2$ and $7f_h/2$ in $M_{\nu}$ and $M_{\mu}$, although there are no closed orbits with the areas corresponding to $f_h/2$, $3f_h/2$, $5f_h/2$ and $7f_h/2$. 
These peaks might be explained by the semiclassical network model with tuning tunneling parameters. 
For example, for the peaks at $f_h/2$ in $M_{\nu}$ and $M_{\mu}$, an electron's effective closed orbital motion with the green area ($A_e/2=|A_h|/2$) in Fig.~\ref{fig8_N}(b) is possible by the tunneling thorough the narrow neck when the magnetic field is strong. Similarly, the small peaks at $3f_h/2$, $5f_h/2$ and $7f_h/2$ in $M_{\mu}$ and $M_{\nu}$ might be understood by the multi-tunneling\cite{Pippard62,Falicov66}.


At $P=0.2$, there are the peaks at $f_h\simeq 0.00667, 2f_h$, $3f_h$ and $4f_h$ in $M_{\nu}$ and $M_{\mu}$, as shown in Fig.~\ref{fig29} (c). The value of $f_h$ corresponds to the area of a hole pocket ($|A_h|/A_{\rm BZ}\simeq 0.0666$) or the sum of the area of two electron pockets ($A_e=|A_h|$), as shown in Fig.~\ref{fig8_N}(c). 
The large peaks at $f_h/2$ in $M_{\nu}$ and $M_{\mu}$ also appear, which corresponds to the area [$A_e/(2A_{\rm BZ})\simeq 0.0333$] of a small electron pocket. 
These are expected in the LK formula. 
Moreover, there are peaks at $3f_h/2$, $5f_h/2$ and $7f_h/2$ in Fig.~\ref{fig29} (c).  These peaks might be explained by the network model for the magnetic breakdown.

Note that the peaks at $3f_h/2$ and $5f_h/2$ in $M_{\nu}$ are large, whereas these are very small in $M_{\mu}$, as shown in Fig.~\ref{fig29} (c). The peaks at $3f_h/2$ and $5f_h/2$ are maximized at $P=0.2$ and $P=0.1$ (near the Lifshitz transition), respectively, as shown in Fig.~\ref{fig30}. 
We consider that the enhancement of the peak at $3f_h/2$ in $M_{\nu}$ is closely related to the periodicities of 
the crossings (green arrows) of the Landau levels and $\mu$ in Figs. \ref{fig35}, \ref{fig36} and \ref{fig36_b}.

The fundamental period of $1/(f_h/2$) in $M_{\nu}$ at $P=0$ and 0.2 comes from the crossings of every second negatively-sloped lines, every second positively-sloped lines and $\mu$. At $P=1.0$ it comes from the crossings of every negatively-sloped lines, every second positively-sloped lines and $\mu$, where negatively-sloped lines (electron pocket's Landau levels) 
are effectively degenerated doubly. That fundamental period is divided into two periods 
[$(1/2.24)(1/(f_h/2))$ and $(1/1.81)(1/(f_h/2))$ in 
Fig. \ref{fig35}, $(1/3)(1/(f_h/2))$ 
and $(2/3)(1/(f_h/2))$ in Fig. \ref{fig36} and $(1/4.82)(1/(f_h/2))$ and $(1/1.81)(1/(f_h/2))$ in Fig. \ref{fig36_b}]. 
These two periods are changed by the spacing of the separations of negatively-sloped lines. 
When $P=0.2$, the ratio of two periods ($1/3:2/3)$ is a simple one, as shown in Fig. \ref{fig36}. Then, the FTI at $3f_h/2$ in $M_{\nu}$ becomes large. This is explained in Appendix \ref{AppendixE_0}. Thus, the enhancement of $3f_h/2$ near the Lifshitz transition is caused by the commensurate separation of the Landau levels. On the other hand, since the separation of electron pockets' Landau levels hardly affects the fundamental period of $1/(f_h/2)$, the FTI at $f_h/2$ is not maximized near the Lifshitz transition. 


At $P=1.0$, there is a large peak at $f_{h}/2\simeq 0.0239$ in the 
FTIs [Fig.~\ref{fig29}(d)]. This frequency corresponds to each area of an electron pocket [$A_e/(2A_{\rm BZ})\simeq 0.0240$] in Fig.~\ref{fig8_N}(d). 
Note that the peaks at $f_h$ and $3f_h$ in $M_{\nu}$ and  $M_{\mu}$ are very small, as shown in Fig.~\ref{fig29}(d). 


%

We have already found\cite{KH2017} the smallness of these peaks in $M_{\mu}$.
In that paper we could not explain the smallness, which would be large in the LK formula with free electrons. Here, we show that the smallness can be explained by the LK formula with the difference of the phase factors for the electron pockets and a hole pocket. 
The phase factor for a hole pocket is $\gamma_h=1/2$,
as in the usual free holes with parabolic energy dispersion, whereas that for two electron pockets
is $\gamma_e=0$, which is the cases in the Dirac fermions\cite{Mikitik}. 
Since the system is overtilted Dirac fermions at $P=1.0$, 
taking $\gamma_e=0$ is not completely justified but it is plausible to take $\gamma_e=0$ due to 
not opening the gap at the Dirac points. 
Then, the LK formula for $P=1.0$ is given by Eq. (\ref{LK_5}) in Appendix \ref{AppendixE}. 
By using the value of $D_4(\varepsilon^0_{\rm F})/D_3(\varepsilon^0_{\rm F})$ in Fig. \ref{fig9_N} (d), we obtain $\widetilde{A}_h/\widetilde{A}_e=0.832$ from Eq. (\ref{DOS}). 
In that case, by setting the value of the FTI at $f_h/2$ in $M_{\mu}$ as $|M_0\widetilde{A}_e|^2$, we can determine the values of Eqs. (\ref{LK_8}) and (\ref{LK_9}), which are indicated by the blue diamonds in Fig.~\ref{fig29}(d). 
These almost coincide with
the FTIs at $f_h/2$, $3f_h/2$, $2f_h$, $5f_h/2$, $7f_h/2$ and $4f_h$ in $M_{\mu}$. However, the FTIs at $f_h$ and $3f_h$ are deviated from these fitting (blue diamonds). 
We guess that the deviations at $f_h$ and $3f_h$ may be due to the magnetic breakdown or the numerical errors.


%


\section{Conclusions}

We study the dHvA oscillations in the two-dimensional compensated metal with overtilted Dirac cones near the Lifshitz transition by using the tight-binding model of $\alpha$-(BEDT-TTF)$_2$I$_3$. 
It is shown that in this system the dHvA oscillations under the condition of the fixed electron number are caused by the periodical crossings of the Landau levels and the chemical potential. This is in contrast with the dHvA oscillations in the two-dimensional free electrons whose origin is periodical jumps of the chemical potential.

At $P\sim 0.2$~kbar and $10.4$ T$\leq H\leq 20.7$ T, we find the enhancements 
of the dHvA oscillations with the frequencies of $3f_h/2$ and $5f_h/2$ under the condition of the fixed electron number. 
These enhancements happen in the pressure region 
where the topology of the Fermi surface changes from one electron pocket to two electron pockets (Lifshitz transition). 
These enhancements are caused by the commensurate separation of two Landau levels with the phase factor ($\gamma_e\simeq 0$) seen in Dirac fermions. 
These will be observed in the two-dimensional overtilted Dirac fermions if the situation near the Lifshitz transition is realized by applying the pressures, doping, etc.


\appendix

\section{Lifshitz and Kosevich formula}
\label{LKformula_dHvA}


Although 
the magnetization should be calculated under the condition of the fixed electron number or the fixed electron filling, $\nu$, (canonical ensemble), Lifshitz and Kosevich\cite{shoenberg,LK} have calculated it under the condition of the fixed chemical potential, $\mu$,  (grand canonical ensemble). This is because the calculation in the grand canonical ensemble is justified if $\mu$ does not depend on the magnetic field or depends very weakly on the magnetic field.

They have derived the LK formula\cite{shoenberg,LK} for the free electron model by using the semiclassical quantization rule\cite{Onsager}. Recently, it has been also shown that the LK formula can be used for the Dirac fermions\cite{Igor2004PRL,Igor2011,Sharapov}. The LK formula at $T=0$ for the two-dimensional multi closed Fermi surface with the area $(A_i)$ is given by
\begin{eqnarray}
M^{\rm LK}&=&M_0\sum_{i}
\widetilde{A}_i\sum_{l=1}^{\infty}\frac{1}{l}\sin\left[2\pi l\left(\frac{F_i}{H}-\gamma_i\right)
\right],\label{LK_0} \\
M_0&=&-\frac{e}{2\pi^2 c\hbar},\\
\widetilde{A}_i&=&\frac{|A_i|}{\frac{\partial A_i(\varepsilon^0)}{\partial \varepsilon^0}\big|_{\varepsilon^0=\mu}}, \label{Ai}
\end{eqnarray} 
where $i$ is the index for the closed orbit and the frequency ($F_i$) is given by 
\begin{equation}
F_i=\frac{c \hbar |A_i|}{2 \pi e}, 
\end{equation}
where 
$A_i>0$ and $A_i<0$ are for the electron pocket 
and for the hole pocket, respectively. Note that $\widetilde{A}_i$ is positive both for the electron pocket and for the hole pocket in the situation studied in this paper. 
When we use $h$ instead of $H$ in Eq. (\ref{LK_0}), we get 
\begin{equation}
\frac{F_i}{H}=\frac{f_i^{}}{h}, 
\end{equation}
where 
\begin{equation}
f_i=\frac{|A_i|}{A_{\rm BZ}}. 
\end{equation}
In Eq. (\ref{LK_0}), $\gamma_i$ is the phase of the oscillation, which comes from the phase of the Landau levels. For example, $\gamma=1/2$ and $\gamma=0$ are 
for the free electron model and the Dirac fermions\cite{Mikitik}, respectively. We consider the case of the single free electron pocket with the area of $A_e$. Eq. (\ref{Ai}) becomes 
\begin{eqnarray}
M^{\rm LK}&=&M_0
\widetilde{A}_e\sum_{l=1}^{\infty}\frac{1}{l}\sin\left[2\pi l\left(\frac{F_e}{H}-\frac{1}{2}\right)
\right].\label{LK_02}
\end{eqnarray}

\begin{figure}[bt]
\begin{flushleft} \hspace{0.5cm}(a)
 \end{flushleft}\vspace{-0.3cm}
\includegraphics[width=0.29\textwidth]{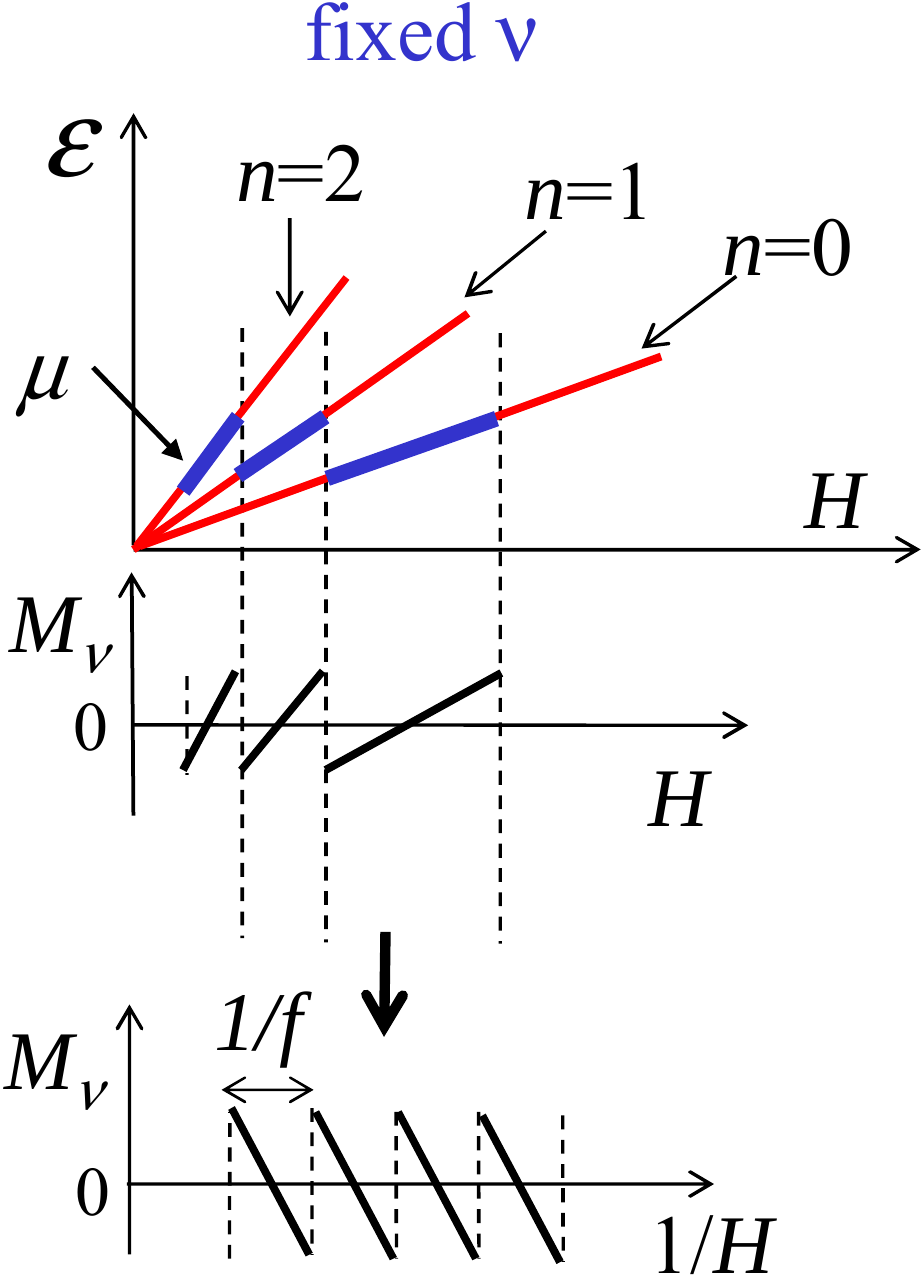}\vspace{-0.1cm}
\begin{flushleft} \hspace{0.5cm}(b) \end{flushleft}\vspace{-0.3cm}
\includegraphics[width=0.29\textwidth]{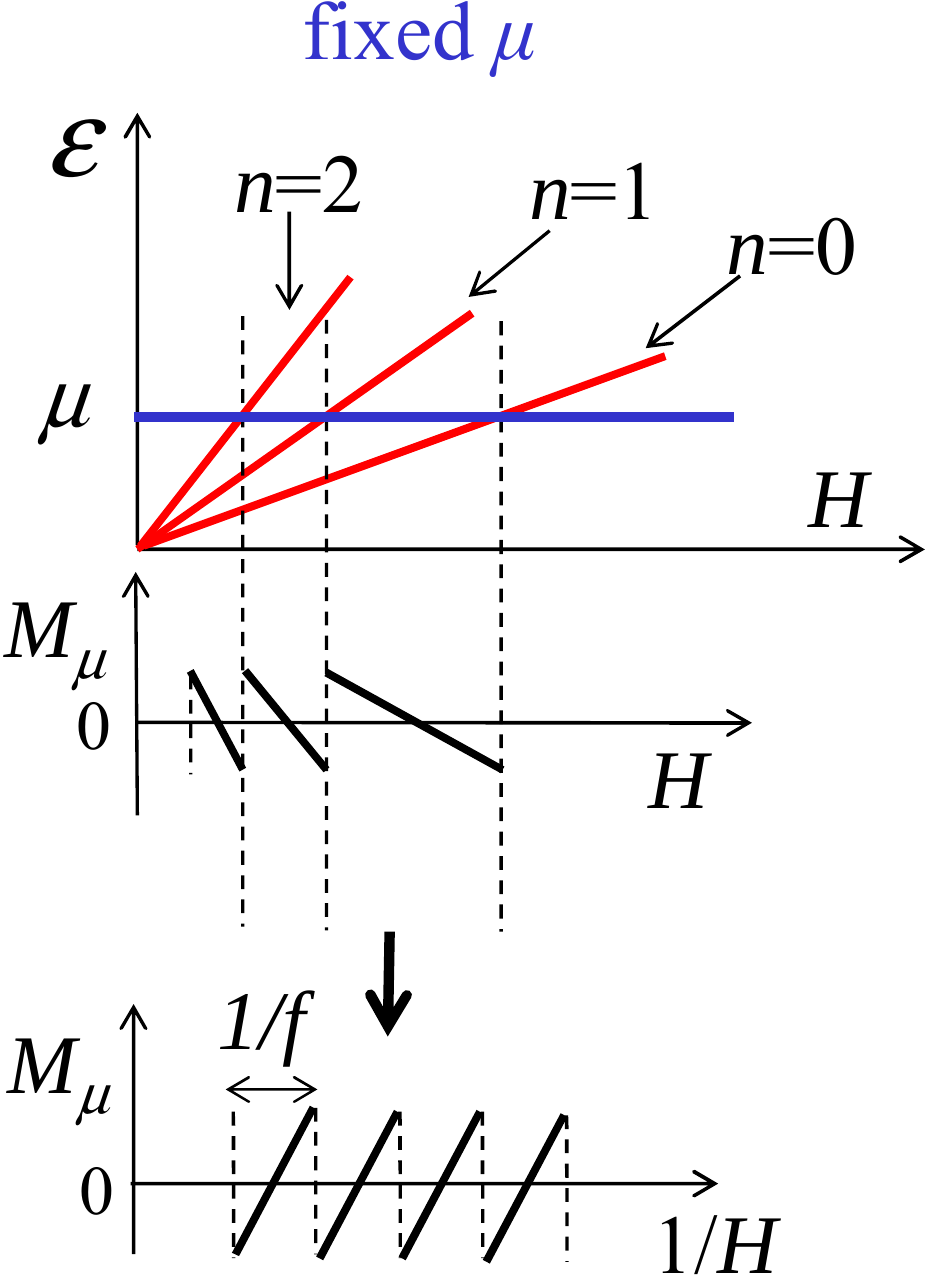}\vspace{0.1cm}
\caption{
In the two-dimensional system with an electron pocket, schematic figures of the Landau levels (red lines), the chemical potential (blue lines) and the dHvA oscillations (black lines). (a) and (b) are for the conditions of 
the fixed electron number and of the fixed chemical potential, respectively. 
}
\label{fig37}
\end{figure}

Under the condition of the fixed $\nu$, 
 the highest Landau level is partially filled, i.e., $\mu$ is pinned at the Landau level at $T=0$, as shown in Fig. \ref{fig37}(a). As the magnetic field is increased, the degeneracy of each Landau level increases, and $\mu$ jumps periodically as a function of $1/H$. These jumps are the origins of the dHvA oscillations. Under the condition of the fixed $\mu$, the Landau levels and $\mu$ crosses periodically. Then, the dHvA oscillations appear, as shown in Fig. \ref{fig37}(b). The wave form of $M_{\nu}$ is inverted from that of $M_\mu$ [Eq. (\ref{LK_02})].

\section{Fourier transform intensities}
\label{AppendixD}
In order to analyze the oscillations in the magnetization, 
we calculate the Fourier transform intensities numerically  as follows.
By choosing the center ($h_c$) and the finite range ($2L$), we calculate 
\begin{equation}
  \mathrm{FTI}(f, \frac{1}{h_c},L) =\left|\frac{1}{2L} 
  \int_{\frac{1}{h_c}-L}^{\frac{1}{h_c}+L} M (h) e^{2\pi i \frac{f}{h}} d \left(\frac{1}{h}\right)\right|^2, \label{FTI}
\end{equation}
where we take $f=j/(2L)$ with integer $j$ ($0\leq j\leq 256$ is used in this study).

For example, in the system with one electron pocket, from Eq. (\ref{LK_02}) the FTIs at $f_e, 2f_e, 3f_e, \cdots$ in $M^{\rm LK}$ become 
\begin{eqnarray}
{\rm FTI}(f=lf_e)&=&
\bigg|\frac{M_0\widetilde{A}_e}{l}\bigg|^2, \label{LK_03}
\end{eqnarray}
where $l=1, 2, 3, \cdots$. 


\section{The time reversal invariant momentum}
\label{TRIM}

The time reversal invariant momentum (TRIM),
$\mathbf{p}_{\textrm{TRIM}}=\hbar \mathbf{k}_{\textrm{TRIM}}$,
is given by
the reciprocal lattice vectors ($\mathbf{G}_j$), where $j=1, 2$ ($j=1, 2, 3$) if we study a two-dimensional system (a three-dimensional system), as
\begin{equation}
\mathbf{k}_{\mathrm{TRIM}} = \sum_{j} \frac{1}{2} n_j \mathbf{G}_j,
\end{equation}
where $n_j=0$ or $1$.
In the two-dimensional system there are four TRIM's in the first Brillouin zone.
Note that $\mathbf{k}_{\mathrm{TRIM}}$ and $-\mathbf{k}_{\mathrm{TRIM}}$
are equivalent.
We study the system having the inversion symmetry,
i.e.
\begin{equation}
{\hat {\cal H}}={\hat P}{\hat {\cal H}} {\hat P}^{-1}, \label{php}
\end{equation}
where ${\hat P}$ is the parity operator and ${\hat P}^{-1}={\hat P}$ and ${\hat {\cal H}}$ is the Hamiltonian. 
The eigenstate and the energy are given by
\begin{equation}
{\hat {\cal H}} \lvert \mathbf{k} \rangle=\epsilon_{\mathbf{k}} \lvert \mathbf{k} \rangle.
  \label{eq1}
\end{equation}
By using Eq. (\ref{php}), we obtain
\begin{align}
 {\hat {\cal H}} \lvert \mathbf{k} \rangle &={\hat P} {\hat {\cal H}} {\hat P}^{-1} \lvert \mathbf{k} \rangle \\
   &={\hat P} {\hat {\cal H}} \lvert-\mathbf{k} \rangle \\
   &={\hat P} \epsilon_{-\mathbf{k}} \lvert-\mathbf{k} \rangle\\
   &=\epsilon_{-\mathbf{k}} \lvert \mathbf{k} \rangle.
   \label{eq2}
\end{align}
 From Eq.~(\ref{eq1}) and Eq.~(\ref{eq2}) we obtain
\begin{equation}
\epsilon_{\mathbf{k}}=\epsilon_{-\mathbf{k}}.
\label{eqepsilonk}
\end{equation}
 We assume that the energy band $\lvert \mathbf{k} \rangle$ does not cross the
other band at the TRIM's. This assumption is not satisfied when 
a spin-orbit coupling is taken into account in the system having the time-reversal symmetry. 
 Here we have employed the model neglecting the spin as well as the spin-orbit coupling. 
It means that the band is simply doubled and the result is not changed, if the spin is taken into account.

The gradient of the band at $\mathbf{k}$ and $-\mathbf{k}$ is obtained as 
\begin{align}
 \nabla \epsilon_{-\mathbf{k}} &= \lim_{\mathbf{\delta k} \to \mathbf{0}} 
    \frac{ \epsilon_{-\mathbf{k}+\mathbf{\delta k}} -  \epsilon_{-\mathbf{k}}}{\mathbf{\delta k} }\\
    &= \lim_{\mathbf{\delta k} \to \mathbf{0}} 
    \frac{ \epsilon_{\mathbf{k}-\mathbf{\delta k}} -  \epsilon_{\mathbf{k}}}{\mathbf{\delta k} } \\
    &= - \nabla \epsilon_{\mathbf{k}},
\end{align}
i.e.,
\begin{equation}
\nabla\epsilon_{\mathbf{k}} = - \nabla\epsilon_{-\mathbf{k}}.
\end{equation}
Therefore, if the band does not cross the other band at $\mathbf{k}=\mathbf{0}$, we obtain
\begin{align}
 \nabla\epsilon_{\mathbf{0}} = - \nabla\epsilon_{\mathbf{0}} =0.
\end{align}
Similarly, by using
\begin{equation}
  \epsilon_{\mathbf{k}_{\mathrm{TRIM}}+\mathbf{q}} = \epsilon_{\mathbf{k}_{\mathrm{TRIM}}-\mathbf{q}},
\end{equation}
we obtain
\begin{equation}
\nabla\epsilon_{\mathbf{k}_{\mathrm{TRIM}}}=0.
\label{eqtrim}
\end{equation}
We conclude that if the system has the inversion symmetry, 
the TRIM may become any of a local maximum point, a local minimum point, an inflection point, or a saddle point. 

\section{Amplitudes of Fourier coefficients of the saw-tooth wave}
\label{AppendixE_0}	

The saw-tooth dependence with period $1/(f/2)$ of the magnetization as a function of
$1/h =x$ is given by 
\begin{equation}
 M^0(x) = 
  c x  \hspace{1cm} \mathrm{if} -\frac{1}{f} < x < \frac{1}{f}
\end{equation}
and
\begin{equation}
 M^0(x) = M^0(x+ \frac{2m}{f}),  \hspace{1cm} m=\pm 1, \pm 2, \cdots. 
\end{equation}
where $c$ is a constant. 

The saw-tooth function, $M^0(x)$, is given by the Fourier series
as
\begin{equation}
 M^0(x) = a^0_0+\sum_{\ell=1}^{\infty} \left[ a^0_{\ell} 
\cos (2\pi \ell \frac{f}{2} x) + b^0_{\ell} 
\sin(2\pi \ell \frac{f}{2} x)\right],
\end{equation}
where the coefficients, $a^0_{\ell}$ and $b^0_\ell$, are given by
\begin{align}
 a^0_\ell &=0, \\
 b^0_\ell &= \frac{2c}{f \pi} \frac{(-1)^{\ell+1}}{\ell}.
\end{align}

\begin{figure}[bt]
\begin{flushleft} \hspace{0.5cm}(a) \end{flushleft}\vspace{-0.6cm}
\includegraphics[width=0.5\textwidth]{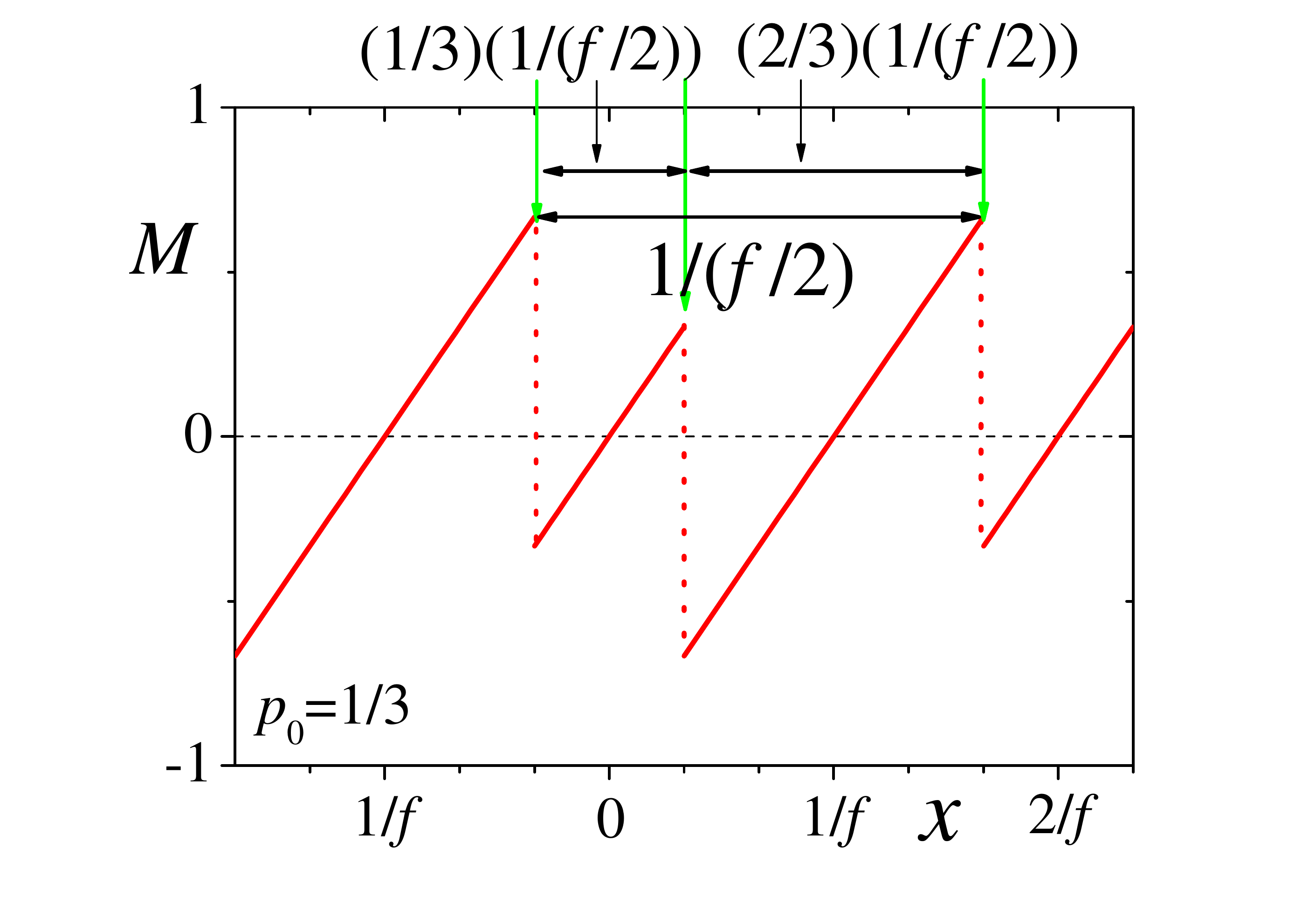}\vspace{-0.0cm}
\begin{flushleft} \hspace{0.5cm}(b) \end{flushleft}\vspace{-0.5cm}
\includegraphics[width=0.5\textwidth]{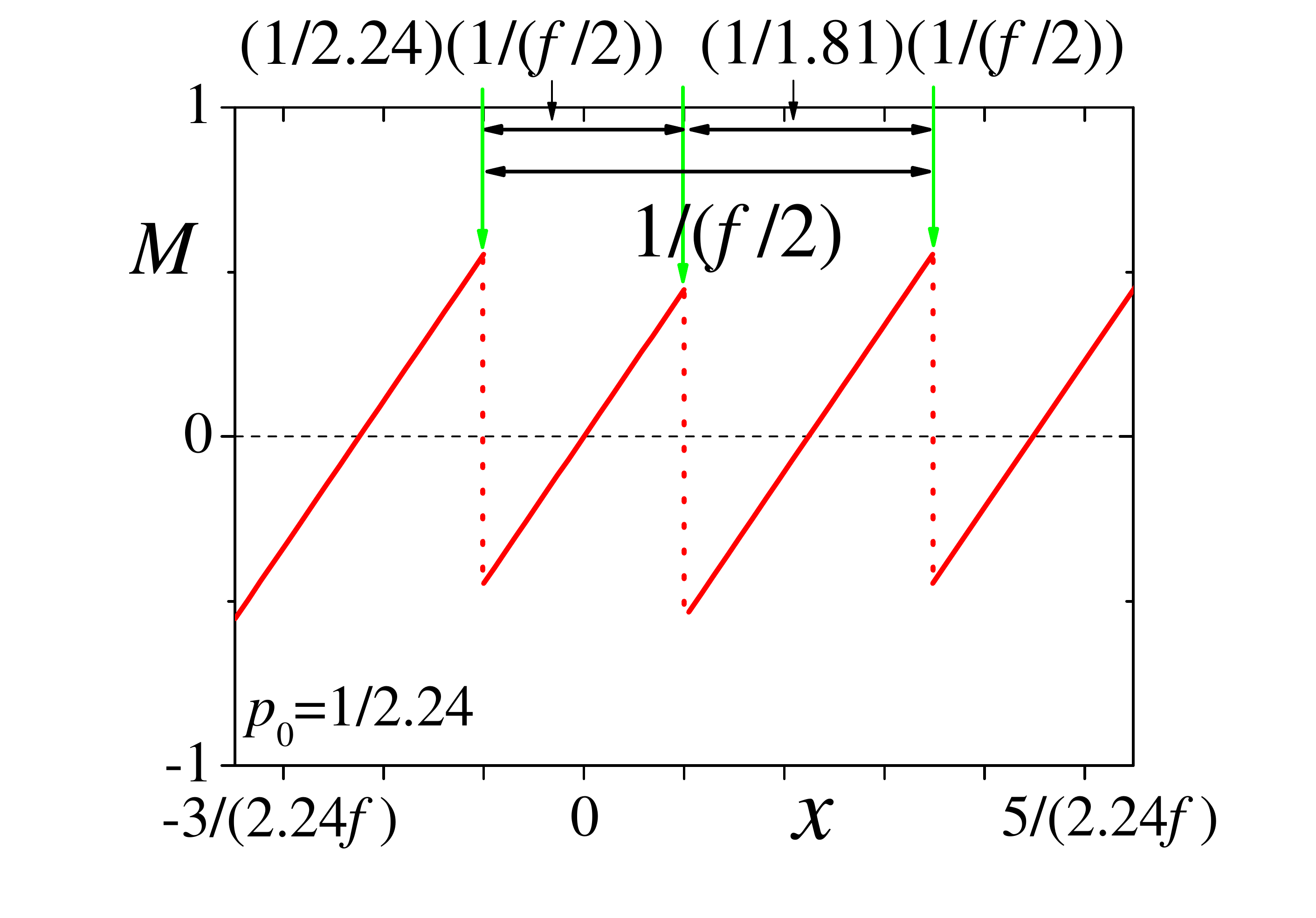}\vspace{-0.2cm}
\caption{
$M(x)$ as a function of $x$ at $p_0=1/3$ (a) and at $p_0=1/2.24$ (b), where we take $c=1$ and $q_0=1$. 
}
\label{fig_19}
\end{figure}

We now consider the modified saw-tooth dependence
\begin{equation}
 M(x) = \left\{ 
 \begin{array}{ll}
 c x, & \mathrm{if } \ -\frac{p_0}{f} < x< \frac{p_0}{f} \\
 c \left(x-\frac{q_0}{f} \right), 
& \mathrm{if } \   \frac{p_0}{f} 
< x < \frac{2-p_0}{f}
\end{array}
 \right.
\end{equation}
where we set $0<p_0\leq 1$, as shown in Figs. \ref{fig_19} (a) and (b). 
Then, we obtain
\begin{equation}
a_0 = \frac{2c(1-p_0)(1-q_0)}{f},
\end{equation}
\begin{equation}
 a_\ell = -\frac{2c(1-q_0)}{f \ell \pi} \sin(\ell p_0 \pi), 
\end{equation}
and
\begin{equation}
 b_\ell = - \frac{2c}{f \ell \pi} 
\cos(\ell p_0 \pi ).\label{bl}
\end{equation}

If $q_0=1$, we obtain $a_0=a_\ell=0$ and 
only sine components exist. 
In this case ($q_0=1$), $|b_3|$ in  
Eq. (\ref{bl}) is maximized at 
$p_0=1/3$ and $2/3$. 
When $p_0=1/3$, the fundamental period [$1/(f/2)$] is divided into 
$(1/3)(1/(f/2))$ and $(2/3)(1/(f/2))$, as shown in 
Fig. \ref{fig_19} (a). When $p_0=1/2.24$, the fundamental period is divided into 
$(1/2.24)(1/(f/2))$ and $(1/1.81)(1/(f/2))$, as shown in 
Fig. \ref{fig_19} (b). 
The similar situations are realized in $M_{\nu}$ at $P=$0, 0.2 and 1.0 [see Figs. \ref{fig35} (b), \ref{fig36} (b) and \ref{fig36_b} (b)]. The saw-tooth pattern of $M_{\nu}$ is
modified commensurately ($p_0=1/3$) at $P=0.2$ near the Lifshitz transition, but incommensurately at $P=0$ and 1.0. 
Therefore, the FTI at $3f_h/2$ in $M_{\nu}$, 
which is proportional to $|b_3|^2$, is enhanced at $P=0.2$.



\section{LK formula with a hole pocket and two small electron pockets}
\label{AppendixE}

We consider the case of one hole pocket with the area of $|A_h|$ and two same electron pockets with each area of $A_e/2$, where $A_e=|A_h|$. If we ignore the effect of the magnetic breakdown, the LK formula of Eq. (\ref{LK_0}) becomes  
\begin{eqnarray}
\frac{M^{\rm LK}}{M_0}&=&
\widetilde{A}_h\sum_{l=1}^{\infty}\frac{1}{l}\sin\left[2\pi l\left(\frac{f_h}{h}-\gamma_h\right)
\right]\nonumber \\
&+&2\widetilde{A}_e\sum_{l=1}^{\infty}\frac{1}{l}\sin\left[2\pi l\left(\frac{f_h}{2h}-\gamma_e\right)
\right],\label{LK_2}	
\end{eqnarray}
where we set $f_e/2=f_h/2$ for two electron pockets. 
By putting $\gamma_h=1/2$ and $\gamma_e=0$ into Eq. (\ref{LK_2}), 
we obtain 
\begin{eqnarray}
\frac{M^{\rm LK}}{M_0}&=&
\widetilde{A}_h\sum_{l=1}^{\infty}(-1)^{l}\frac{1}{l}\sin\left[2\pi l\left(\frac{f_h}{h}\right)
\right]\nonumber \\
&+&2\widetilde{A}_e\sum_{l=1}^{\infty}\frac{1}{l}\sin\left[2\pi l\left(\frac{f_h}{2h}\right)
\right] \\
&=&
\widetilde{A}_e \sum_{m=1}^{\infty}\biggr\{
\bigg[ (-1)^m\bigg(\frac{\widetilde{A}_h}{\widetilde{A}_e}\bigg) +1\bigg]\frac{1}{m}\sin\left[2\pi m\left(\frac{f_h}{h}\right)\right] \nonumber \\
&+&\frac{2}{2m-1}\sin\left[2\pi(2m-1)\left(\frac{f_h}{2h}\right) \right] 
\biggr\}.
\label{LK_5}
\end{eqnarray}
From Eq. (\ref{LK_5}), the FTIs are given by 
\begin{eqnarray}
{\rm FTI}(f=mf_h)&=&
\bigg|\frac{M_0\widetilde{A}_e}{m}\bigg[ (-1)^m\bigg(\frac{\widetilde{A}_h}{\widetilde{A}_e}\bigg) +1\bigg]\bigg|^2, \label{LK_8}\\
{\rm FTI}(f=m\frac{f_h}{2})&=&\bigg|\frac{2M_0\widetilde{A}_e}{2m-1}\bigg|^2,
\label{LK_9}
\end{eqnarray}
where $m=1, 2, 3, \cdots$.

Since the density of states is proportional to $\partial A_i(\varepsilon^0)/\partial \varepsilon^0$, 
by using Eq. (\ref{Ai}) we obtain 
\begin{eqnarray}
\frac{\widetilde{A}_h}{\widetilde{A}_e}=
\frac{|A_h|}{\frac{\partial A_h(\varepsilon^0)}{\partial \varepsilon^0}\big|_{\varepsilon^0=\mu}}
\frac{\frac{\partial A_e(\varepsilon^0)}{\partial \varepsilon^0}\big|_{\varepsilon^0=\mu}}{|A_e|}
=\frac{D_4(\varepsilon^0)}{ D_3(\varepsilon^0)}.\label{DOS}
\end{eqnarray}
In a special case of $\widetilde{A}_h/\widetilde{A}_e=1$ in Eq. (\ref{LK_8}), 
the FTIs at $f_h$, $3f_h$, $5f_h, \dots$ vanish.



\end{document}